\documentclass[10pt,preprint2]{aastex}

\newcommand{\esoq}{ESO\,215-G?009}
\newcommand{\mcg}{MCG--04-02-003}
\newcommand{\atg}{ATCA\,J061608--574552}
\newcommand{\hi}{H\,{\sc i}}
\newcommand{\hii}{H\,{\sc ii}}
\newcommand{\mlr}{${\cal M}_{\rm HI}/L_{\rm B}$}
\newcommand{\mtmr}{${\cal M}_{\rm HI}/{\cal M}_{\rm tot}$}

\newcommand{\tmlr}{${\cal M}_{\rm tot}/L_{\rm B}$}
\newcommand{\mls}{~${\cal M}_{\sun}/L_{\sun,{\rm B}}$}
\newcommand{\mmls}{${\cal M}_{\sun}/L_{\sun,{\rm B}}$}
\newcommand{\mB}{$m_{\rm B}$}

\newcommand{\MB}{$M_{\rm B}$}
\newcommand{\AB}{$A_{\rm B}$}
\newcommand{\AG}{$A_{\rm G}$}
\newcommand{\LB}{$L_{\rm B}$}
\newcommand{\Lsun}{~$L_{\sun,{\rm B}}$}

\newcommand{\Msun}{~${\cal M}_{\sun}$}

\newcommand{\FHI}{$F_{\rm HI}$}
\newcommand{\Speak}{$S_{\rm peak}$}
\newcommand{\whalf}{$w_{\rm 50}$}
\newcommand{\wxx}{$w_{\rm 20}$}
\newcommand{\MHI}{${\cal M}_{\rm HI}$}

\newcommand{\Mtot}{${\cal M}_{\rm tot}$}
\newcommand{\kms}{~km\,s$^{-1}$}
\newcommand{\kkms}{km\,s$^{-1}$}
\newcommand{\jks}{~Jy~km\,s$^{-1}$}
\newcommand{\jjks}{Jy~km\,s$^{-1}$}
\newcommand{\vmax}{$v_{\rm rot, max}$}
\newcommand{\vsys}{$v_{\rm sys}$}
\newcommand{\pI}{Paper~I}

\slugcomment{}

\shorttitle{High \mlr{} Galaxies in the Local Universe}
\shortauthors{Warren, Jerjen \& Koribalski}

\begin{document}
\title{Stellar and Gas properties of High \hi{} Mass-to-Light Ratio Galaxies in the Local Universe}
\author{Bradley E. Warren\altaffilmark{1} and Helmut Jerjen}
\affil{Research School of Astronomy and Astrophysics, Australian National 
       University, Mount Stromlo Observatory, Cotter Road, Weston ACT 2611, 
       Australia}
\email{bewarren@mso.anu.edu.au, jerjen@mso.anu.edu.au\\}
\and
\author{B\"arbel S. Koribalski}
\affil{Australia Telescope National Facility, CSIRO, PO Box 76, Epping NSW 1710, Australia}
\email{Baerbel.Koribalski@csiro.au}

\altaffiltext{1}{Affiliated with the Australia Telescope National Facility, CSIRO.}

\begin{abstract}
We present a multi-wavelength study ({\em BVRI} band photometry and \hi{} line interferometry) of nine late-type galaxies selected from the HIPASS {\it Bright Galaxy Catalog} on the basis of apparently high \hi{} mass-to-light ratios (3\mls{} $<$ \mlr{} $<$ 27\mls{}).  We found that most of the original estimates for \mlr{} based on available photographic magnitudes in the literature were too high, and conclude that genuine high \hi{} mass-to-light ratio ($>$5\mls{}) galaxies are rare in the Local Universe.  Extreme high \mlr{} galaxies like \esoq{} appear to have formed only the minimum number of stars necessary to maintain the stability of their \hi{} disks, and could possibly be used to constrain galaxy formation models.  They may to have been forming stars at a low, constant rate over their lifetimes.  The best examples all have highly extended \hi{} disks, are spatially isolated, and have normal baryonic content for their total masses but are deficent in stars.  This suggests that high \mlr{} galaxies are not lacking the baryons to create stars, but are underluminous as they lack either the internal or external stimulation for more extensive star formation.
\end{abstract}


\section{Introduction}
\label{sec:intro}

The number of low mass dark matter halos predicted by models of a CDM dominated Universe far exceeds the quantity of observed dwarf galaxies, typically by several orders of magnitude \citep*[see][]{kau93,moo99,kly99}.  In this context we consider a galaxy to be a dark matter halo that contains baryons.  Consequently the slope of dark matter mass functions generally rises much more steeply than observed galaxy luminosity functions (\citealp*{tre02,hil03,bla03}; but see also \citealp{bla04}).

While there are some physical processes that could help narrow this discrepancy \citep*{kly99,sha04}, the theoretical low mass halo frequency is not reduced enough to reconcile them with observations.  So we are left with the conclusions that either the current most favored cosmological models significantly over-estimate the number of low mass dark matter halos present in the Local Universe, or the observations have failed to find the vast majority of low mass galaxies to date.  If the latter were true, then it is important to look at why dwarf galaxies could be missed and how we could detect them.

Two reasons why galaxies might not have been found yet in optical surveys are that they could exhibit low stellar densities ($<$ 1\Msun{}\,pc$^{-2}$) or most of their baryons are in invisible form.  In the extreme case, they do not contain any baryonic matter at all and are in fact ``empty'' dark matter halos.  These possibilities could be dark matter halos in which the star formation from accreted gas has been halted, suppressed, never began, or there were simply no baryons to form stars to begin with.  Such objects lacking stars might be considered as being ``dark galaxies,'' which could be easily missed in surveys biased towards optical wavelengths \citep{dis76}.  However, dark galaxies have yet to be found in the blind \hi{} surveys such as HIPASS \citep{kor04,doy05}.

If they do exist, large numbers of low mass dark galaxies could naturally steepen the mass functions from observations, which are mostly derived from optical or near-infrared galaxy luminosity functions without consideration of other baryonic matter.  Recent large scale observations of neutral hydrogen gas (\hi{}) are now allowing mass functions to be derived based on non-stellar properties.  \citet{zwa03} produced one of the most extensive \hi{} mass functions based on a catalogue of the 1000 \hi{}-brightest galaxies in the Southern hemisphere and found a similar low mass end slope to other observational studies, again in contradiction with $\Lambda$CDM models.

Galaxies that have failed to convert most of their primordial gas into stars, and thus retained a high proportion of \hi{}, may well provide a partial solution.  While not entirely ``dark'' these galaxies are hard to detect optically, but may be detectable through 21cm line observations.  The \hi{} mass-to-light ratio (\mlr{}, which compares the \hi{} mass to the {\em B} band luminosity) of these objects could be significantly higher than the typical ratios measured for late-type galaxies, so that they would be in a lower mass bin in a luminosity function than they would be for a baryonic mass function.  If high \mlr{} galaxies existed in significant numbers then they could help correct the discrepancy in two ways, by including more previously unknown galaxies, and by shifting known galaxies to higher mass bins than they would be placed in with purely optical results.

An example of an extreme \hi{} mass-to-light ratio object, which could be described as a ``dim'' galaxy, is the nearby dwarf irregular \esoq{} with \mlr{} = $22 \pm 4$\mls{} \citep*[][ hereafter \pI{}]{war04}.  This faint low surface brightness dwarf irregular was found to be spatially isolated (1.7~Mpc from the nearest neighbor), with a low current star formation rate ($\la 2.5 \times 10^{-3}$\Msun{}\,yr$^{-1}$).  It has an extended regularly rotating \hi{} disk, which can be traced out to over six times the Holmberg radius of the optical galaxy, making it one of the most extended \hi{} envelopes relative to the optical extent.

\pI{} included an analysis of the \hi{} gas surface density of \esoq{}.  The azimuthally averaged surface density at all radii was below the critical gas surface density needed for large scale star formation as defined by the \citet{too64} stability criteria \citep{ken89,mar01}.  It was proposed in \citet{ver02} that a large fraction of low mass halos may form \citeauthor{too64} stable gas disks and become ``dark'' galaxies, possibly 95\% of objects with halo masses of $\la 10^{10}$\Msun{}.  If so \esoq{} would be just the tip of the iceberg.  If we can find more galaxies similar to \esoq{}, where the gas density after gravitational collapse is too low for efficient star formation, it may go some way to explaining the discrepancy between the dark matter halo mass function and the observed galaxy luminosity function.

To continue our study of the stellar and gas properties of galaxies with high \mlr{} we have selected a sample of nine galaxies from the BGC in the approximate range 3\mls{} $<$ \mlr{} $<$ 27\mls{}. In this paper, \S~\ref{sec:sample} looks at what was previously known about the sample galaxies. \S~\ref{sec:obs} summarizes our 21cm and optical observations.  \S\S~\ref{sec:radio} and \ref{sec:optp} present the results of the \hi{} line observations and optical photometry, respectively.  \S~\ref{sec:dis-indi} compares the properties of individual galaxies.  \S~\ref{sec:dis} contains the discussion of these results and the possible implications, while \S~\ref{sec:conc} gives our conclusions.

\section{Galaxy Selection}
\label{sec:sample}

\begin{deluxetable}{lccccccccc} 
\tabletypesize{\scriptsize}
\tablecaption{Summary of Previously Measured Galaxy Properties.
  \label{tab:prop}}
\tablewidth{0pt}
\tablehead{\colhead{Name} & \colhead{Center} & \colhead{Galactic} & \multicolumn{3}{c}{Bright Galaxy Catalog} & \colhead{LEDA} & \colhead{SFD98} & \multicolumn{2}{c}{BGC + LEDA} \\

 \colhead{HIPASS Name} & \colhead{$\alpha$(J2000.0)} & \colhead{$l$} & \colhead{\vsys{}} & \colhead{$D$} & \colhead{\FHI{}} & \colhead{\mB{}} & \colhead{\AB{}} & \colhead{\MB{}} & \colhead{\mlr{}} \\

    & \colhead{$\delta$(J2000.0)} & \colhead{$b$} & \colhead{(\kkms{})} & \colhead{(Mpc)} & \colhead{(\jjks{})} & \colhead{(mag)} & \colhead{(mag)} & \colhead{(mag)} & \colhead{(\mmls{})} \\
 
 \colhead{(1)} & \colhead{(2)} & \colhead{(3)} & \colhead{(4)} & \colhead{(5)} & \colhead{(6)} & \colhead{(7)} & \colhead{(8)} & \colhead{(9)} & \colhead{(10)}}
\startdata
 \mcg{}            & $00^{\rm h}\,19^{\rm m}\,11^{\rm s}$ & 62\fdg6 & $669 \pm 3$ & 9.5 & $16.0 \pm 2.5$ & $15.8 \pm 0.2$ & $0.08 \pm 0.01$ & $-14.1 \pm 0.2$ & $4.8 \pm 1.6$ \\
 HIPASS\,J0019--22 & --22\degr\,40\arcmin\,14\arcsec{} & --81\fdg4 \\ \\

 ESO\,473-G024     & $00^{\rm h}\,31^{\rm m}\,23^{\rm s}$ & 75\fdg7 & $540 \pm 4$ & 7.6 & $7.2 \pm 1.8$ & $16.2 \pm 0.2$ & $0.08 \pm 0.01$ & $-13.3 \pm 0.2$ & $3.0 \pm 1.3$ \\
 HIPASS\,J0031--22 & --22\degr\,46\arcmin\,02\arcsec{} & --83\fdg7 \\ \\

 ESO\,121-G020     & $06^{\rm h}\,15^{\rm m}\,53^{\rm s}$ & 266\fdg5 & $577 \pm 5$ & 4.1 & $14.1 \pm 2.9$ & $17.0 \pm 0.4$ & $0.17 \pm 0.03$ & $-11.3 \pm 0.4$ & $11 \pm 6$ \\
 HIPASS\,J0615--57 & --57\degr\,43\arcmin\,24\arcsec{} & --27\fdg3 \\ \\

 ESO\,428-G033     & $07^{\rm h}\,25^{\rm m}\,49^{\rm s}$ & 244\fdg2 & $1727 \pm 4$ & 19.5 & $12.8 \pm 2.7$ & $17.9 \pm 0.5$ & $1.10 \pm 0.18$ & $-14.7 \pm 0.5$ & $10 \pm 7$ \\
 HIPASS\,J0725--30B & --30\degr\,55\arcmin\,05\arcsec{} & --6\fdg9 \\ \\
 
 \esoq{}           & $10^{\rm h}\,57^{\rm m}\,30^{\rm s}$ & 284\fdg1 & $598 \pm 2$ & 4.2 & $104.4 \pm 11.5$ & $16.4 \pm 0.4$ & $0.95 \pm 0.15$ & $-12.6 \pm 0.4$ & $24 \pm 12$ \\
 HIPASS\,J1057--48 & --48\degr\,11\arcmin\,02\arcsec{} & 10\fdg5 \\ \\
 
 ESO\,572-G009     & $11^{\rm h}\,53^{\rm m}\,23^{\rm s}$ & 284\fdg1 & $1745 \pm 3$ & 20.4 & $8.5 \pm 1.9$ & $17.4 \pm 0.2$ & $0.16 \pm 0.03$ & $-14.3 \pm 0.2$ & $10 \pm 4$ \\
 HIPASS\,J1153--18 & --18\degr\,09\arcmin\,59\arcsec{} & 42\fdg6 \\ \\
 
 ESO\,505-G007     & $12^{\rm h}\,03^{\rm m}\,30^{\rm s}$ & 289\fdg5 & $1785 \pm 4$ & 20.8 & $20.5 \pm 3.2$ & $17.7 \pm 0.2$ & $0.36 \pm 0.06$ & $-14.2 \pm 0.2$ & $27 \pm 9$ \\
 HIPASS\,J1203--25 & --25\degr\,28\arcmin\,22\arcsec{} & 36\fdg1 \\ \\
 
 IC\,4212          & $13^{\rm h}\,12^{\rm m}\,09^{\rm s}$ & 312\fdg0 & $1484 \pm 2$ & 18.1 & $47.5 \pm 4.6$ & $16.4 \pm 1.1$ & $0.19 \pm 0.03$ & $-15.0 \pm 1.1$ & $23 ^{ +45}_{ -16}$ \\
 HIPASS\,J1311--06 & --06\degr\,58\arcmin\,31\arcsec{} & 55\fdg5 \\ \\
 
 ESO\,348-G009     & $23^{\rm h}\,49^{\rm m}\,23^{\rm s}$ & 349\fdg8 & $648 \pm 4$ & 8.4 & $13.4 \pm 2.2$ & $16.7 \pm 0.7$ & $0.06 \pm 0.01$ & $-13.0 \pm 0.7$ & $9 \pm 7$ \\
 HIPASS\,J2349--37 & --37\degr\,46\arcmin\,23\arcsec{} & --73\fdg2 \\
\enddata
\end{deluxetable}

\begin{figure*} 
  \includegraphics[width=8cm]{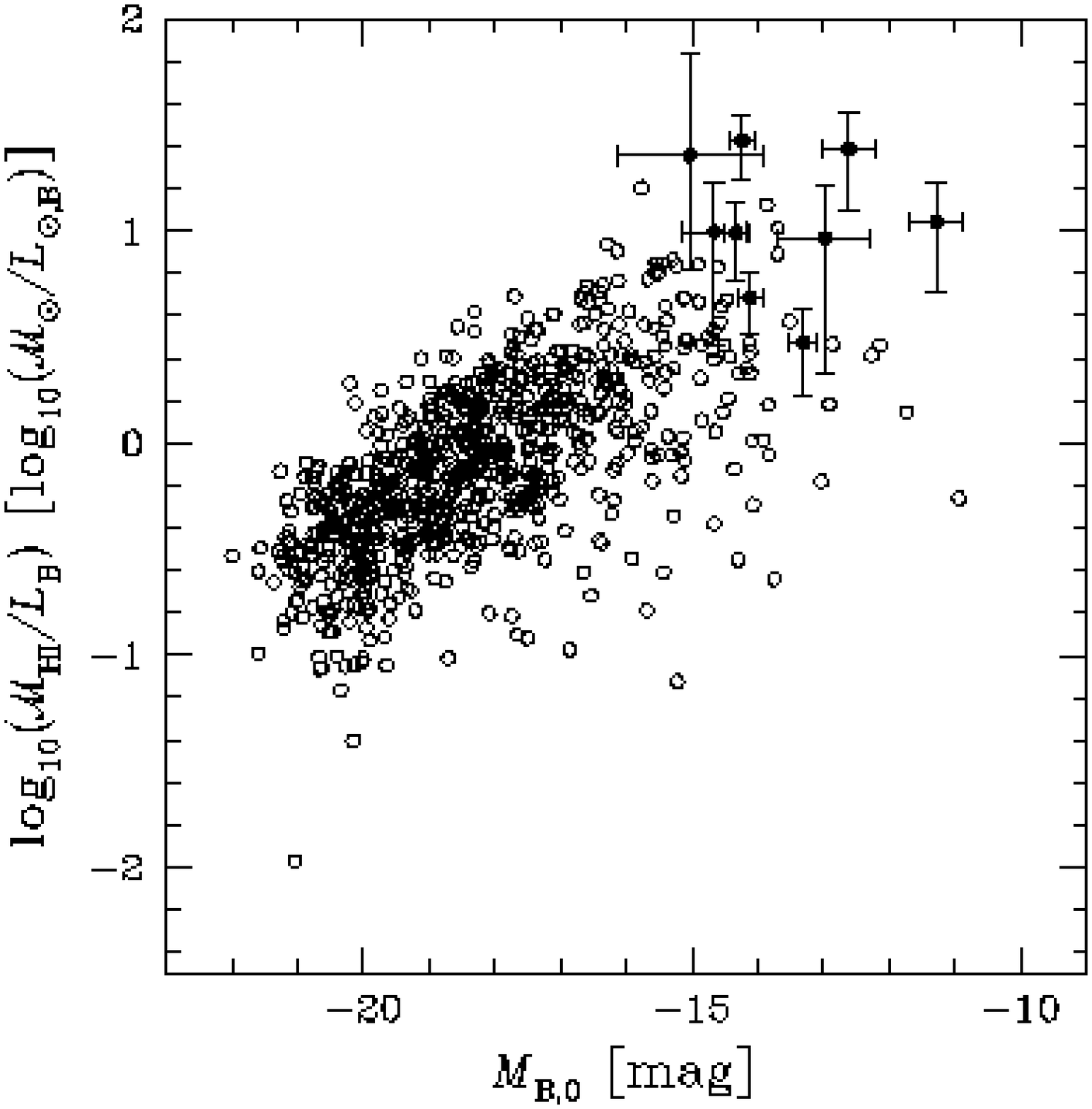}
\caption{\hi{} mass-to-light ratio versus absolute photographic {\em B} 
   magnitude for 789 galaxies in the HIPASS Bright Galaxy Catalog 
   \citep{kor04} that have blue apparent magnitudes listed in LEDA. The 
   filled circles with error bars highlight the positions of the nine galaxies examined 
   as part of this study.
\label{fig:mlmbt}}
\end{figure*}

The HIPASS Bright Galaxy Catalog \citep[][ hereafter BGC]{kor04} lists the 1000 \hi{}-brightest extragalactic sources (by \hi{} peak flux density) in the Southern hemisphere ($v_{\rm sys} < 8\,000$\kms{}).  Photometric and structural parameters for the BGC's {\it optical} counterparts were obtained in 2002 from the Lyon-Meudon Extragalactic Database \citep[LEDA,][ and references therein, now moved to HyperLEDA]{pat97} to study the statistical properties and various scaling relations of these galaxies.  First estimates of \hi{} mass-to-light ratios were obtained for 789 BGC galaxies that had mean apparent {\em B} band photographic magnitudes from LEDA, using the equation:  
\begin{equation}  
\frac{{\cal M}_{\rm HI}}{L_{\rm B}} = 1.5 \times 10^{-7} F_{\rm HI}~ 10^{0.4(m_{\rm B}-A_{\rm B})}~~\frac{{\cal M}_{\sun}}{L_{\sun,{\rm B}}} ,
\label{eqn:mlr}
\end{equation}
where \MHI{} is the \hi{} mass in solar units, \LB{} is the {\em B} band luminosity in solar units, \FHI{} is the integrated \hi{} flux density in \jjks{}, \mB{} is the apparent {\em B} magnitude, and \AB{} is the {\em B} band Galactic extinction. Extinction correction from the host galaxy is not included for reasons explained in \S~\ref{sec:dis-extinct}.  Fig.~\ref{fig:mlmbt} shows the log(\mlr{}) distribution for these galaxies as a function of their absolute {\em B} magnitude:
\begin{equation} 
    M_{\rm B,0} = m_{\rm B} - A_{\rm B} - 5\log(D) - 25 ~~{\rm mag} ,
\label{eqn:abmag}
\end{equation}
where the galaxy distances, $D$ (Mpc), were calculated from the Local Group velocities given in the BGC.  Throughout this paper we adopt H$_0$ = 75\kms{}\,Mpc$^{-1}$.

The relation between these two quantities seems to suggest that many of the low luminosity galaxies listed in the BGC have high \mlr{}, up to 27\mls{}, well above typical values for late-type galaxies of less than 1\mls{} \citep[median \mlr{} of 0.78\mls{} in][ for type Sm/Im galaxies]{rob94}.  The nine filled circles mark the positions of the galaxies subject to our detailed follow up observations, including \esoq{} (\pI{}).  The galaxies were chosen for various reasons, mostly because of a high estimated \mlr{}, but also for reasons of unusual morphology, and after initial ATCA follow up observations showed some unexpected results (see \S~\ref{sec:structure}).

Previously measured properties of those nine galaxies are summarised in Table~\ref{tab:prop}.  The columns are as follows: (1) commonly used galaxy name and HIPASS source name; (2) J2000.0 Right Ascension and Declination as given in the RC3 \citep{dev91}; (3) Galactic longitude and latitude; (4) \hi{} systemic velocity as given in the BGC; (5) galaxy distance derived from the velocity relative to the barycentre of the Local Group as given in the BGC; (6) total integrated \hi{} flux density as given in the BGC; (7) apparent {\em B} band photographic magnitude as listed in LEDA; (8) \citet*[ hereafter SFD98]{sch98} Galactic dust extinction in the {\em B} band; (9) absolute {\em B} band magnitude calculated as in eqn.~\ref{eqn:abmag} using the LEDA magnitude and SFD98 extinction; (10) preliminary estimate of \hi{} mass-to-light ratio calculated using eqn.~\ref{eqn:mlr} from the LEDA, BGC and SFD98 data.

\section{Observations} 
\label{sec:obs}

Each galaxy was observed in two different wavelength regimes. Optical CCD photometry was obtained with the Australian National University (ANU) 2.3-meter Telescope at the Siding Spring Observatory.  \hi{} (21cm) line data were obtained with the Australia Telescope Compact Array (ATCA).

\begin{deluxetable}{lcccccccc} 
\tabletypesize{\scriptsize}
\tablecaption{Summary of Observations for each Galaxy. 
\label{tab:obs}}
\tablewidth{0pt}
\tablehead{
   \colhead{Name} & \multicolumn{3}{c}{Optical} & \colhead{\phantom{0000}} & \multicolumn{4}{c}{Radio} \\
   \colhead{} & \colhead{Band} & \colhead{Exposure Time} & \colhead{Seeing} & & \colhead{Arrays} & \colhead{Time On Source} & \colhead{Central Freq.} & \colhead{Phase Cal.} \\
   \colhead{} & \colhead{} & \colhead{(seconds)} & \colhead{(arcsec)} & & \colhead{} & \colhead{(hours)} & \colhead{(MHz)} & \colhead{} \\
  \colhead{(1)}  & \colhead{(2)} & \colhead{(3)} & \colhead{(4)} & & \colhead{(5)} & \colhead{(6)} & \colhead{(7)} & \colhead{(8)} }
\startdata
 \mcg{}        & {\em B} & 3000 ($10\times300$) & 2\farcs2 & & H75B & $\sim1.5$ & 1417 & PKS\,0023--263 \\
               & {\em V} & 2400 ($8\times300$) & 2\farcs0 & & H168B & $\sim8.6$ & 1417 & PKS\,0023--263 \\
               & {\em R} & 1800 ($6\times300$) & 1\farcs9 \\
               & {\em I} & 1800 ($6\times300$) & 2\farcs2 \\ \\

 ESO\,473-G024 & {\em B} & 3000 ($10\times300$) & 2\farcs5 & & H75B & $\sim1.5$ & 1417 & PKS\,0023--263 \\
               & {\em V} & 2400 ($8\times300$) & 3\farcs3 & & H168B & $\sim1.0$ & 1417 & PKS\,0023--263 \\
               & {\em R} & 1800 ($6\times300$) & 3\farcs3 \\
               & {\em I} & 1800 ($6\times300$) & 2\farcs0 \\ \\

 ESO\,121-G020 & {\em B} & 3000 ($10\times300$) & 2\farcs1 & & 750D & $\sim10.5$ & 1417 & PKS\,0407--658 \\
               &  &  &  & & 1.5B & $\sim10.9$ & 1417 & PKS\,0407-658 \\
               & {\em R} & 1800 ($6\times300$) & 1\farcs9 & & EW352 & $\sim2.4$ & 1416 & PKS\,0537--441 \\ \\

 ESO\,428-G033 & {\em B} & 3000 ($5\times600$) & 2\farcs2 & & 750D & $\sim10.6$ & 1412 & PKS\,0614--349 \\
               & {\em V} & 2400 ($4\times600$) & 2\farcs0 & & 1.5B & $\sim10.3$ & 1412 & PKS\,0614--349 \\
               & {\em R} & 1800 ($3\times600$) & 1\farcs9 & & EW352 & $\sim1.1$ & 1414 & PKS\,0614--349 \\
               & {\em I} & 1800 ($3\times600$) & 3\farcs0 & & EW367B & $\sim0.8$ & 1414  & PKS\,0614--349 \\ \\

 \esoq{}       & {\em B} & 3000 ($10\times300$) & 1\farcs9 & & EW352 & $\sim11.6$ & 1417 & PKS\,1215--457 \\
               & {\em V} & 2400 ($8\times300$) & 1\farcs9 & & 750A & $\sim10.6$ & 1417 & PKS\,1215--457 \\
               & {\em R} & 1800 ($6\times300$) & 1\farcs8 & & 6A & $\sim11.5$ & 1417 & PKS\,1215--457 \\
               & {\em I} & 1800 ($6\times300$) & 2\farcs0 \\ \\

 ESO\,572-G009 & {\em B} & 1800 ($3\times600$) & 1\farcs9 & & H75B & $\sim1.0$ & 1412 & PKS\,1127--145 \\
               & {\em V} & 2400 ($8\times300$) & 2\farcs7 \\
               & {\em R} & 1800 ($6\times300$) & 2\farcs2 \\ \\

 ESO\,505-G007 & {\em B} & 3000 ($10\times300$) & 2\farcs0 & & H75B & $\sim1.0$ & 1412 & PKS\,1127--145 \\
               & {\em V} & 1800 ($6\times300$) & 2\farcs2 & & H168B & $\sim1.9$ & 1412 & PKS\,1127--145 \\
               & {\em R} & 1800 ($6\times300$) & 2\farcs2 & & H75B & $\sim9.2$ & 1412 & PKS\,1151--348  \\ \\

 IC\,4212      & {\em B} & 3000 ($5\times600$) & 2\farcs9 & & H75B & $\sim8.8$ & 1413 & PKS\,1308--220 \\
               & {\em V} & 2400 ($4\times600$) & 2\farcs3 \\
               & {\em R} & 1800 ($3\times600$) & 1\farcs9 \\ \\

 ESO\,348-G009 & {\em B} & 3000 ($10\times300$) & 1\farcs7 & & 750D & $\sim10.6$ & 1417 & PKS\,0008--421 \\
               & {\em V} & 2400 ($8\times300$) & 1\farcs4 & & 1.5B & $\sim9.8$ & 1417 & PKS\,0008--421 \\
               & {\em R} & 1800 ($6\times300$) & 1\farcs5 & & EW352 & $\sim1.7$ & 1417 & PKS\,0008--421 \\
               & {\em I} & 1800 ($6\times300$) & 1\farcs4 \\
\enddata
\end{deluxetable}

\subsection{Radio Observations}
\label{sec:obs-rad}

ATCA \hi{} line observations of the selected galaxies were carried out between June 2002 and June 2003.  The galaxies ESO\,121-G020, ESO\,428-G033, and ESO\,348-G009 ($\delta < -30$\degr{}) were observed for $2 \times \sim12$ hours in different East-West arrays, while \esoq{} was observed for $3 \times \sim$12 hours.  For the other five galaxies ($\delta > -30$\degr{}) we used the compact hybrid arrays that include antennas on the Northern spur, resulting in a rather large synthesized beams.  The galaxies \mcg{}, ESO\,505-G007, and IC\,4212 were observed for $\sim$10 hours, while ESO\,572-G009 and ESO\,473-G024 were only observed in snapshot mode ($\sim1 - 2$ hours taken over a 12 hour period).  \hi{} snapshot observations that were initially taken for the other galaxies (except \esoq{} and IC\,4212) were added to the other observations.  Details of the \hi{} observations for each galaxy are given in Table~\ref{tab:obs}. The columns are as follows: (5) ATCA configurations used; (6) approximate time on source for each array; (7) central observing frequency; and (8) phase calibrator.  We used a bandwidth of 8 MHz, divided into 512 channels, resulting in a channel width of 3.3\kms{}.  The velocity resolution of the \hi{} data is $\sim$4\kms{}.  The primary calibrator for all observations was PKS\,1934--638.

Data reduction and analysis were performed with the {\sc MIRIAD} package using standard procedures, with further analysis using {\sc AIPS}, {\sc GIPSY}, and {\sc KARMA}.  Channels affected by Galactic \hi{} emission were discarded where appropriate.  After continuum subtraction, the \hi{} data were Fourier-transformed using ``natural'' weighting and a channel width of 4\kms{}.  The data were cleaned and restored with the synthesized beam (the size of which is given in Table~\ref{tab:radio} for each galaxy).  Primary beam correction was applied.  \hi{} distributions (0th moment) were obtained for all galaxies using cutoffs between 3 and 4$\sigma$ and are shown in Fig.~\ref{fig:himap}, while the corresponding \hi{} spectra are shown in Fig.~\ref{fig:hispectra}.  For ESO\,121-G020, ESO\,428-G033, \esoq{}, and ESO\,348-G009 mean velocity fields and dispersion maps were also produced as they had sufficient resolution.

\subsection{Optical Photometry}
\label{sec:obs-opt}

{\em BVRI} band CCD images were obtained at the 2.3m telescope as a series of 300\,s or 600\,s exposures during observing runs between April 2002 and February 2004 using the Nasmyth Imager (SITe $1124 \times 1024$ thinned CCD).  The imager has a circular field of view with a diameter of 6\farcm62 and a pixel size of 0\farcs59.  Table~\ref{tab:obs} gives a summary of the observations taken for each galaxy in each band.  The columns are as follows: (2) broad band (Cousins) filters used; (3) total observing time in each of the optical bands including the number of individual exposures; and (4) atmospheric seeing in the final optical images.  Most observations were taken at low airmass.  Twilight sky flat fields in all bands and bias images were obtained at the same time.  On each photometric night several \citet{lan92} standard stars were taken together with shallow 120\,s {\em BVRI} images of the galaxy fields to perform the photometric calibration of the deeper images.  

\begin{figure*} 
\begin{tabular}{@{}lll@{}}
 \includegraphics[width=5.5cm]{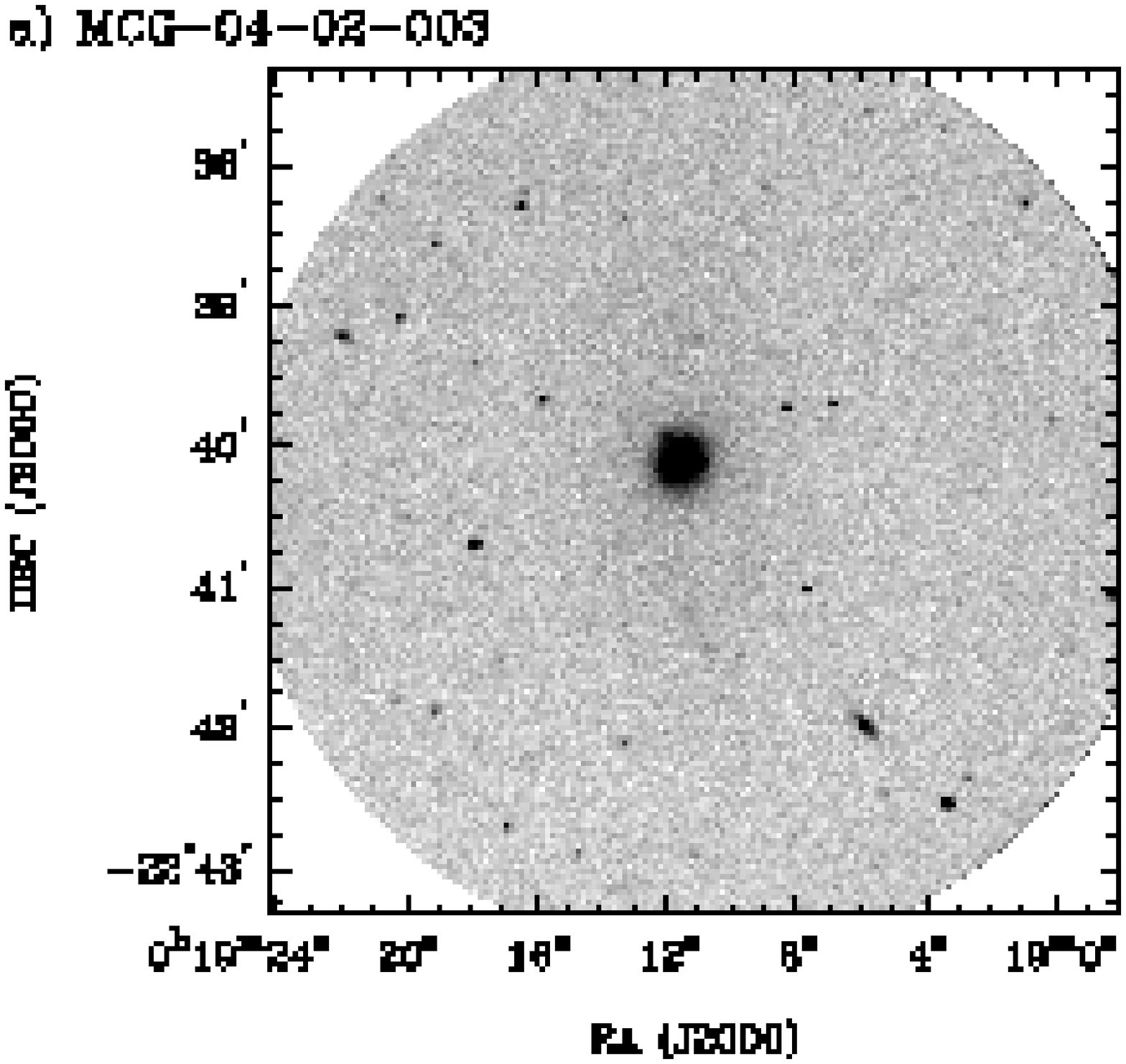} &
 \includegraphics[width=5.5cm]{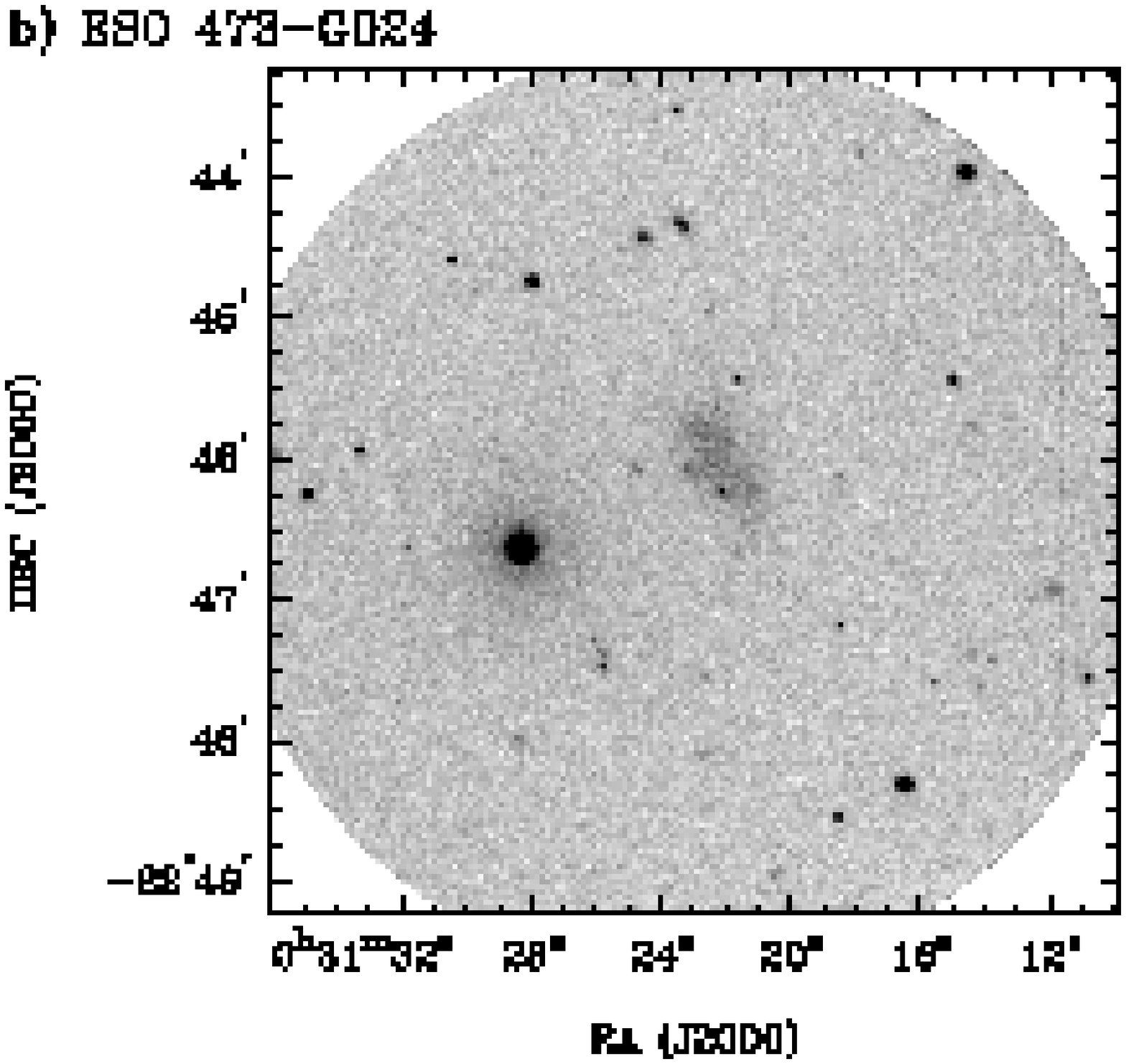} &
 \includegraphics[width=5.5cm]{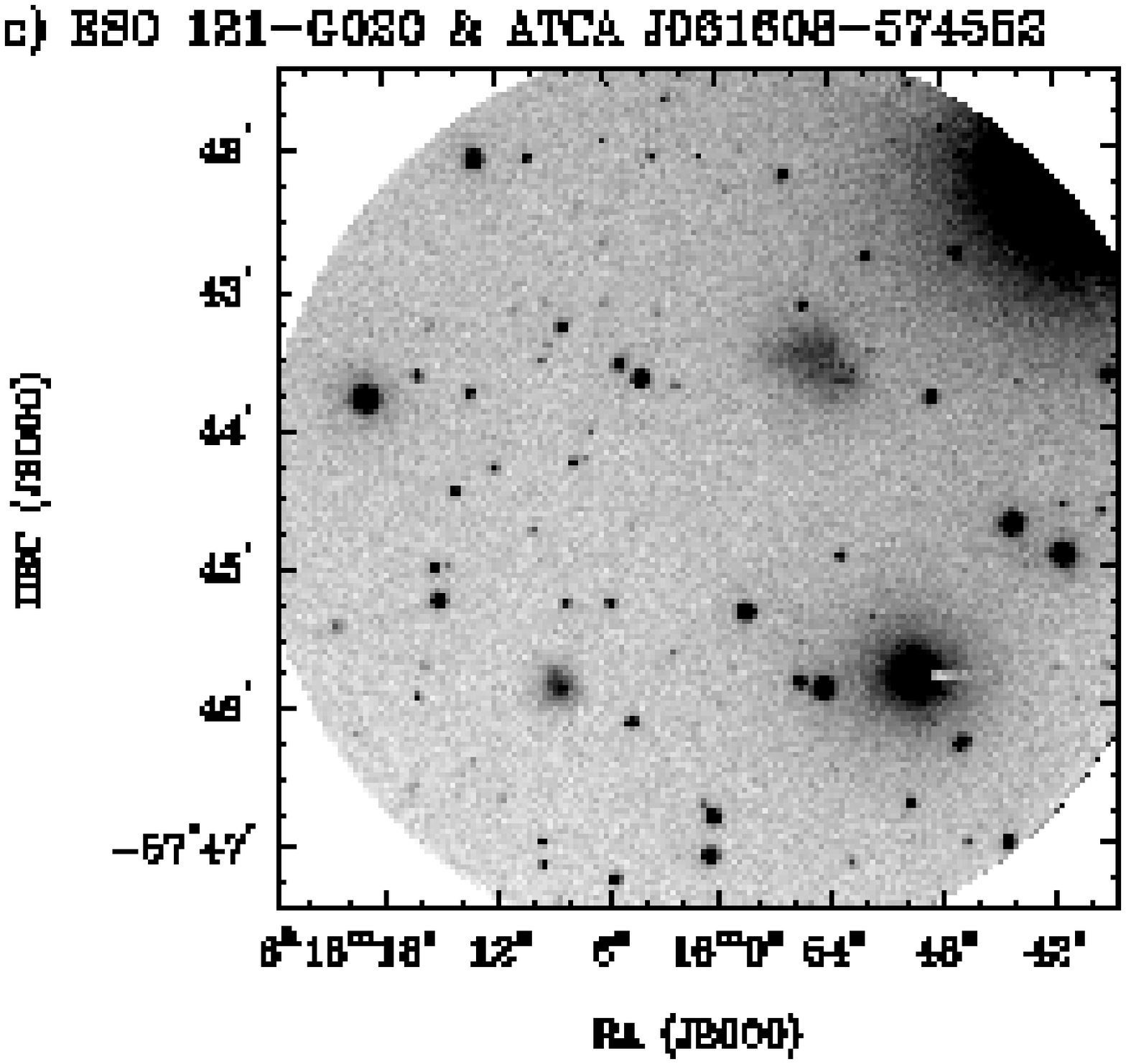} \\ \\

 \includegraphics[width=5.5cm]{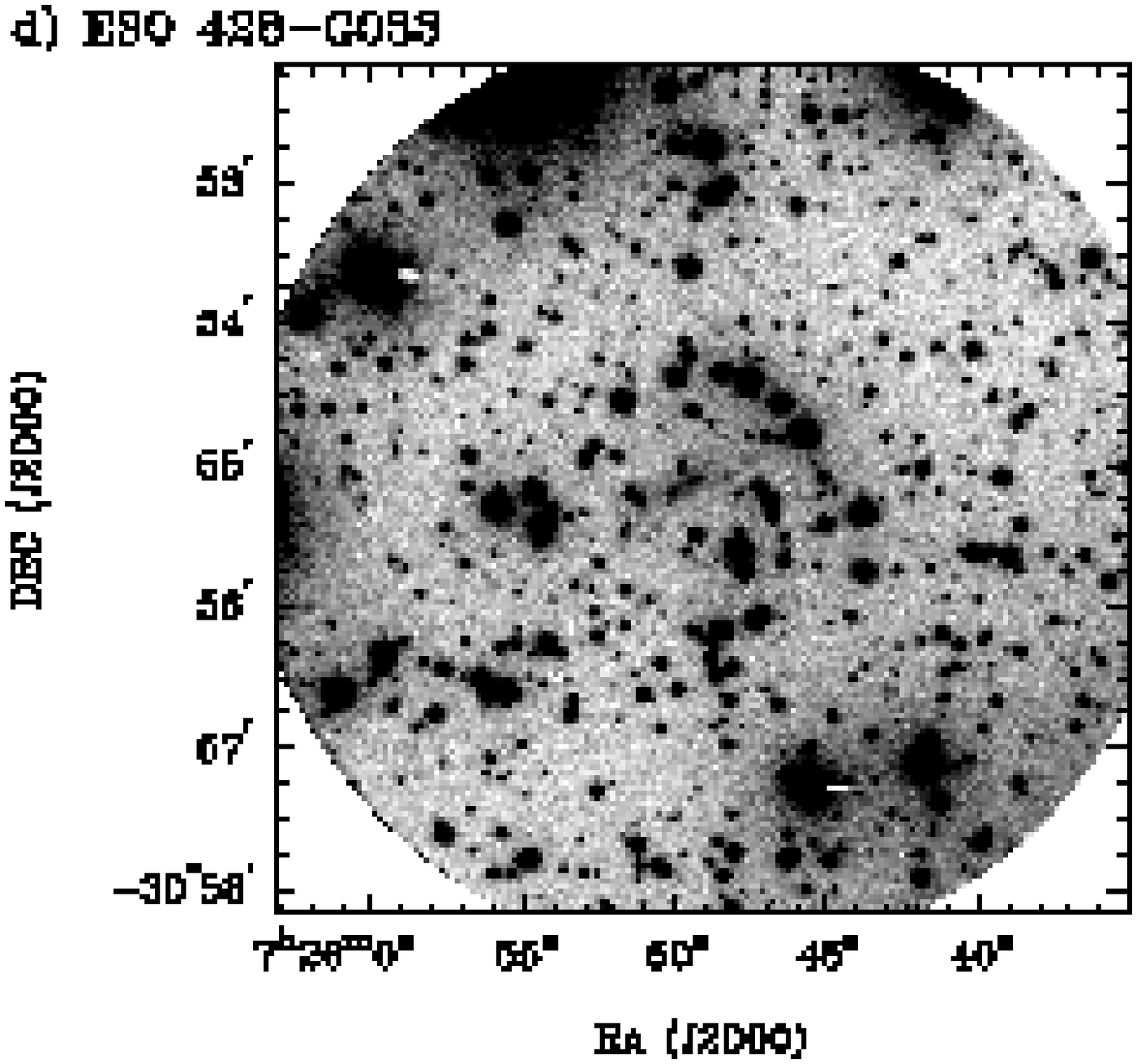} &
 \includegraphics[width=5.5cm]{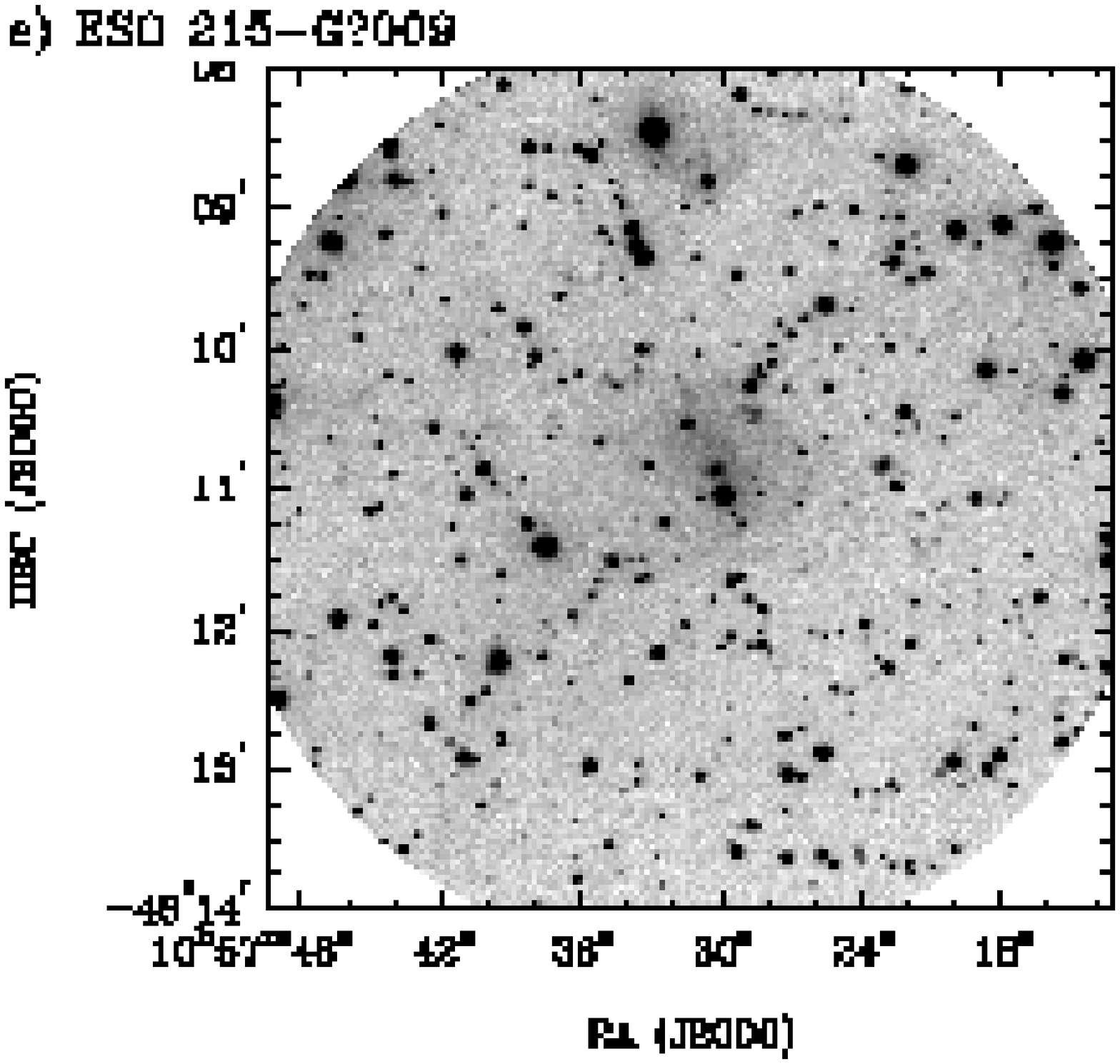} & 
 \includegraphics[width=5.5cm]{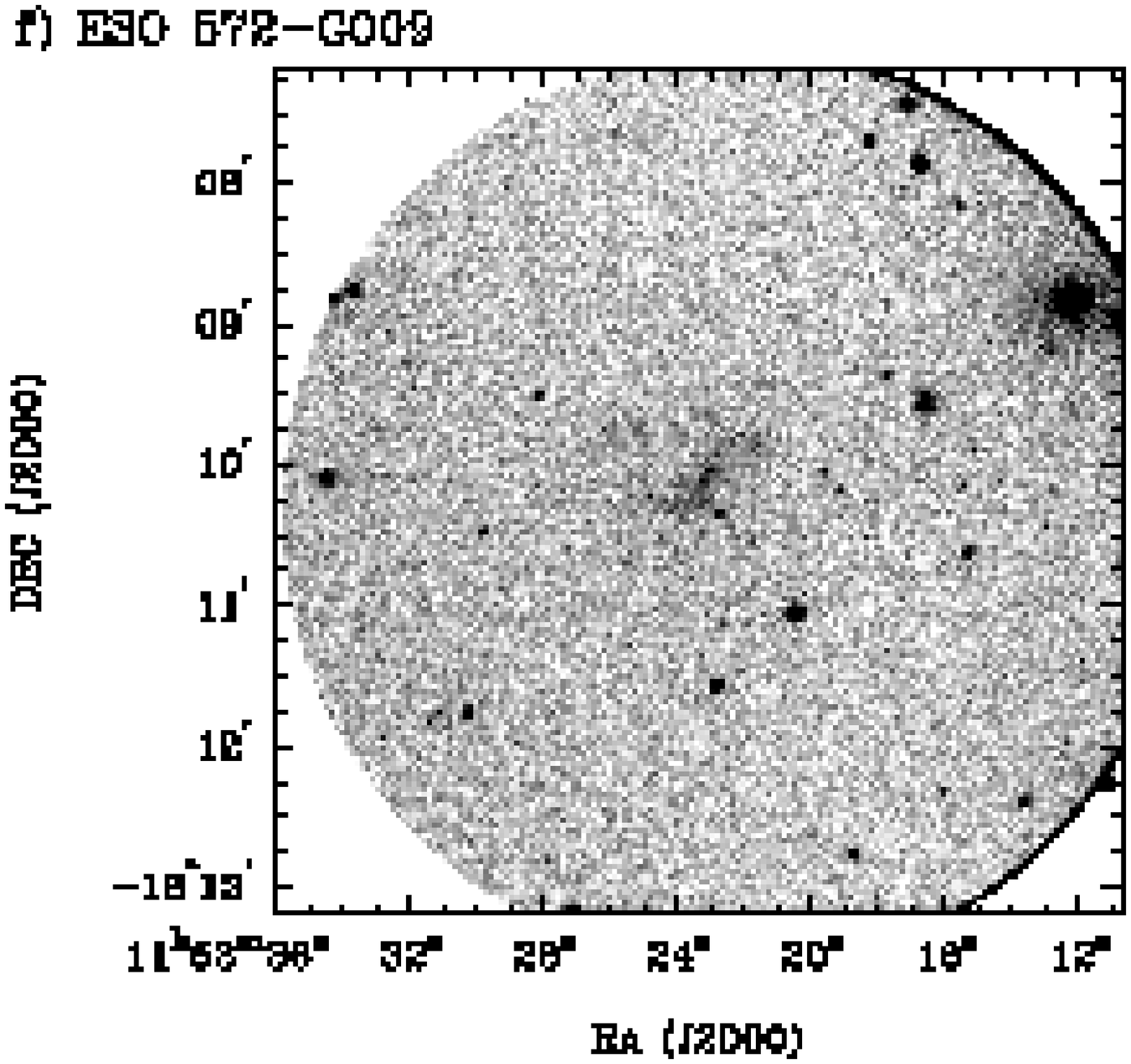} \\ \\

 \includegraphics[width=5.5cm]{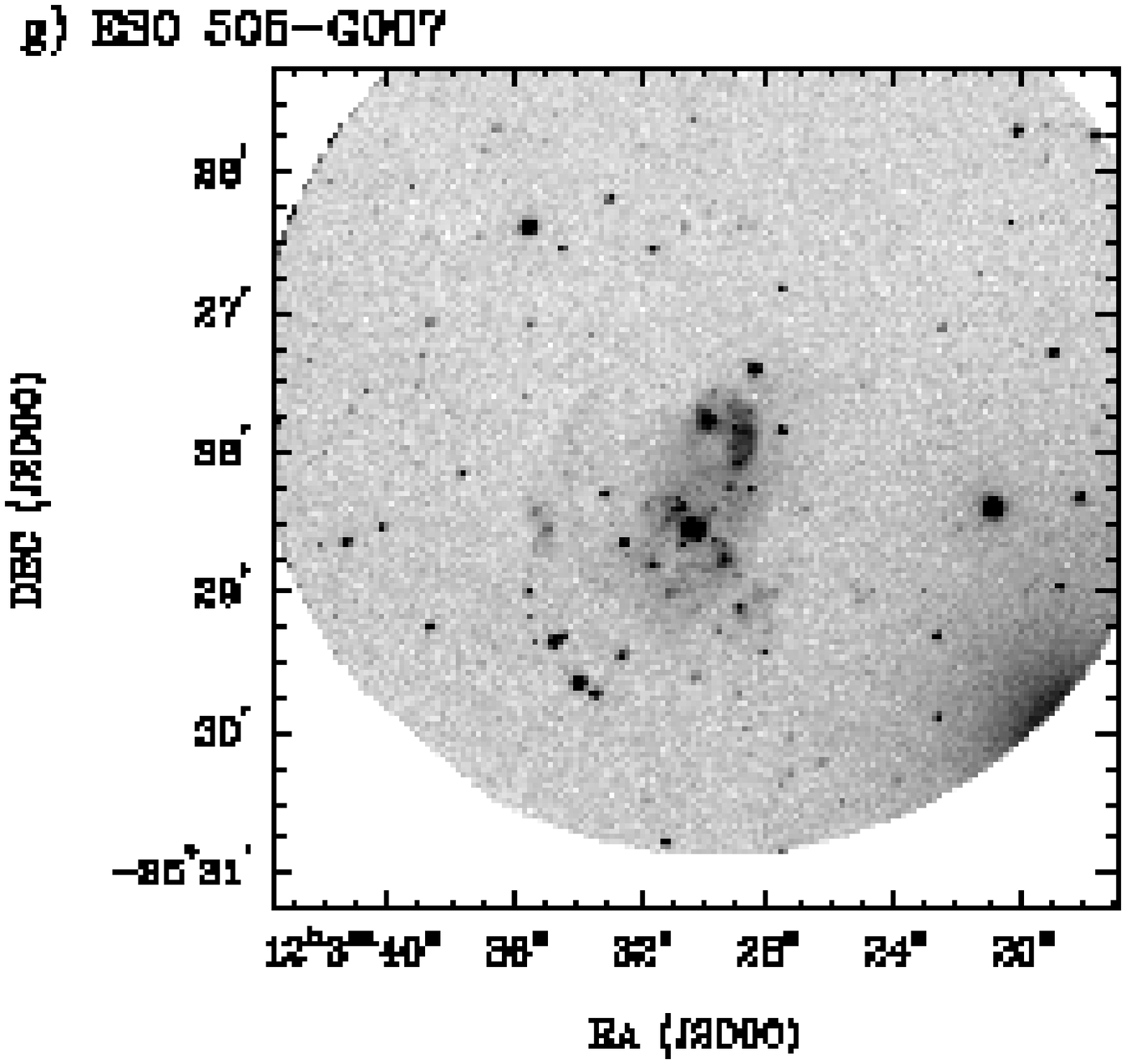} &
 \includegraphics[width=5.5cm]{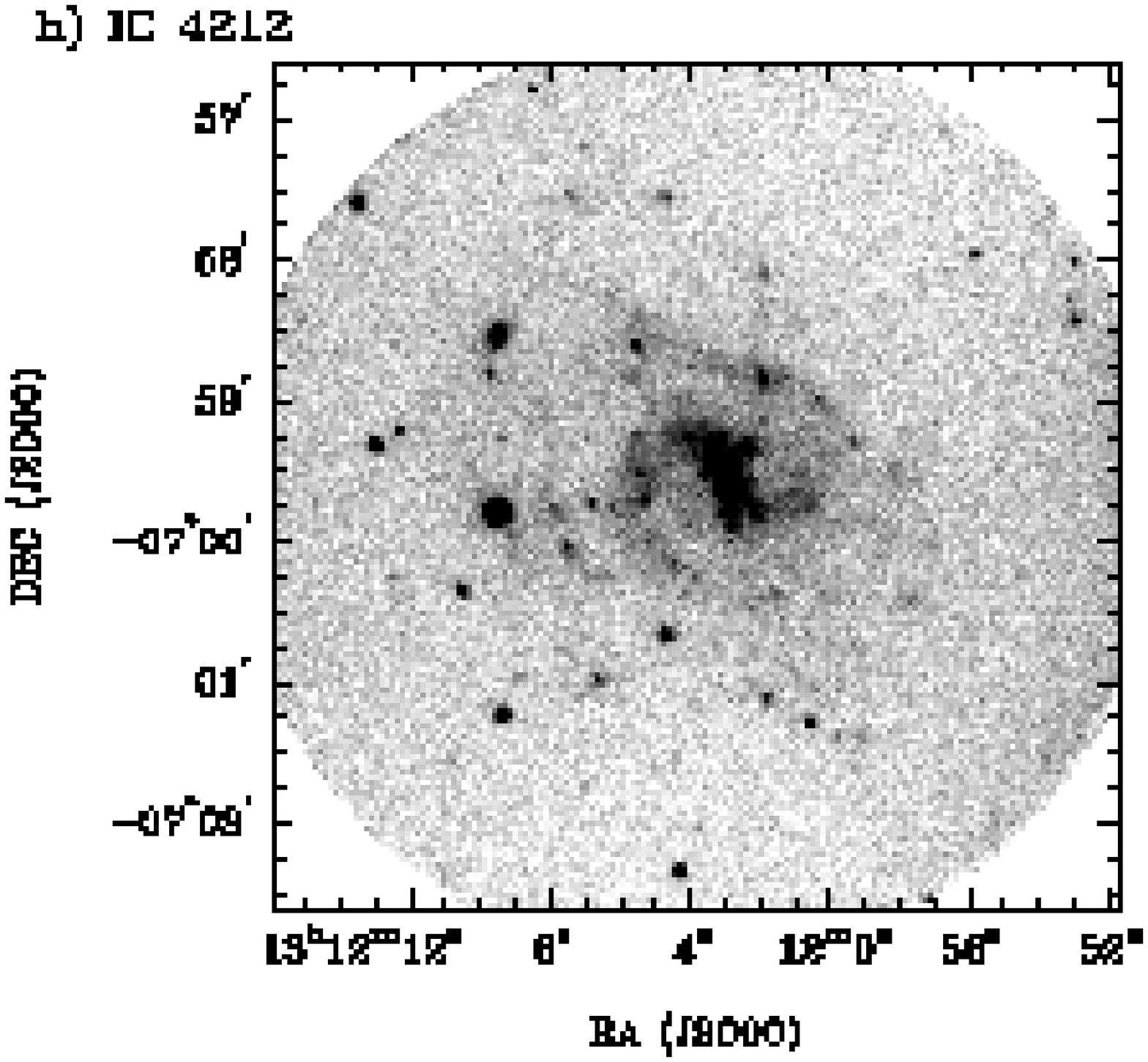} & 
 \includegraphics[width=5.5cm]{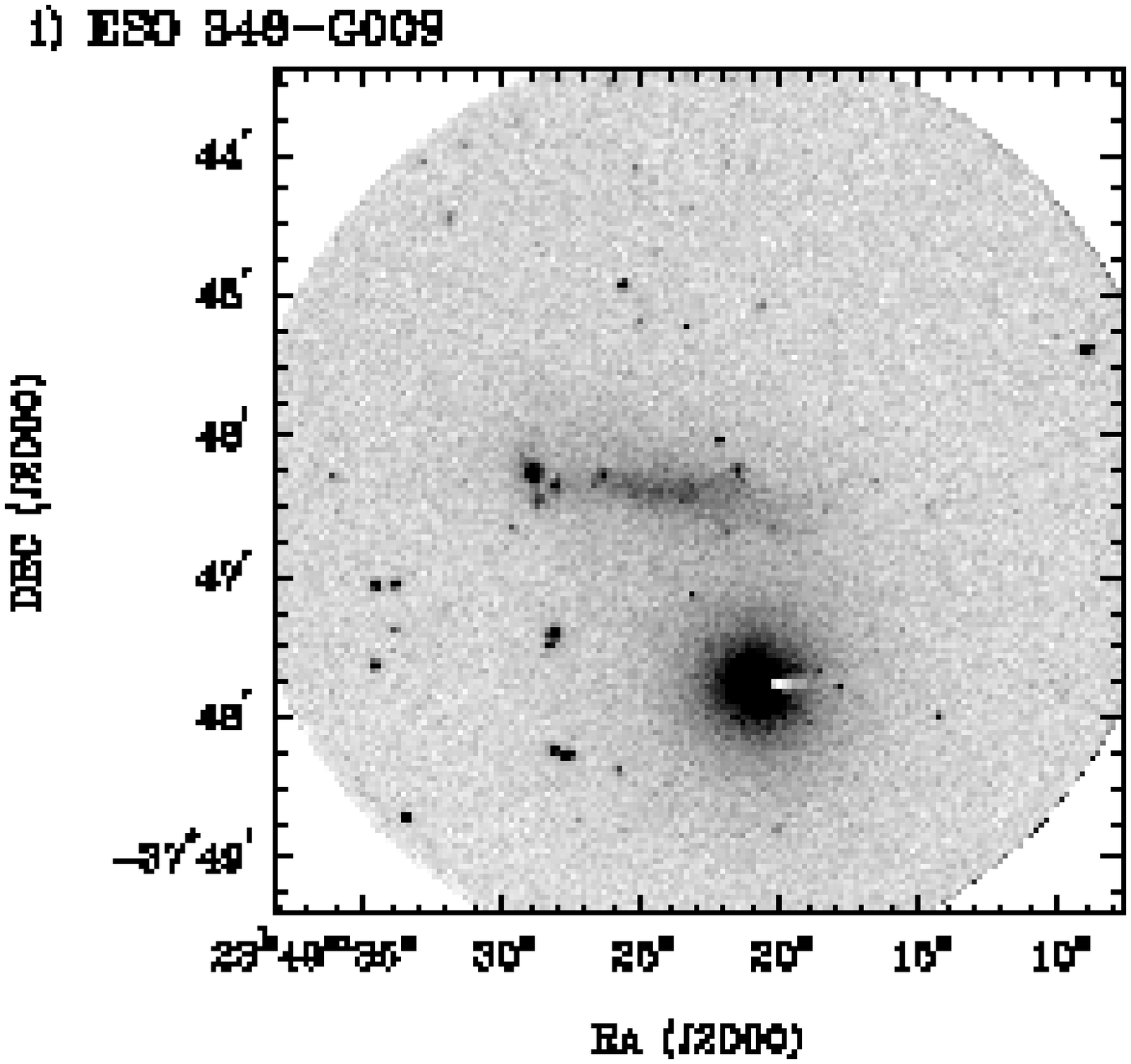} \\
\end{tabular}
\caption{Observed deep {\em B} band images of all galaxies.  The panels show {\em a}) \mcg{}, {\em b}) ESO\,473-G024, {\em c}) ESO\,121-G020 \& \atg{}, {\em d}) ESO\,428-G033, {\em e}) \esoq{}, {\em f}) ESO\,572-G009, {\em g}) ESO\,505-G007, {\em h}) IC\,4212, and {\em i}) ESO\,348-G009.
\label{fig:optimage}}
\end{figure*}

Data reduction, photometric calibration, and analysis were carried out within IRAF using standard procedures.  After overscan subtraction, bias subtraction, and flatfielding, individual sets of images were registered and the sky level was subtracted. The images for each band were then combined into a single image (to increase signal-to-noise, remove cosmic rays, etc.) and the photometric calibration applied.  Fig.~\ref{fig:optimage} shows the resulting master images in the {\em B} band for all nine galaxies.

Foreground stars were removed by replacing them with the surrounding sky so that only the galaxy remained.  Special care was taken to restore the light distribution under any stars superimposed onto the galaxies, e.g. using the mirror image from across the galaxies center.  For more details of this technique see \citet{jer03}.  To illustrate the final result Fig.~\ref{fig:starsub} shows the {\em B} band images of the galaxies after cleaning.

\section{Radio Properties}
\label{sec:radio}

ATCA \hi{} follow-up observations are needed to obtain accurate positions of the targeted HIPASS BGC sources and to reliably identify their optical counterparts.  For those galaxies where we have high angular resolution \hi{} observations (ESO\,121-G020, ESO\,428-G033, \esoq{}, and ESO\,348-G009), we also analyse their \hi{} structure and kinematics, including the galaxy rotation curve.

\subsection{\hi{} Structure}
\label{sec:structure}

\begin{deluxetable}{lcccccc} 
\tabletypesize{\scriptsize}
\tablecaption{ATCA \hi{} Results.
  \label{tab:radio}}
\tablewidth{0pt}
\tablehead{ \colhead{Name} & \colhead{Beam (\hi{})} & \colhead{\Speak{}} & \colhead{\FHI{}} & \colhead{\vsys{}} & \colhead{\whalf{}} & \colhead{\wxx{}} \\

 \colhead{} & \colhead{(arcsec)} & \colhead{(Jy)} & \colhead{(\jjks{})} & \colhead{(\kkms{})} & \colhead{(\kkms{})} & \colhead{(\kkms{})} \\

 \colhead{(1)} & \colhead{(2)} & \colhead{(3)} & \colhead{(4)} & \colhead{(5)} & \colhead{(6)} & \colhead{(7)} }
\startdata
 \mcg{}        & $193\times129$ & $0.232 \pm 0.009$ & $16.2 \pm 0.6$ & $670 \pm 2$ & $112 \pm 2$ & $126 \pm 2$ \\ \\

 ESO\,473-G024 & $359\times205$ & $0.17 \pm 0.02$ & $5.7 \pm 0.9$ & $542 \pm 3$ & $37 \pm 2$ & $50 \pm 3$ \\ \\

 ESO\,121-G020 & $32\times30$   & $0.204 \pm 0.006$ & $9.1 \pm 0.3$ & $583 \pm 2$ & $47 \pm 2$ & $61 \pm 4$ \\
 ~~\atg{}      & $32\times30$   & $0.075 \pm 0.006$ & $2.7 \pm 0.2$ & $554 \pm 4$ & $36 \pm 3$ & $56 \pm 8$ \\ \\

 ESO\,428-G033 & $45\times30$   & $0.179 \pm 0.005$ & $14.5 \pm 0.3$ & $1728 \pm 2$ & $94 \pm 2$ & $110 \pm 2$ \\ \\

 \esoq{}       & $38\times35$   & $2.128 \pm 0.005$ & $122 \pm 4$ & $597 \pm 1$ & $64 \pm 2$ & $90 \pm 4$ \\ \\

 ESO\,572-G009 & $477\times327$ & $0.25 \pm 0.03$ & $7.2 \pm 1.3$ & $1740 \pm 4$ & $36 \pm 2$ & $49 \pm 2$ \\ \\

 ESO\,505-G007 & $413\times304$ & $0.347 \pm 0.009$ & $21 \pm 3$ & $1776 \pm 2$ & $69 \pm 3$ & $88 \pm 5$ \\ \\

 IC\,4212      & $444\times357$ & $0.342 \pm 0.009$ & $46.0 \pm 1.0$ & $1476 \pm 1$ & $158 \pm 2$ & $172 \pm 2$ \\ \\

 ESO\,348-G009 & $44\times27$   & $0.165 \pm 0.005$ & $13.1 \pm 0.3$ & $648 \pm 1$ & $86 \pm 2$ & $100 \pm 3$ \\
\enddata
\end{deluxetable}

\begin{figure*} 
\begin{tabular}{@{}lll@{}}
 \includegraphics[width=5.5cm]{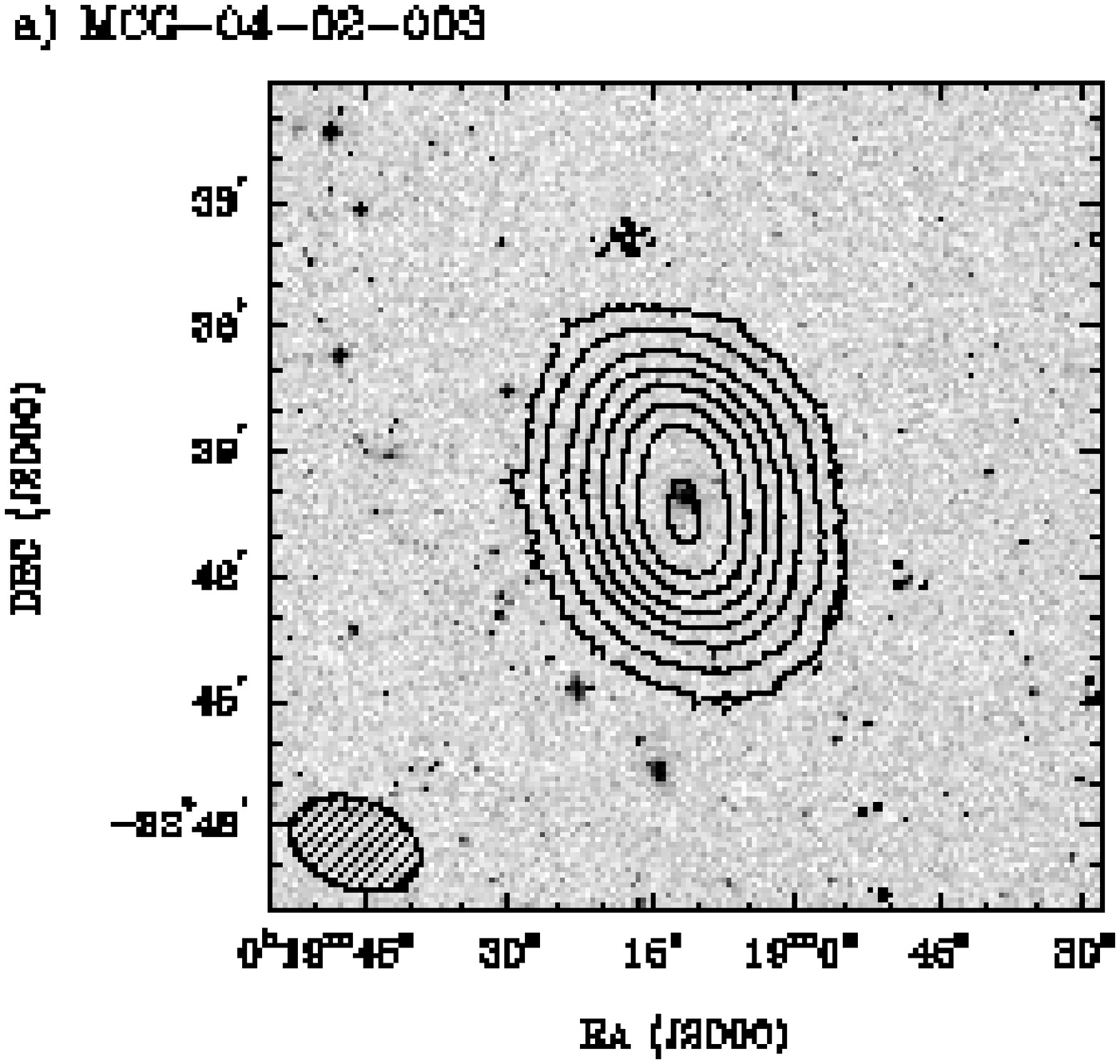} &
 \includegraphics[width=5.5cm]{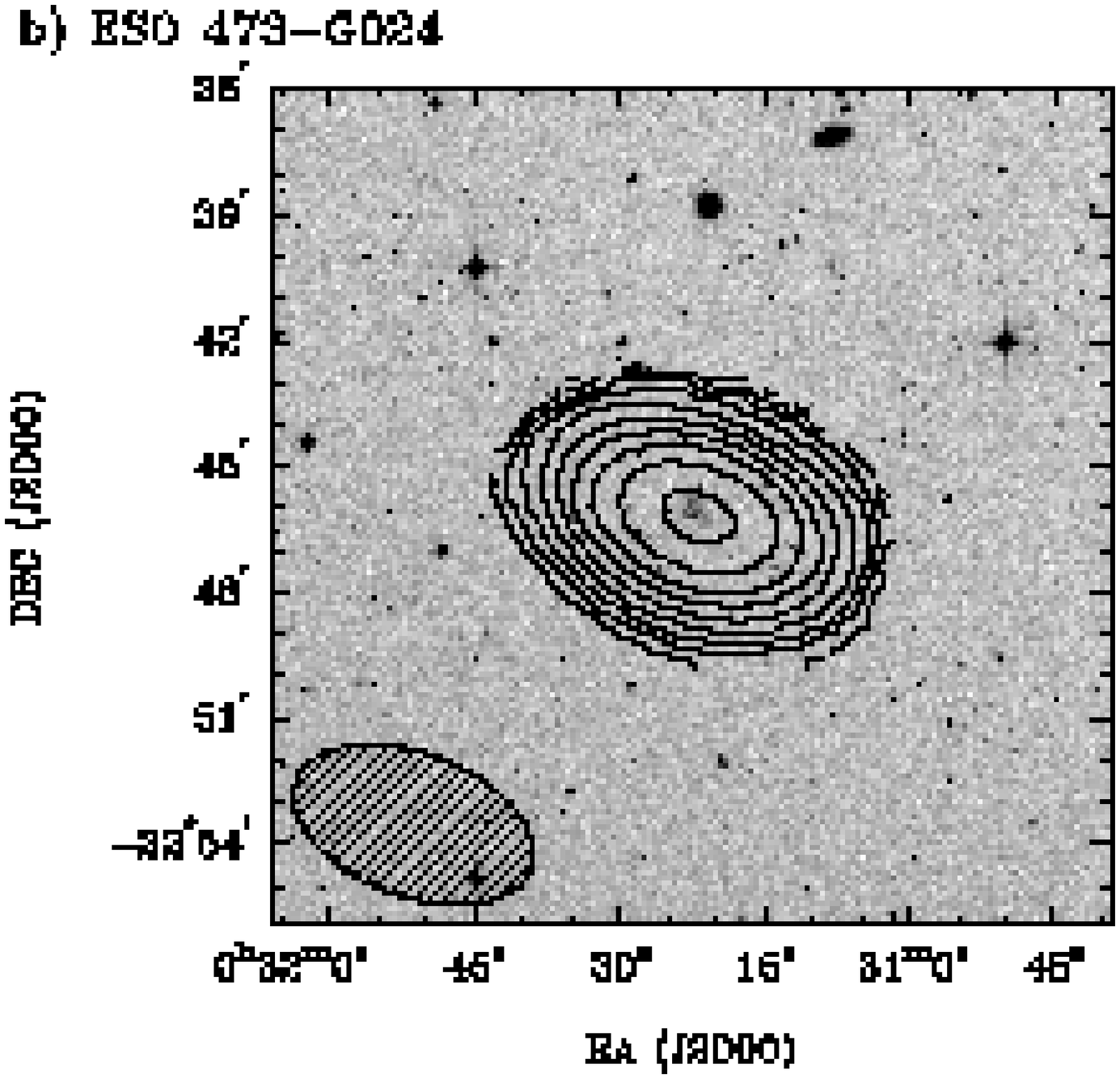} &
 \includegraphics[width=5.5cm]{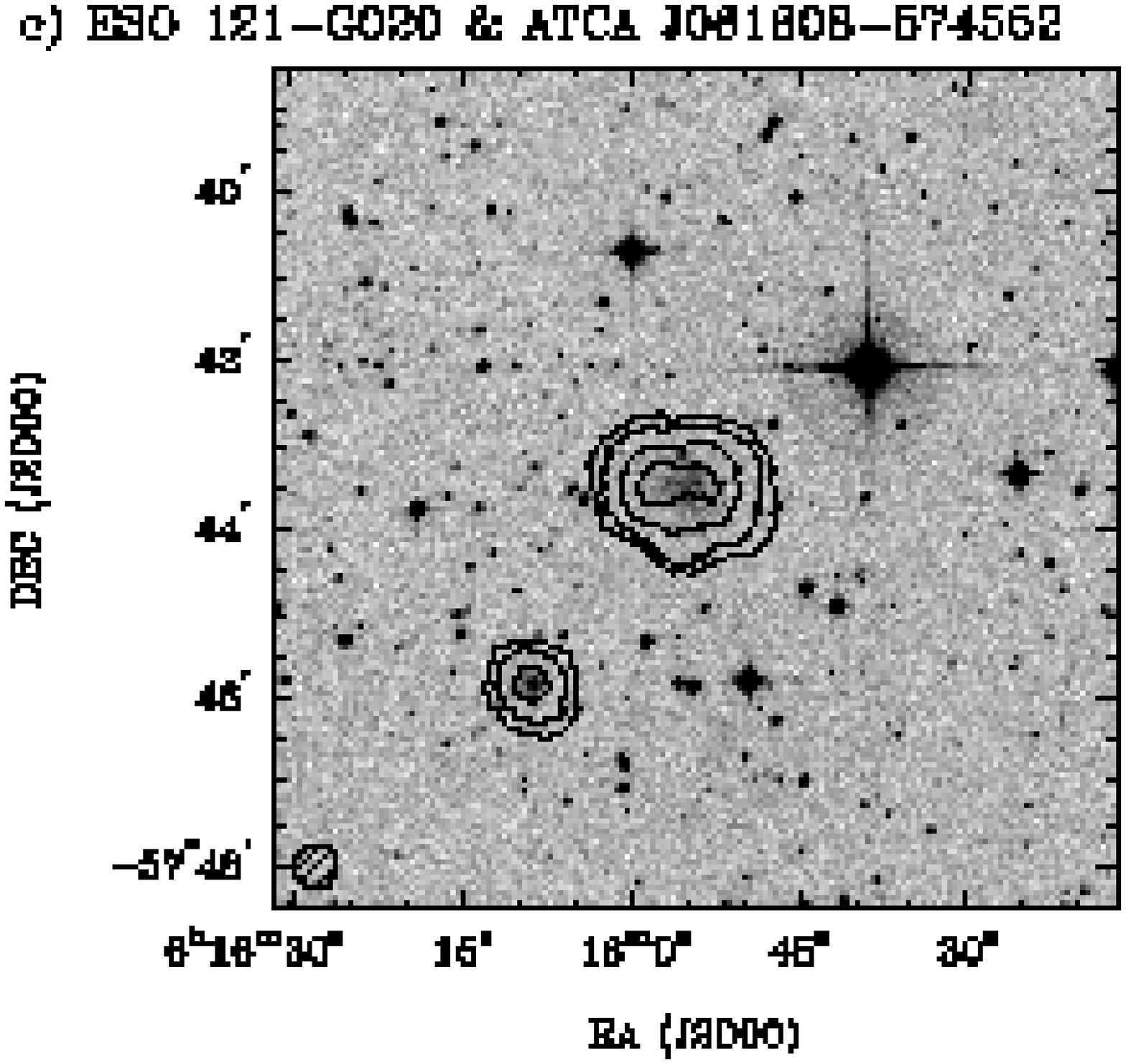} \\ \\

 \includegraphics[width=5.5cm]{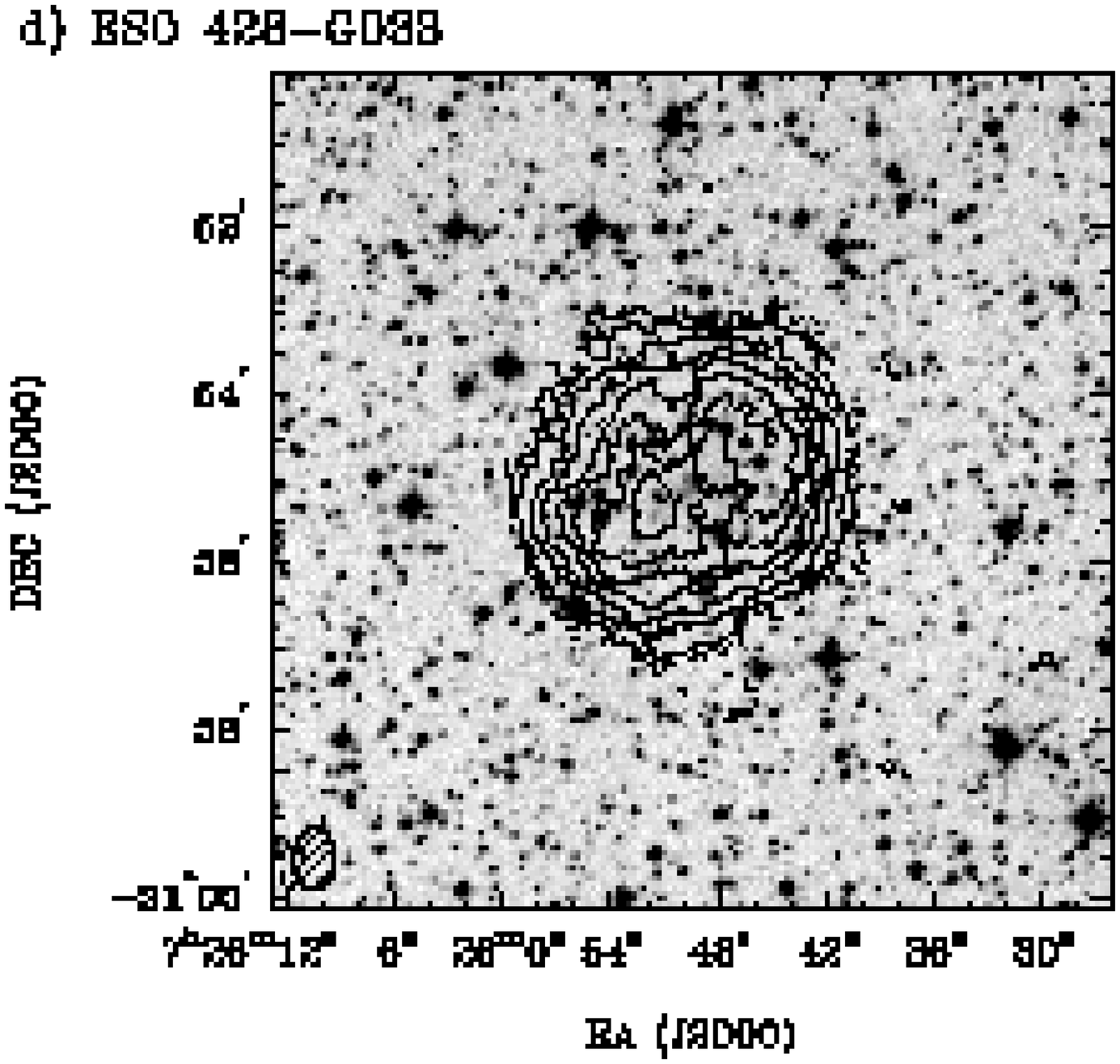} &
 \includegraphics[width=5.5cm]{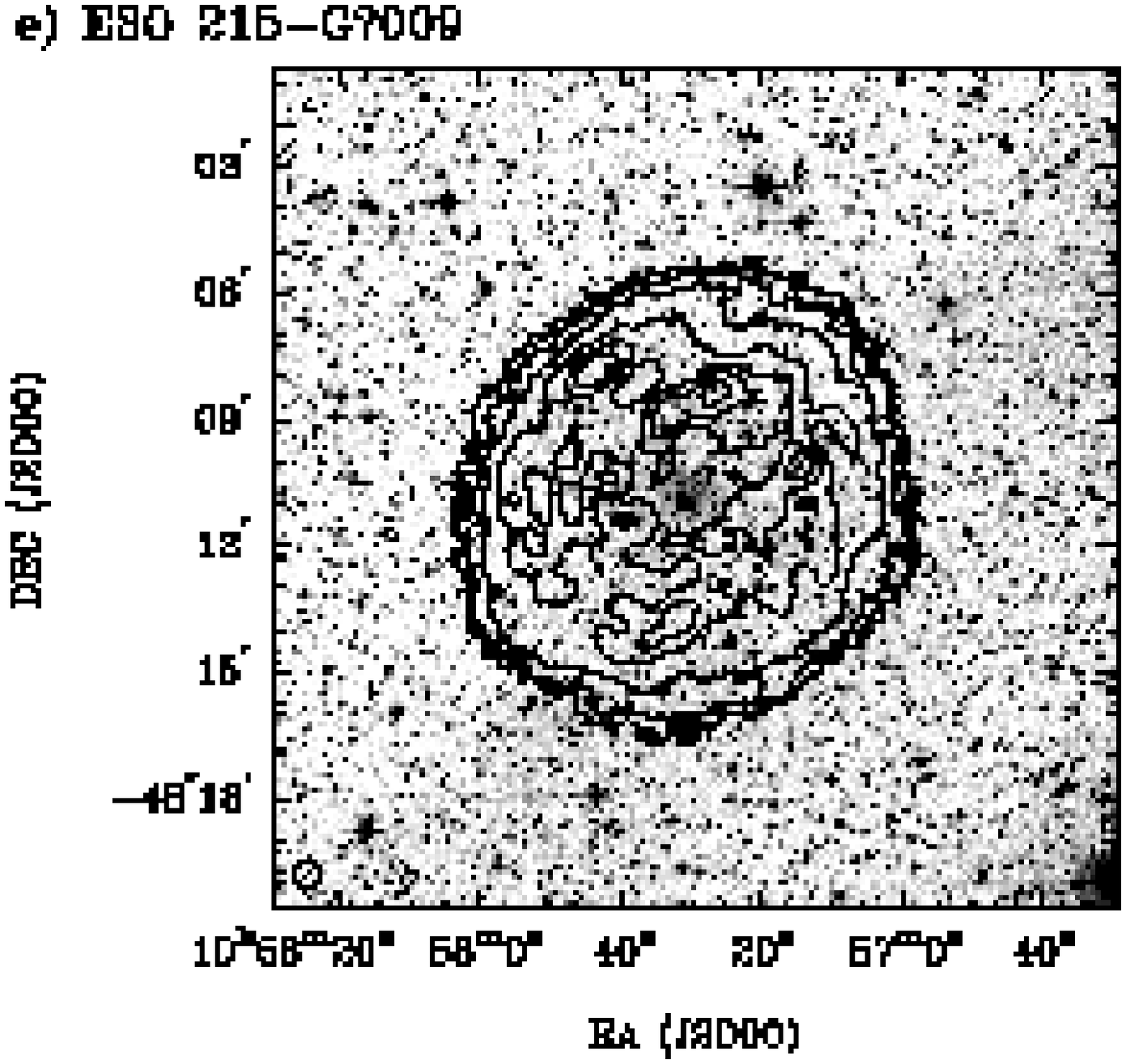} &
 \includegraphics[width=5.5cm]{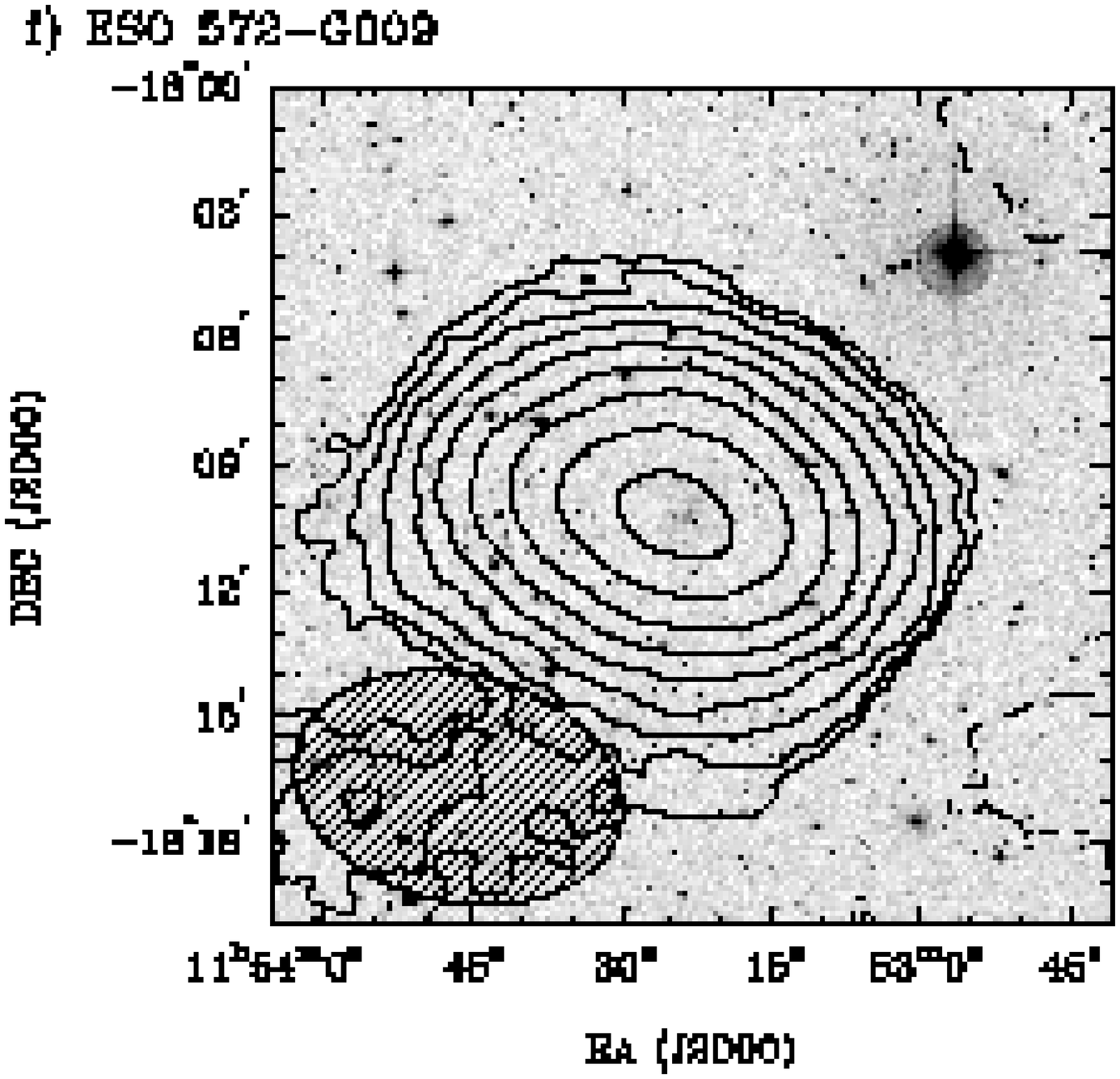} \\ \\

 \includegraphics[width=5.5cm]{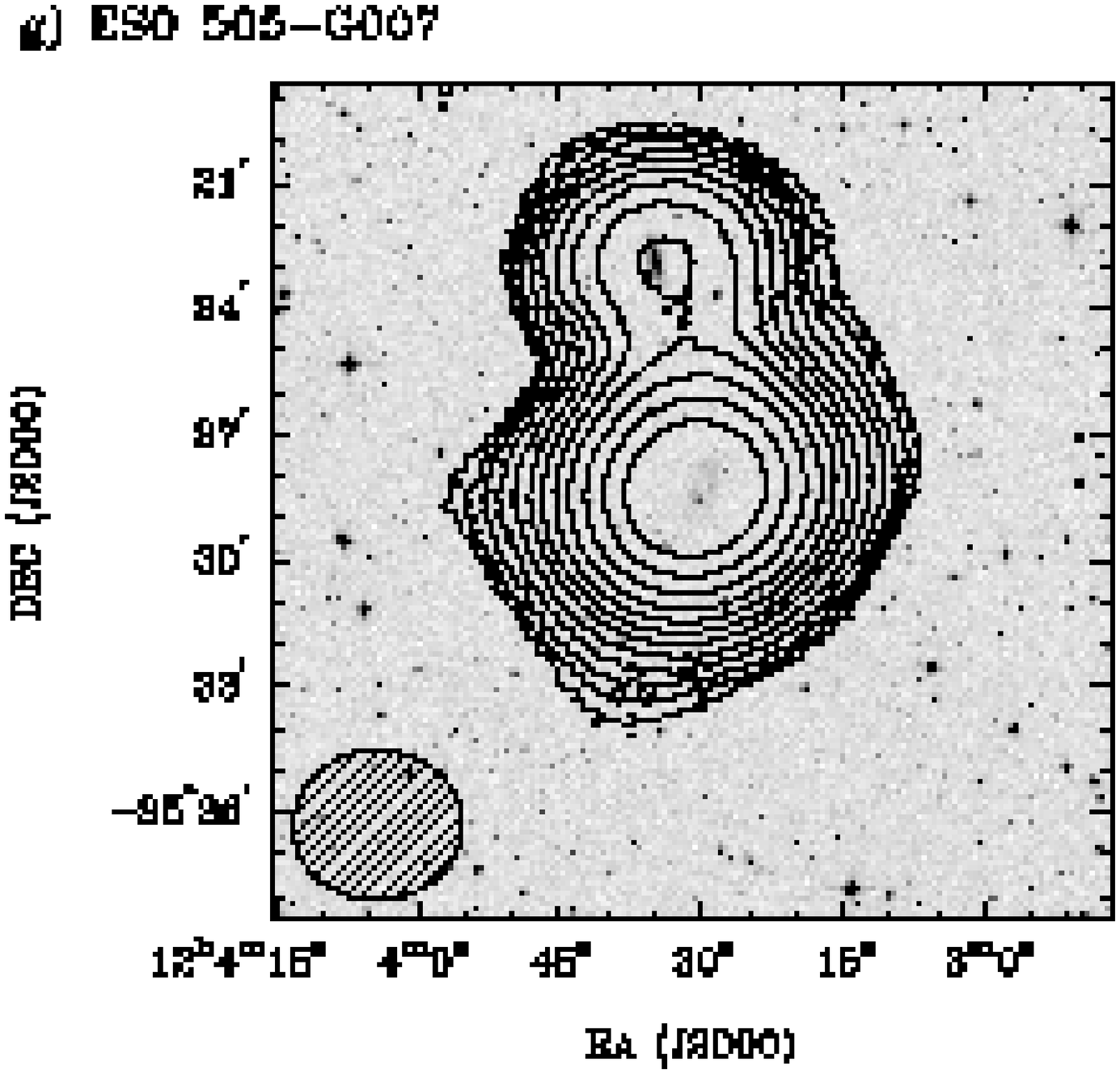} &
 \includegraphics[width=5.5cm]{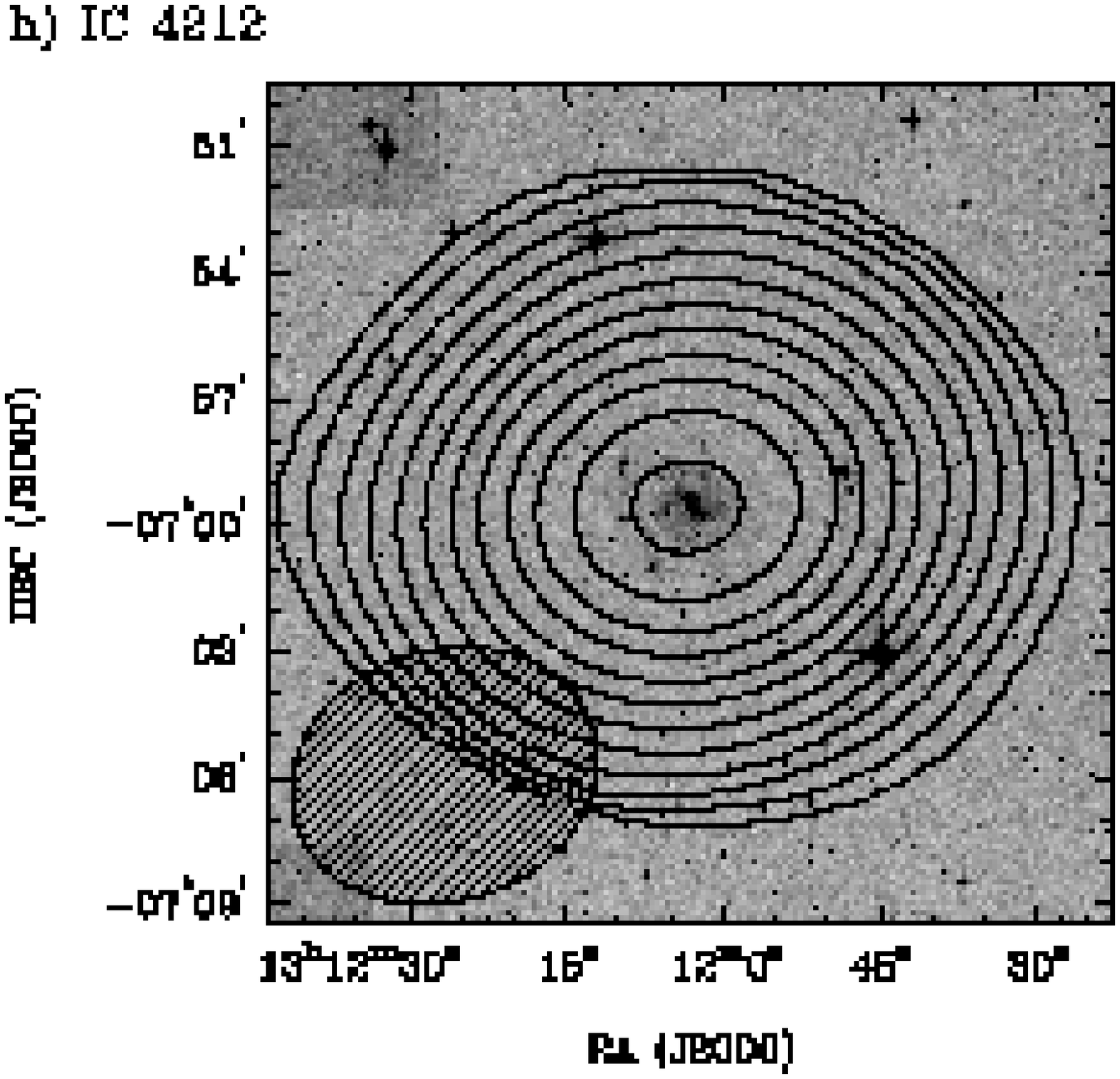} &
 \includegraphics[width=5.5cm]{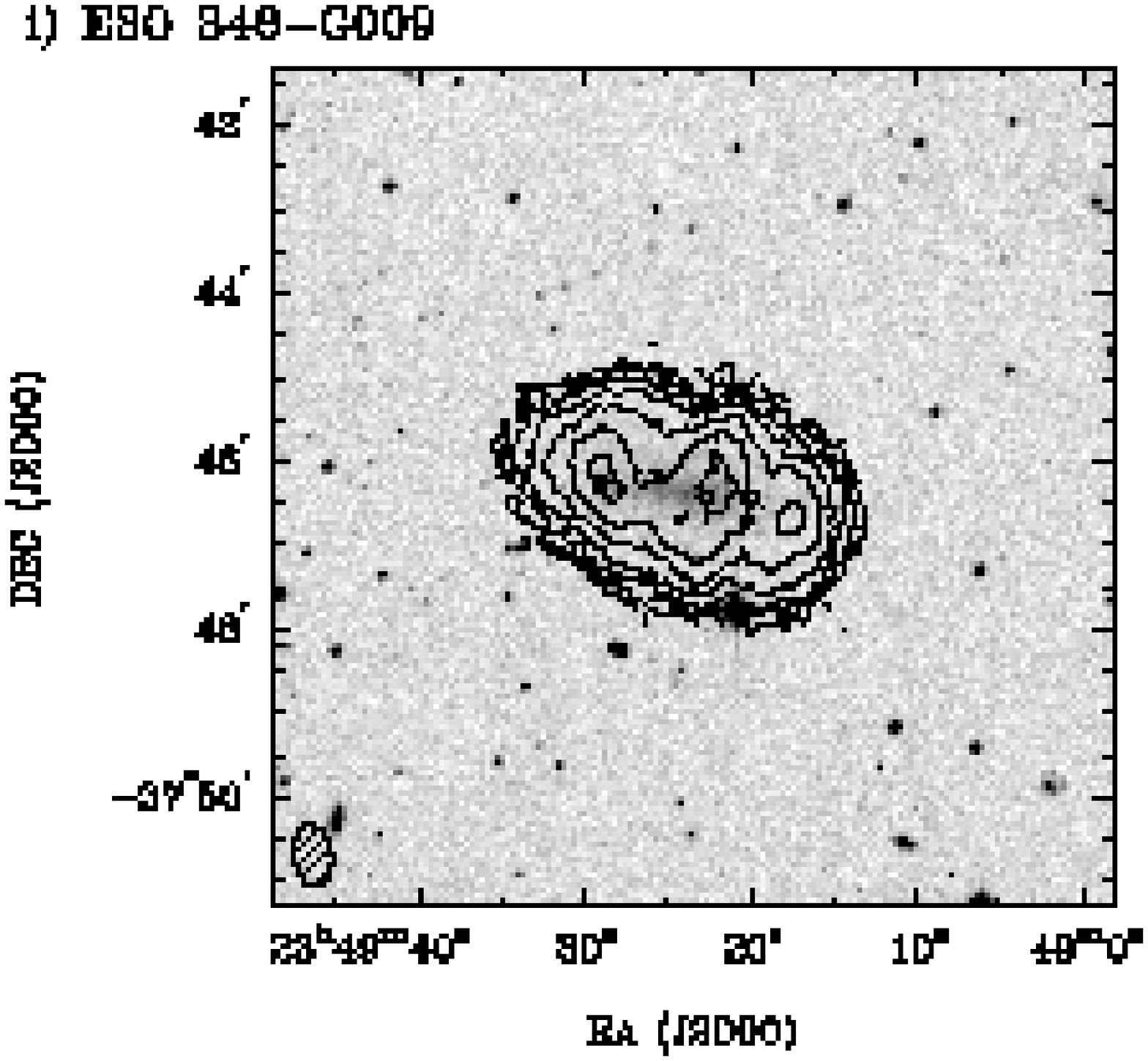} \\
\end{tabular}
\caption{Integrated \hi{} intensity distribution overlaid onto an optical DSS\,II {\em R} band image.  The panels show {\em a}) \mcg{}, {\em b}) ESO\,473-G024, {\em c}) ESO\,121-G020 \& \atg{}, {\em d}) ESO\,428-G033, {\em e}) \esoq{}, {\em f}) ESO\,572-G009, {\em g}) ESO\,505-G007 (with ESO\,505-G008), {\em h}) IC\,4212, and {\em i}) ESO\,348-G009.  East is to the left and North is up for all images in this paper.}
\label{fig:himap}
\end{figure*}

\begin{figure*} 
\begin{tabular}{@{}lll@{}}
 \includegraphics[width=5.5cm]{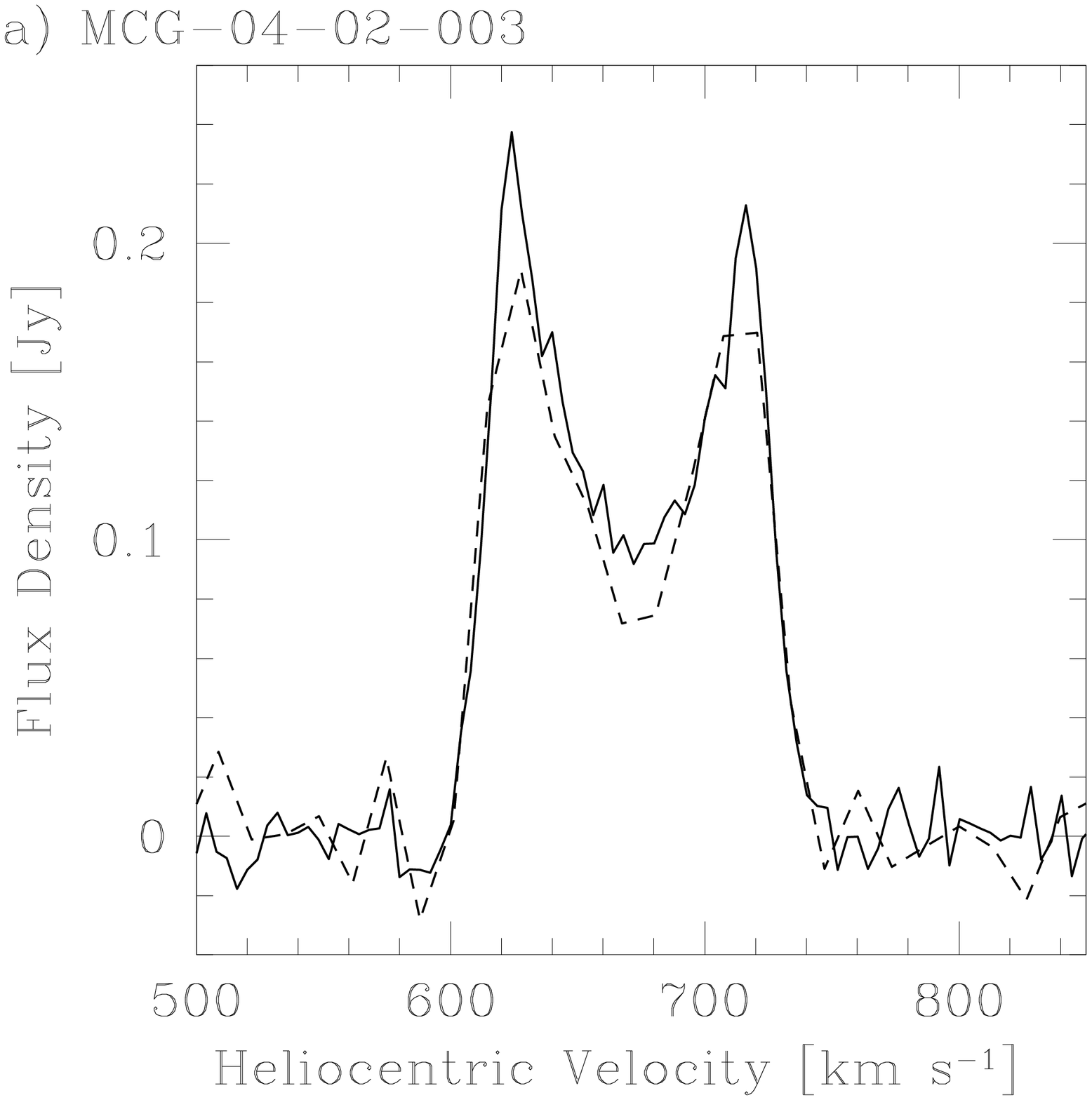} &
 \includegraphics[width=5.5cm]{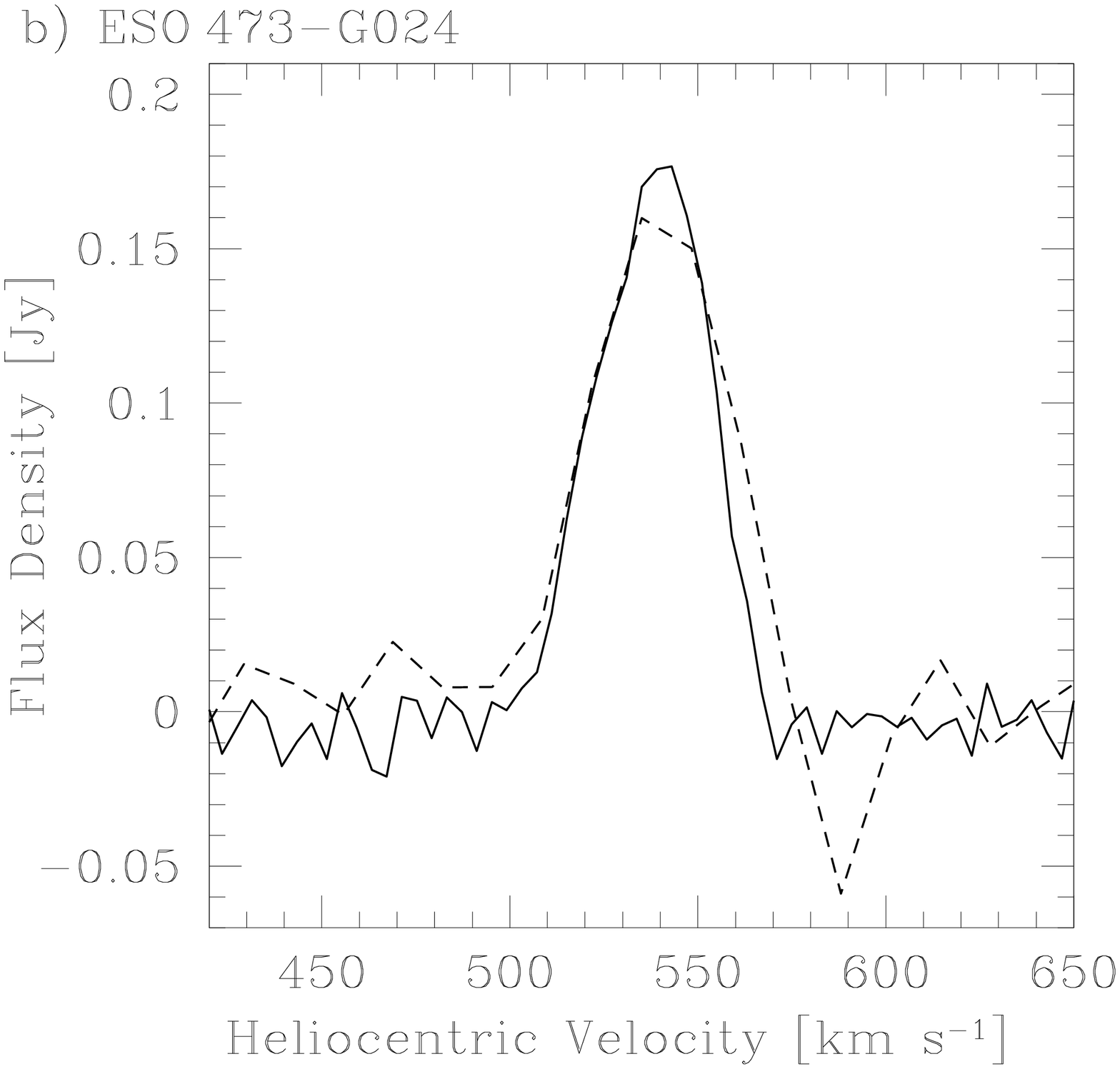} &
 \includegraphics[width=5.5cm]{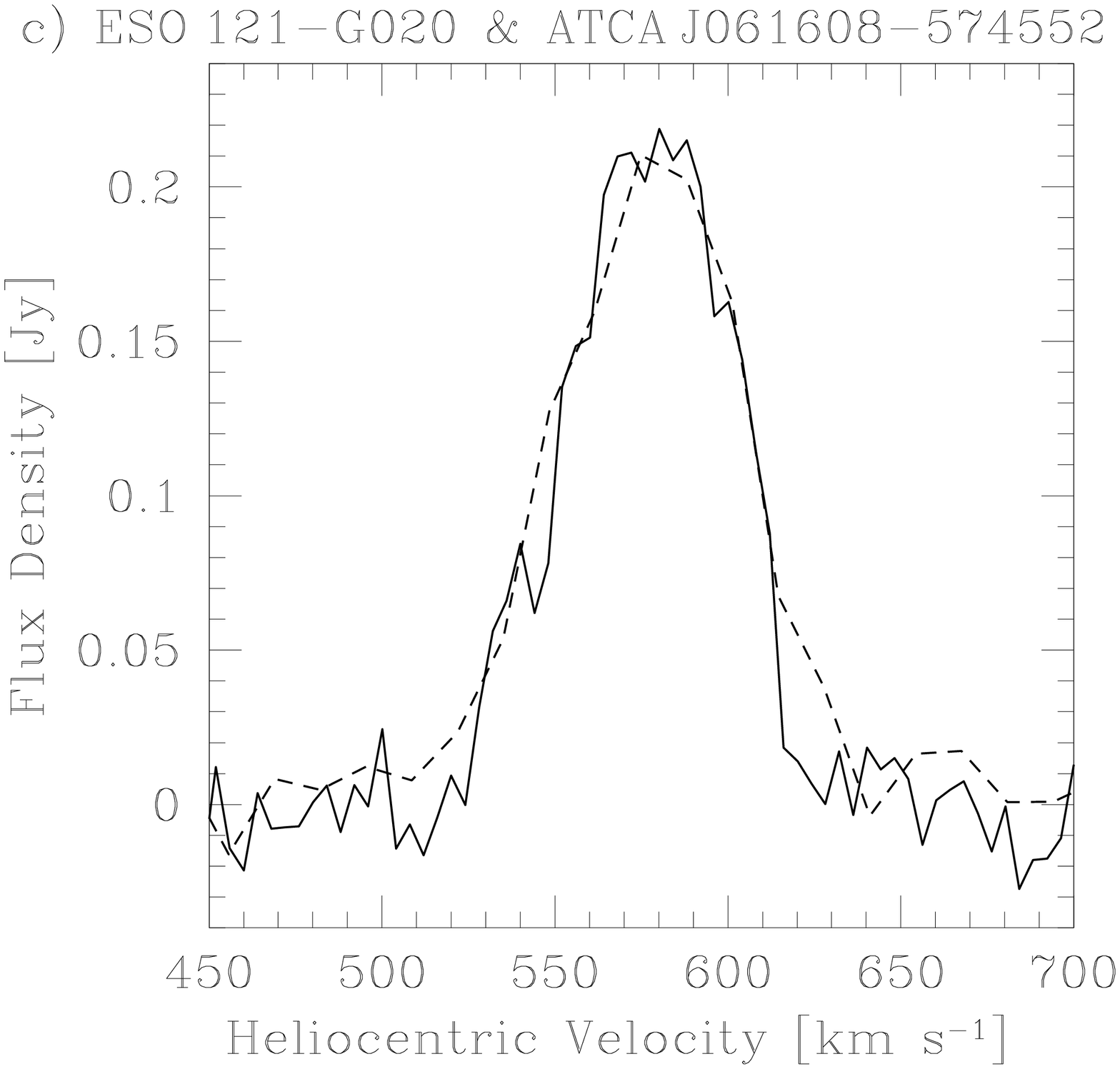} \\ \\

 \includegraphics[width=5.5cm]{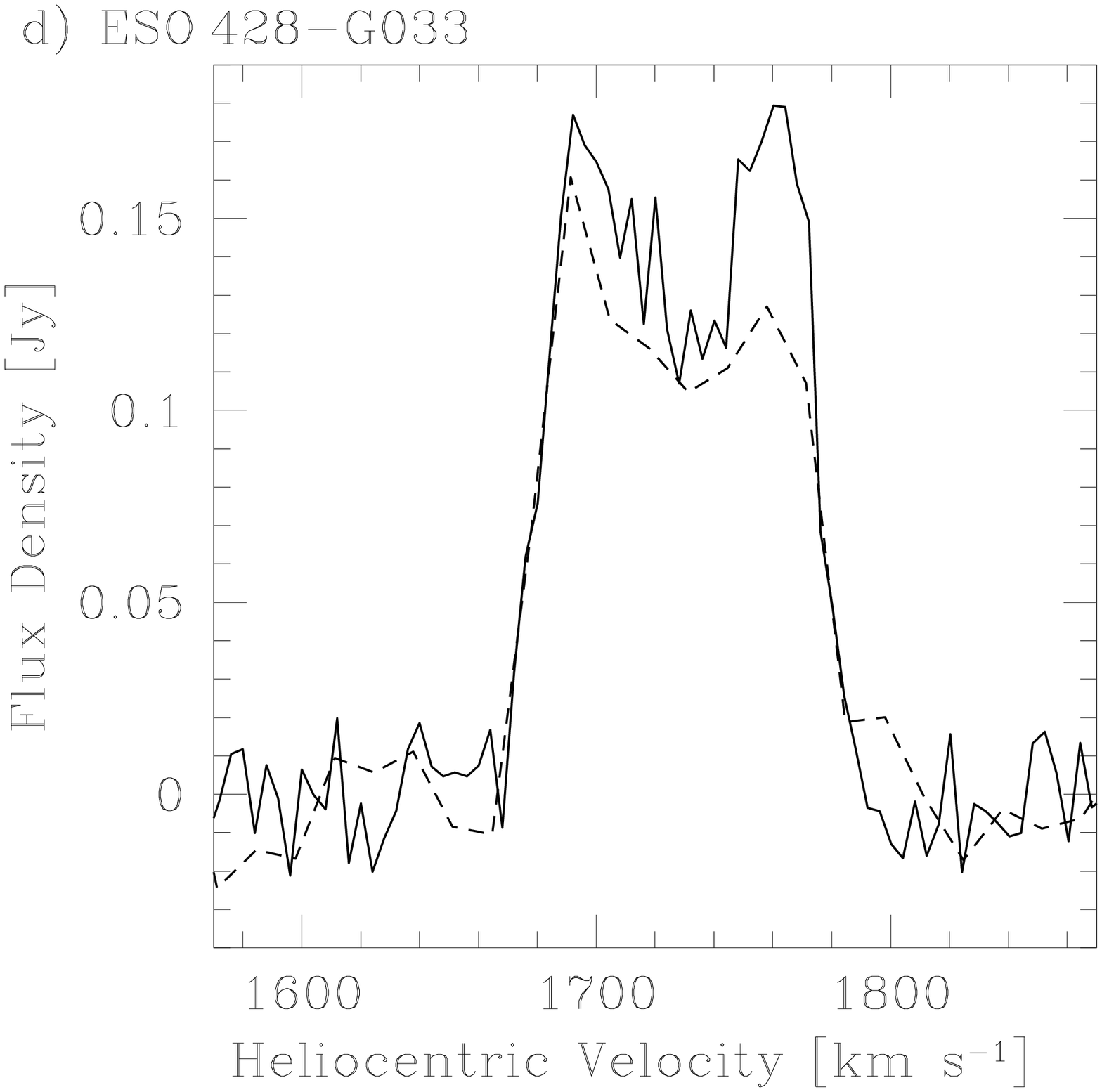} &
 \includegraphics[width=5.5cm]{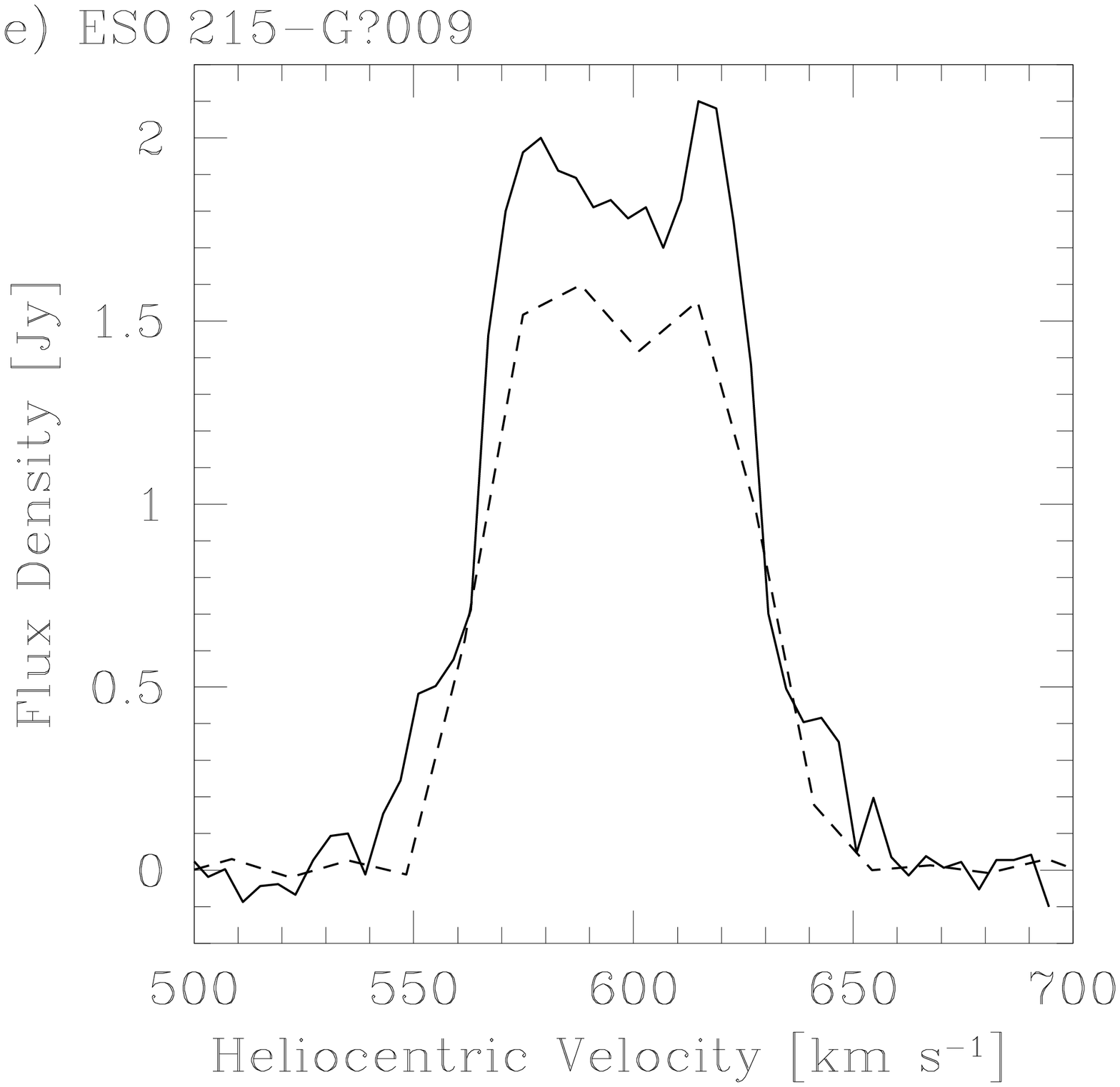} &
 \includegraphics[width=5.5cm]{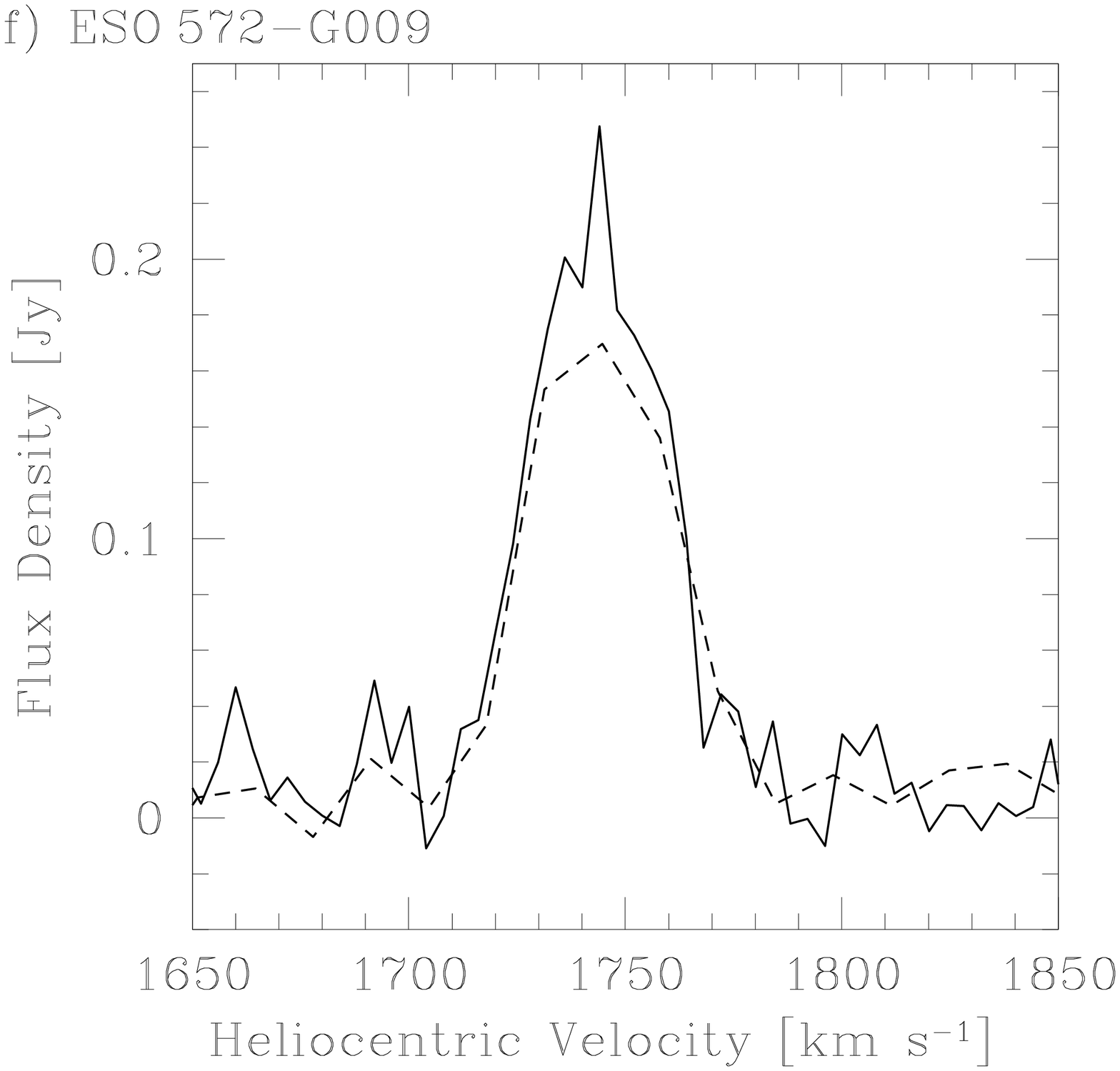} \\ \\

 \includegraphics[width=5.5cm]{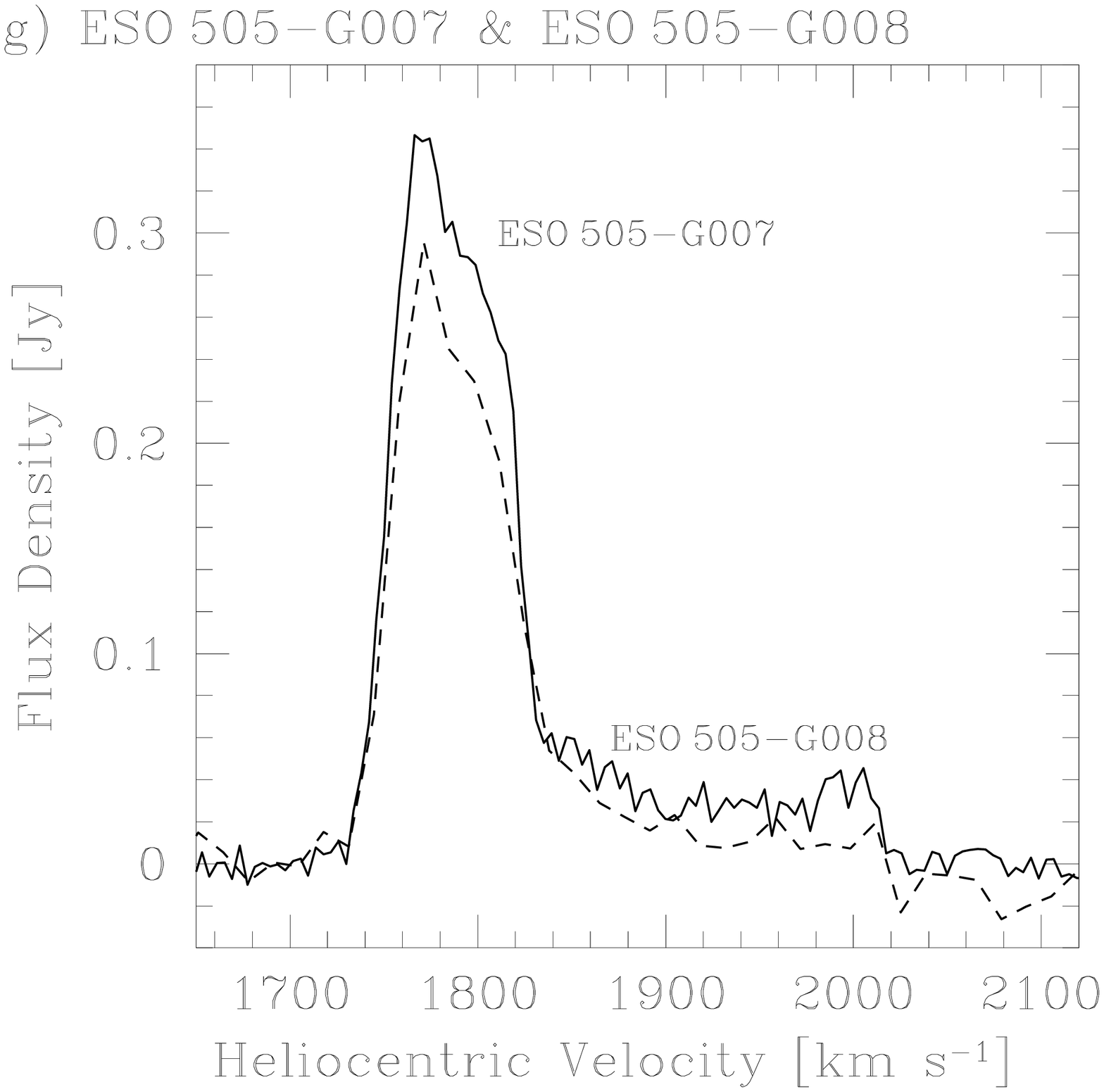} &
 \includegraphics[width=5.5cm]{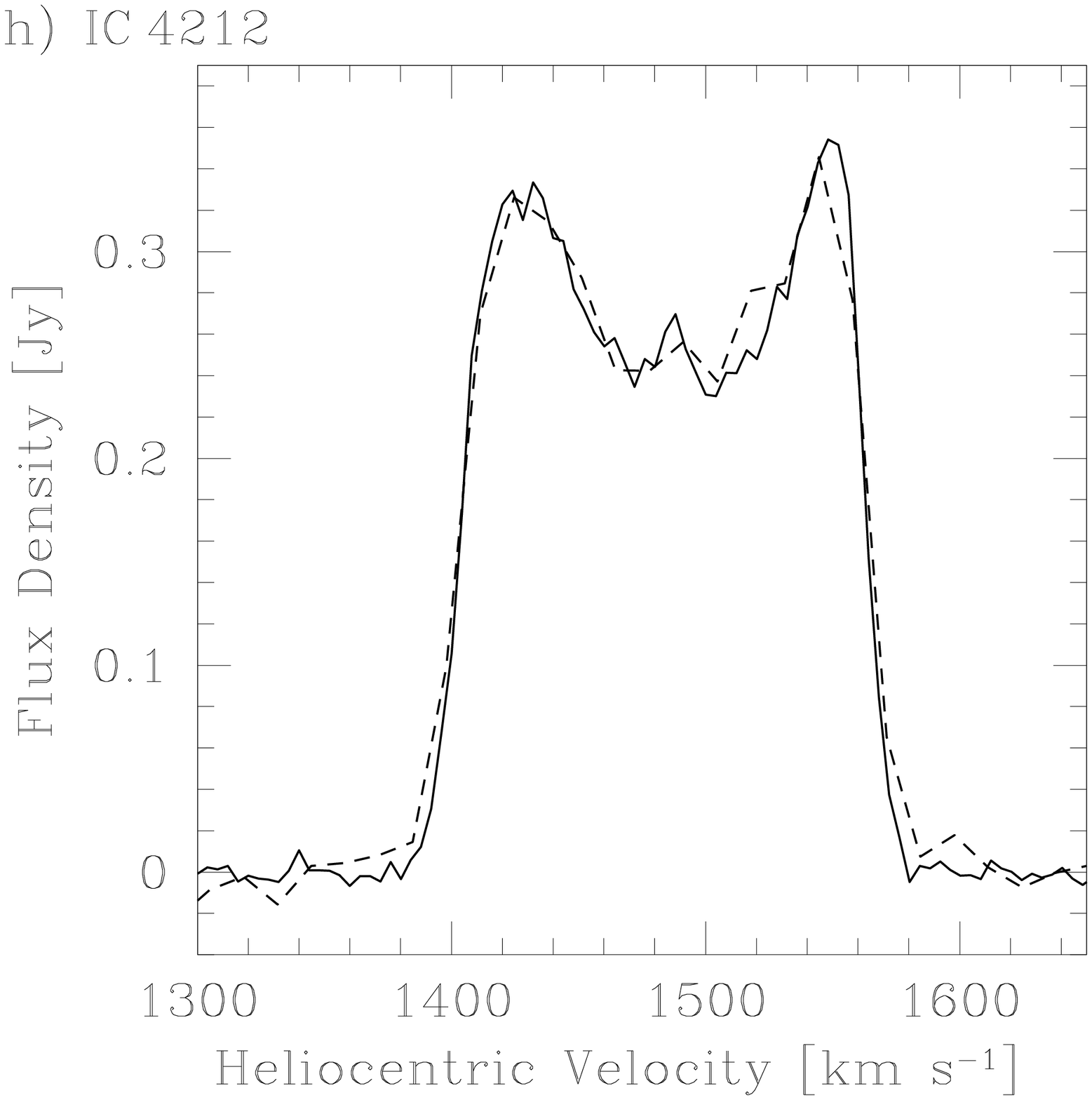} &
 \includegraphics[width=5.5cm]{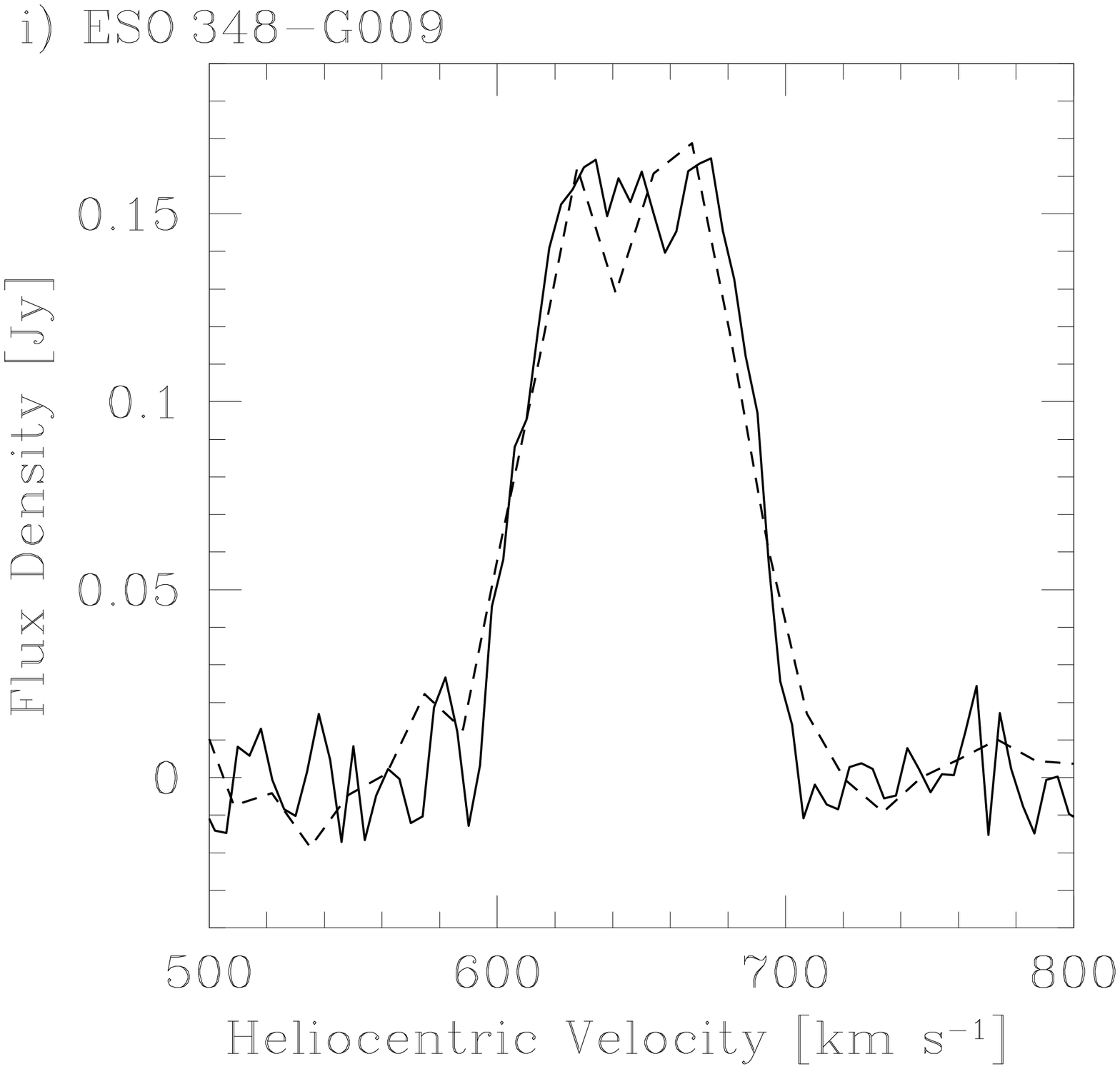} \\
\end{tabular}
\caption{Global \hi{} spectra of the nine galaxies as obtained from the BGC (dashed lines) and the ATCA (solid line).  The panels show {\em a}) \mcg{}, {\em b}) ESO\,473-G024, {\em c}) ESO\,121-G020 \& \atg{} (total spectrum), {\em d}) ESO\,428-G033, {\em e}) \esoq{}, {\em f}) ESO\,572-G009, {\em g}) ESO\,505-G007 (with ESO\,505-G008), {\em h}) IC\,4212, and {\em i}) ESO\,348-G009.
\label{fig:hispectra}}
\end{figure*}

The \hi{} distributions of all observed galaxies are shown in Fig.~\ref{fig:himap} overlaid onto second generation Digitized Sky Survey (DSS\,{\sc II}) {\em R} band images.  The synthesized beam sizes for the individual observations differ significantly due to the use of either East-West or northern hybrid arrays.  The ATCA \hi{} spectra of the observed galaxies are shown in Fig.~\ref{fig:hispectra}, with the BGC spectra plotted for comparison. 

\begin{figure*} 
  \includegraphics[width=8cm]{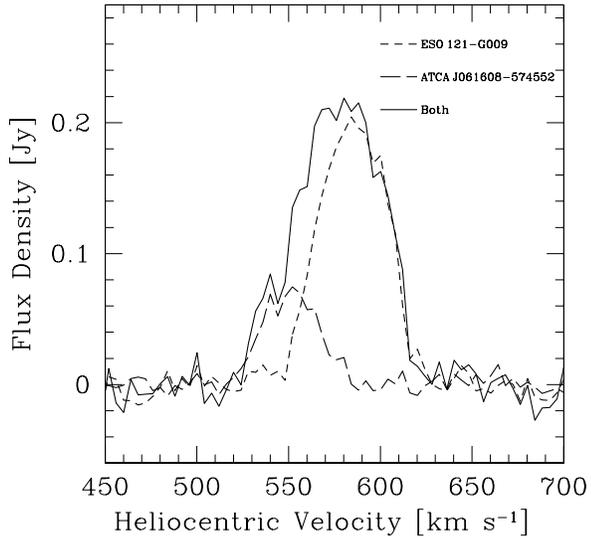}
\caption{\hi{} spectra of ESO\,121-G020 and \atg{} as obtained from the ATCA, showing the spectra for both galaxies (dashed lines) and for the total system (solid line, as shown in Fig.~\ref{fig:hispectra}{\em c}).
\label{fig:121spect}}
\end{figure*}

The ATCA \hi{} observations of the galaxy ESO\,121-G020 reveal a previously uncatalogued galaxy at a projected distance of only 3\arcmin{} (see Fig.~\ref{fig:himap}{\em c}).  The two galaxies are well resolved spatially, and their systemic velocities differ by only $29 \pm 6$\kms{}, less than the 50\% velocity width of either galaxy.  In the following we refer to this companion as \atg{} according to its \hi{} centre position.  The \hi{} spectra of the two galaxies as well as the global spectrum of the galaxy pair is shown in Fig.~\ref{fig:121spect} (Fig.~\ref{fig:hispectra}{\em c} shows the global spectrum only).  \atg{} has no previous optical measurements and we included it in our optical follow-up observations (see \S~\ref{sec:optp}).  Because \atg{} contributes $\sim$20\% to the \hi{} flux density of HIPASS J0615--57, the \hi{} mass to light ratio for ESO\,121-G020 decreases slightly. 

The galaxy ESO\,505-G007 likewise has a neighbor, the catalogued galaxy ESO\,505-G008, which lies at a projected distance of $\sim7$\arcmin{}.  However, the two galaxies are not entirely spatially or spectrally resolved in our \hi{} observations.  ESO\,505-G007 appears to have a low inclination in the optical images and its \hi{} line is relatively narrow.  In contrast, the galaxy ESO\,505-G008 is seen close to edge-on and shows a broad \hi{} line, as seen in Fig.~\ref{fig:hispectra}{\em g}.  By fitting two point sources to the low-resolution \hi{} distribution we determine an \hi{} flux density of $21 \pm 3$\jks{} for ESO\,505-G007, in excellent agreement with the BGC value for HIPASS\,J1203--25 despite the confusion, and $8 \pm 3$\jks{} for ESO\,505-G008. 

The ATCA \hi{} maps of the other seven galaxies match the optical counterparts identified in the BGC.  For \mcg{} (Fig.~\ref{fig:himap}{\em a}) the \hi{} distribution is extended North-South and aligned with the stellar distribution (see \S~\ref{sec:optp}), although little other structure is distinguishable with the large beam.  For IC\,4212 (Fig.~\ref{fig:himap}{\em h}) the source appears extended compared to the large beam, and deconvolution with imfit in {\sc MIRIAD} indicates that the \hi{} extends $\sim$200\arcsec{}.  We find that for all galaxies, except \esoq{} (see \pI{}), the measured ATCA \hi{} flux densities are in agreement with the BGC values (including the combined ESO\,121-G020/\atg{} system). 

Our results for the ATCA \hi{} observations of the nine galaxies are listed in Table~\ref{tab:radio}.  The columns are as follows: (1) galaxy name (with \atg{} included); (2) size of the synthesized beam; (3) \hi{} peak flux density; (4) integrated \hi{} flux density; (5) \hi{} systemic velocity from the \hi{} line; (6) velocity width of the \hi{} line at 50\% of the peak flux density; and (7) velocity width of the \hi{} line at 20\% of the peak flux density.

\subsection{\hi{} Gas Dynamics}
\label{sec:velo}

\begin{deluxetable}{lccccc} 
\tabletypesize{\scriptsize}
\tablecaption{Rotation curve fit for galaxies with high-resolution ATCA \hi{} data.
  \label{tab:rot}}
\tablewidth{0pt}
\tablehead{ \colhead{Name} & \colhead{\vsys{}} & \colhead{{\it PA}} & \colhead{{\em i}} & \colhead{\vmax{}} & \colhead{$r_{\rm max}$} \\

   \colhead{} & \colhead{(\kkms{})} & \colhead{(degrees)} & \colhead{(degrees)} & \colhead{(\kkms{})} & \colhead{(arcsec)} \\

 \colhead{(1)} & \colhead{(2)} & \colhead{(3)} & \colhead{(4)} & \colhead{(5)} & \colhead{(6)}}
\startdata
 ESO\,121-G020 & $584.5 \pm 1.0$ & $262 \pm 2$ & $78 \pm 5$ & $21 \pm 2$ & $80 \pm 10$ \\ \\

 ESO\,428-G033 & $1717 \pm 3$ & $295 \pm 5$ & $11 \pm 5$ & $200^{+160}_{-60}$ & $130 \pm 5$ \\ \\

 \esoq{}       & $597 \pm 1$ & $119 \pm 2$ & $36 \pm 10$ & $51 \pm 8$ & $370 \pm 20$ \\ \\

 ESO\,348-G009 & $646 \pm 2$ & $245 \pm 3$ & $80 \pm 5$ & $50 \pm 5$ & $160 \pm 15$ \\
\enddata
\end{deluxetable}

\begin{figure*} 
\begin{tabular}{@{}ll@{}}
 \includegraphics[width=8cm]{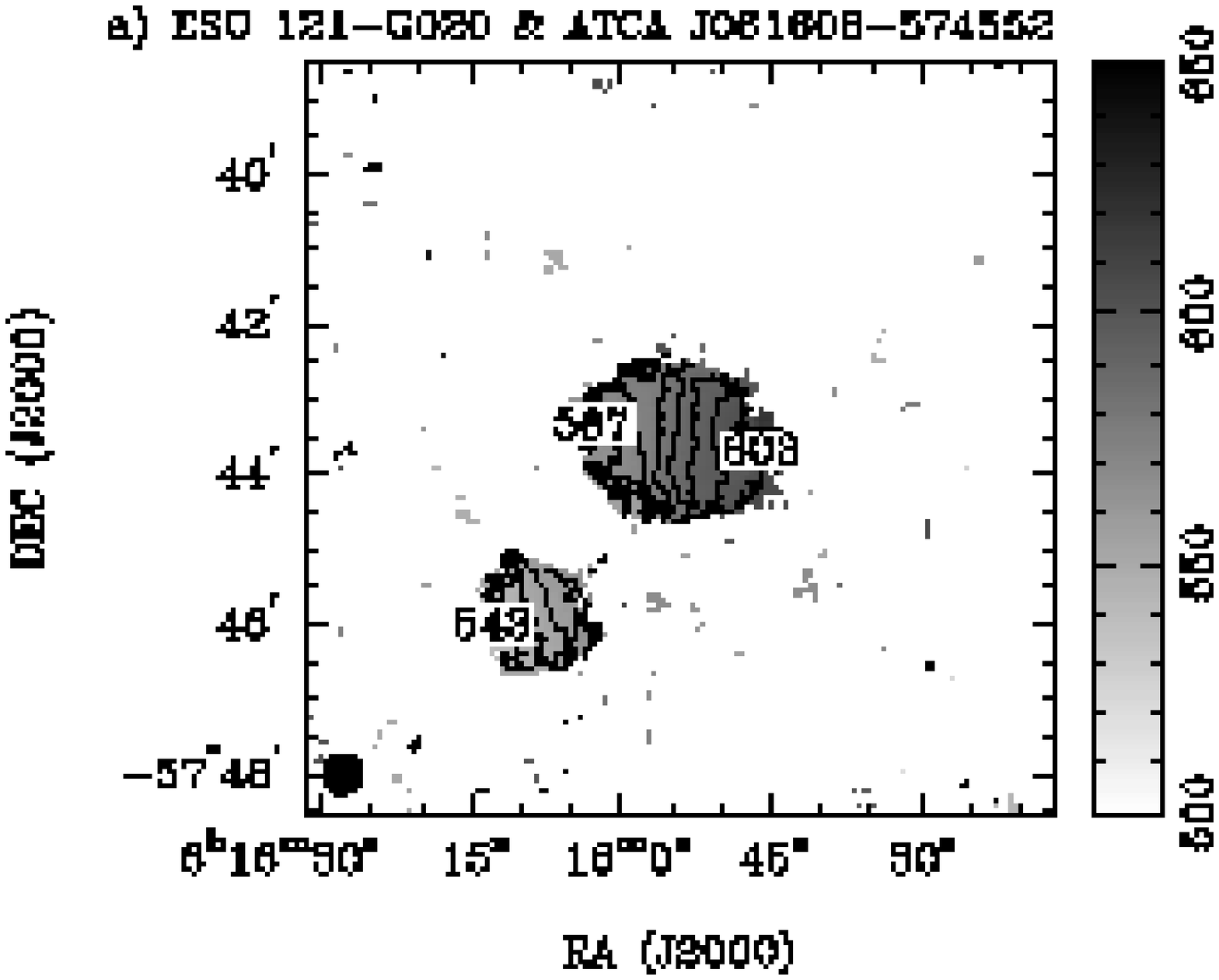} &
 \includegraphics[width=8cm]{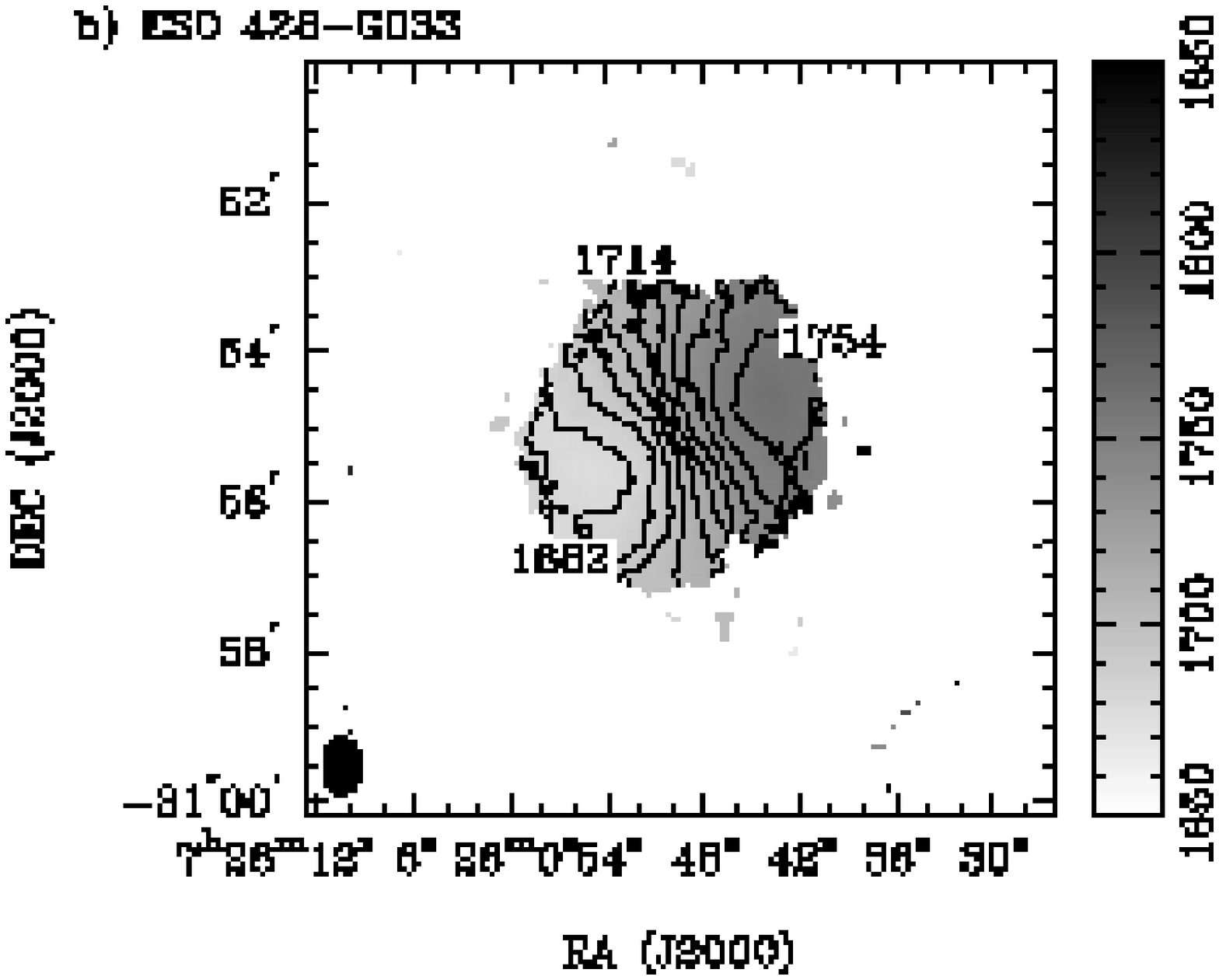} \\ \\

 \includegraphics[width=8cm]{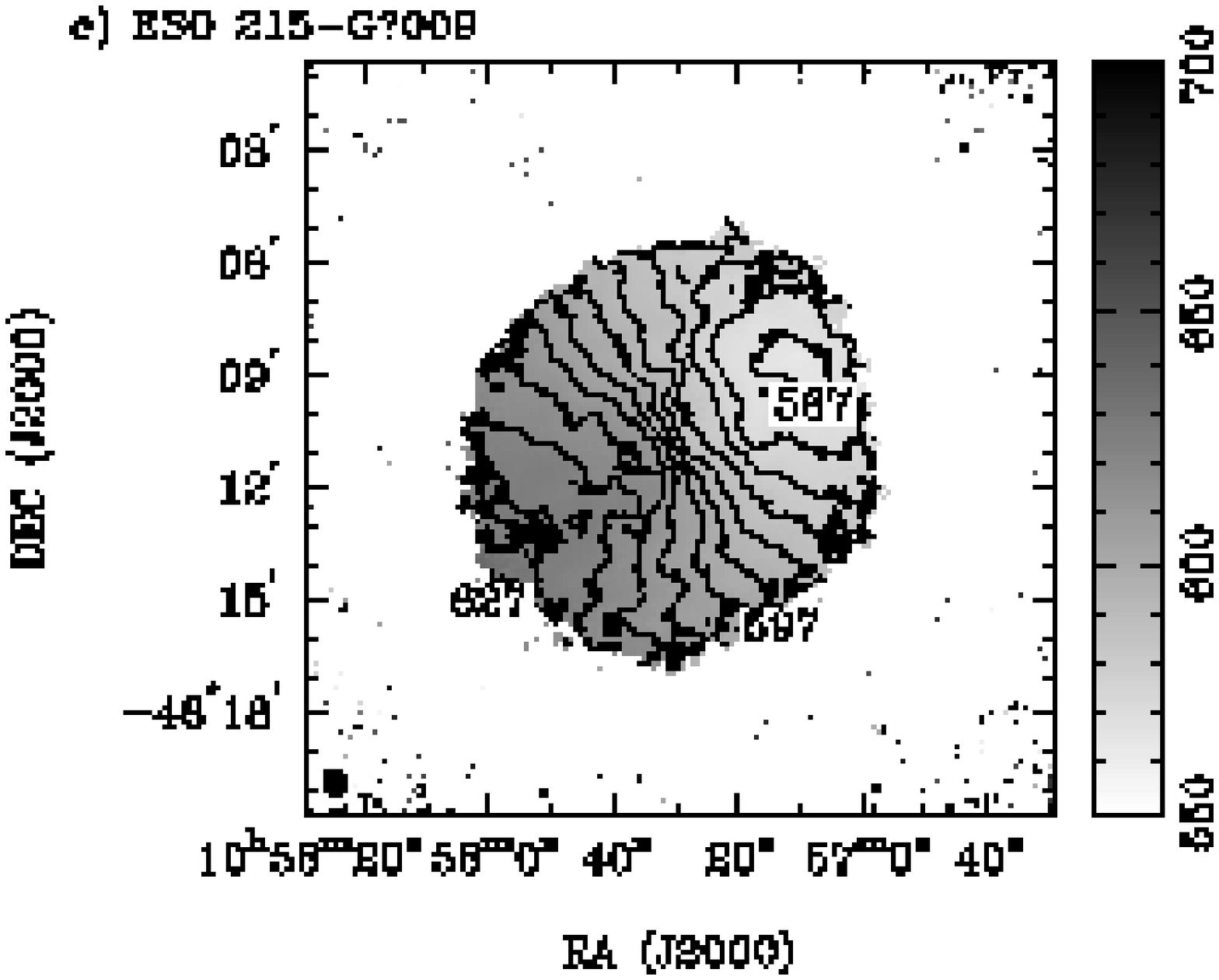} &
 \includegraphics[width=8cm]{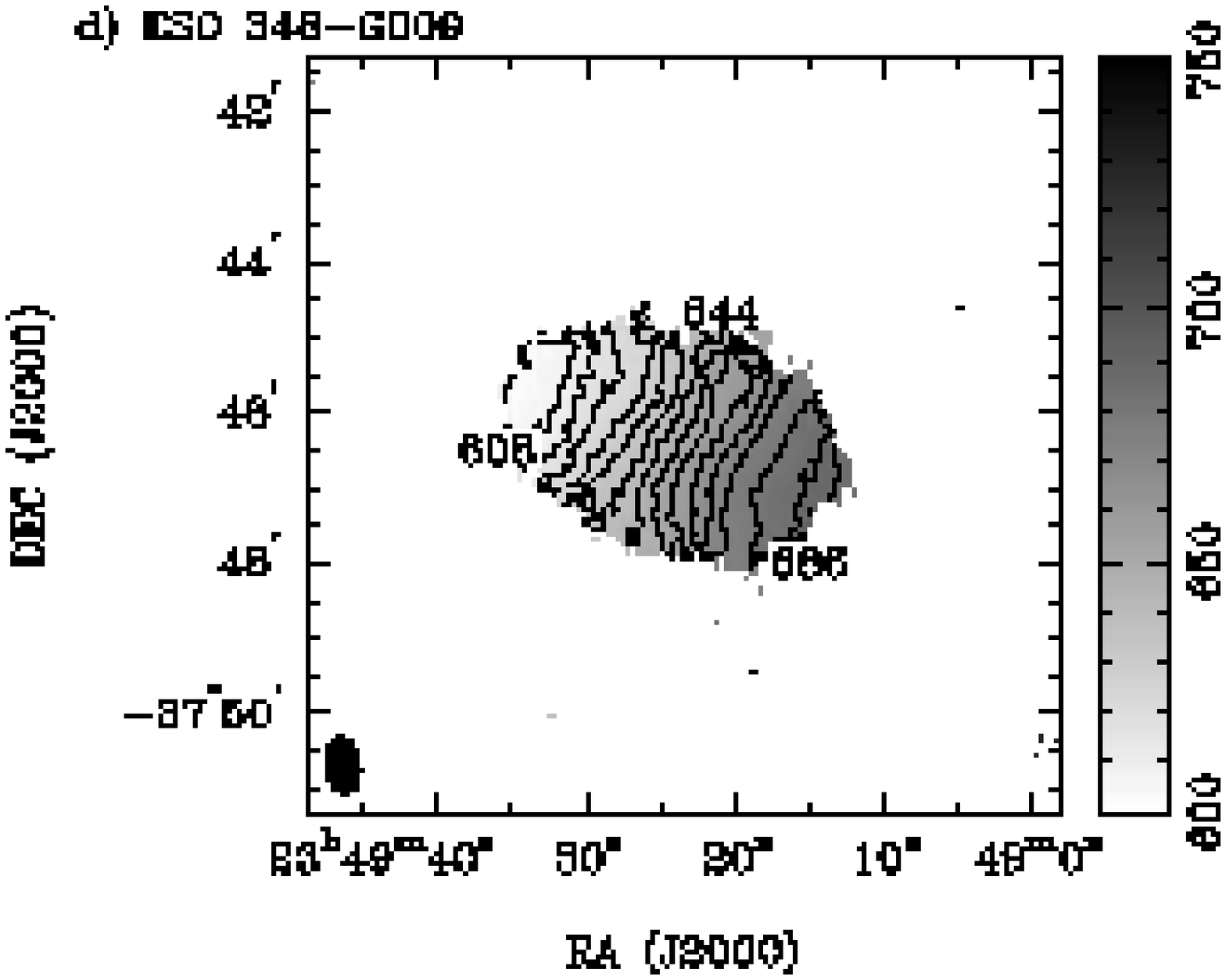} \\
\end{tabular}
\caption{The \hi{} velocity fields (moment 1 maps) for the four galaxy fields with the best resolution.  The panels show {\em a}) ESO\,121-G020 \& \atg{}, {\em b}) ESO\,428-G033, {\em c}) \esoq{}, {\em d}) ESO\,348-G009.  The numbers mark contour levels in \kms{}.
\label{fig:hivfield}}
\end{figure*}

The \hi{} velocity fields (1st moment) of the galaxies ESO\,121-G020/\atg{}, ESO\,428-G033, \esoq{} (\pI{}), and ESO\,348-G009 are shown in Fig.~\ref{fig:hivfield}.  All galaxies show clear signs of rotation.  The velocity field of the galaxy ESO\,428-G033 appears similar to that of \esoq{}, displaying fairly regular, undisturbed gas dynamics.  The extreme velocity contours on both the approaching and receding sides close, suggesting that either the rotation curve turns down at large radii or the galaxy is warped.  The position angle ($PA$) of ESO\,428-G033, as measured from its \hi{} velocity field, is aligned with the bright optical emission, although the nature of the stellar feature is unclear (see \S~\ref{sec:dis-compd}).  Likewise, the position angle of the galaxy ESO\,348-G009 aligns well with the optical disk (see \S\S~\ref{sec:optp} and \ref{sec:dis-comp} for more on the optical details of the galaxies).

The analysis of the rotation curve for \esoq{} using {\sc rocur} in {\sc AIPS} was presented in \pI{}.  For the other three galaxies we instead used the equivalent procedure {\sc rotcur} in {\sc GIPSY} \citep[both use the tilted ring algorithm described by][]{beg89}.  We used the standard procedure of narrowing the free parameters (centre position, \vsys{}, position angle, and inclination) down one at a time until the best fit to all parameters was obtained and a rotation curve could be produced.  All three fits were done with 10\arcsec{} rings (the \esoq{} fit used 12\arcsec{} rings).  After an initial fit which included both sides of the galaxy, the fit was done individually for the approaching and receding sides of the galaxy to check for any asymmetry.  Of the three new fits, the only galaxy which had significant differences between the two sides was ESO\,428-G033, where the inclination fit was lower on the approaching side, which may have been a result of the very low inclination (this was included in the uncertainties).

\begin{figure*} 
\begin{tabular}{@{}ll@{}}
 \includegraphics[width=8cm]{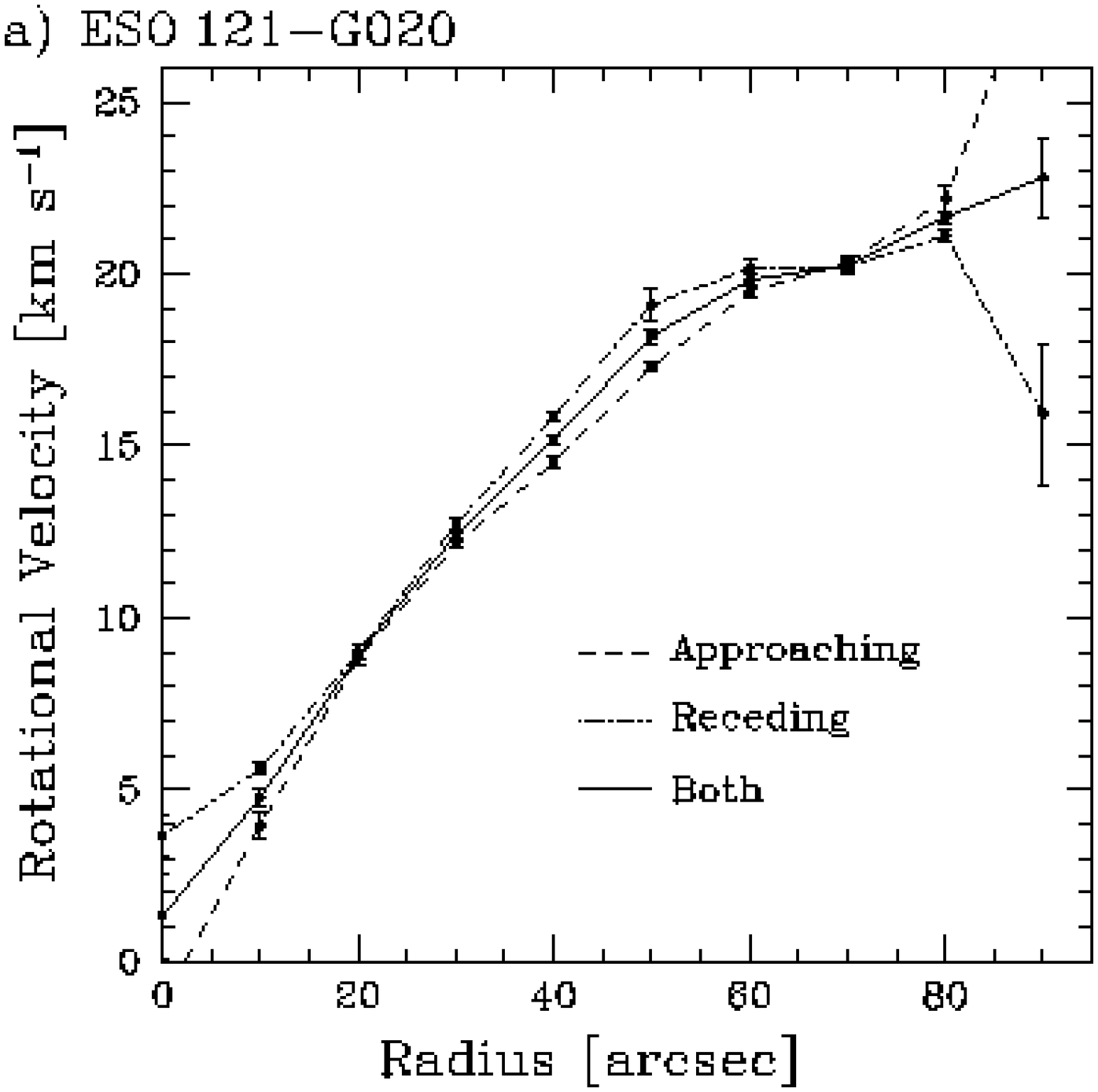} &
 \includegraphics[width=8cm]{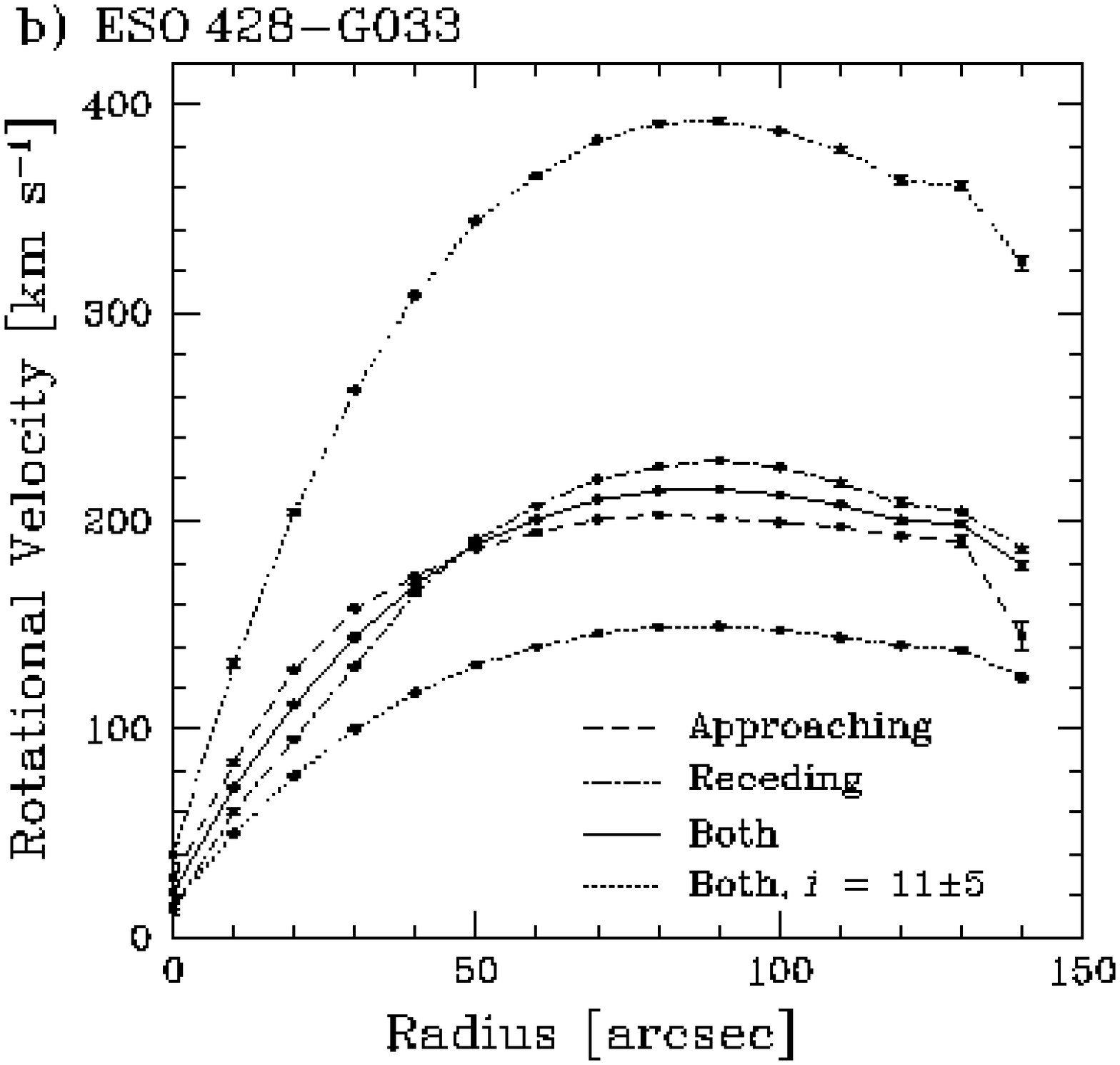} \\ \\

 \includegraphics[width=8cm]{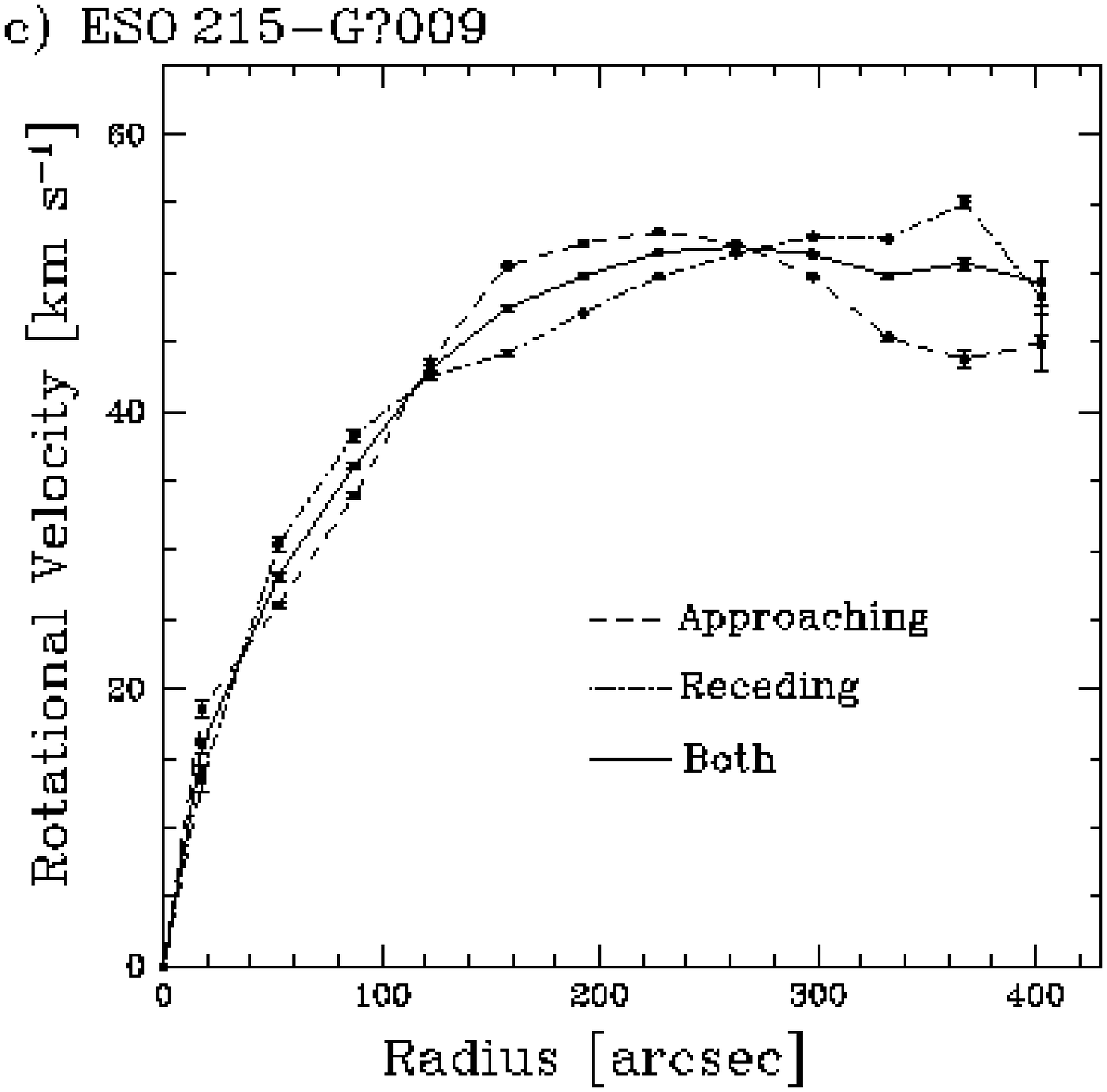} &
 \includegraphics[width=8cm]{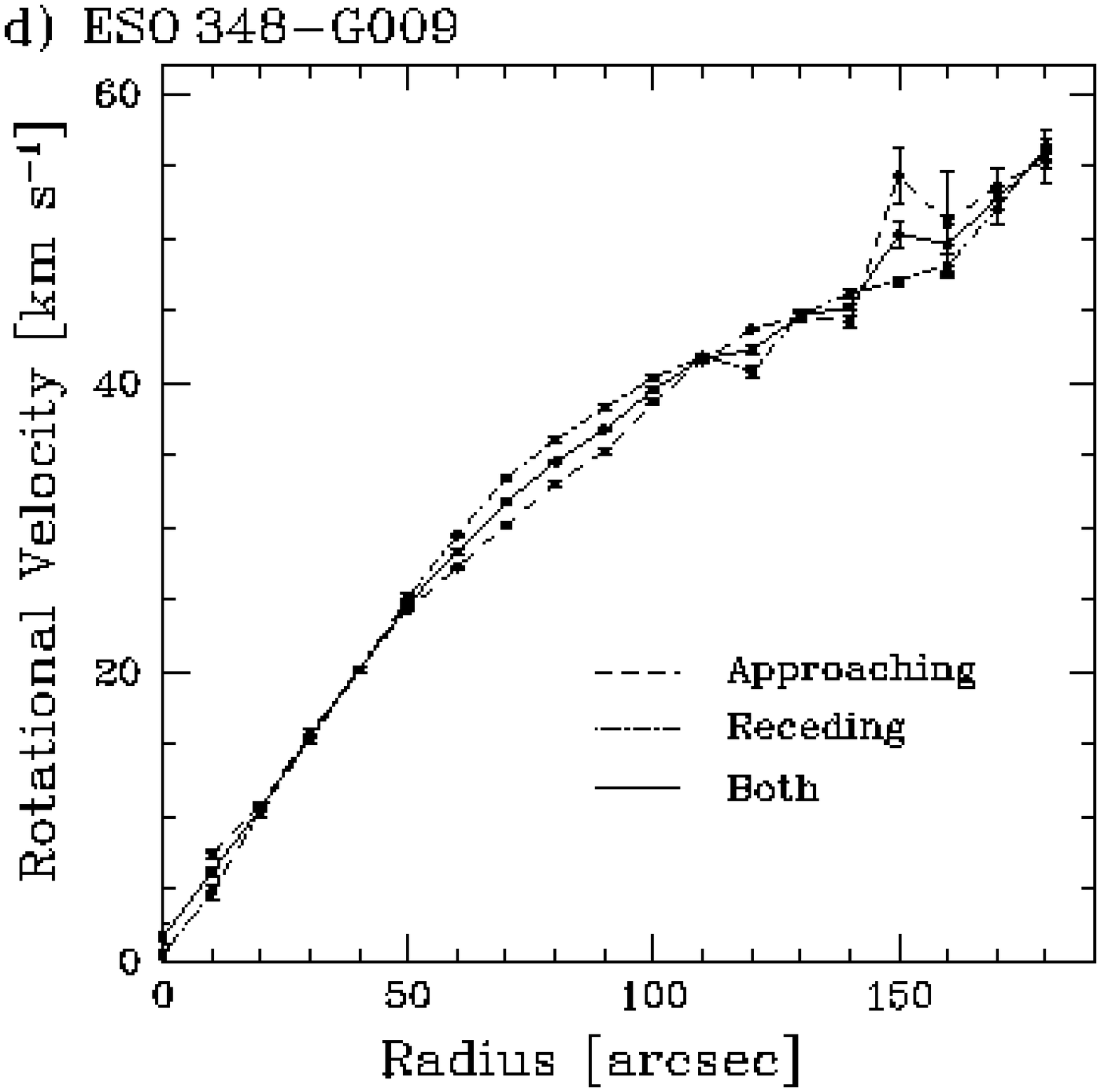} \\
\end{tabular}
\caption{The \hi{} rotation curves as derived from the mean \hi{} velocity field for the four sample galaxies to which they were fit.  The panels show {\em a}) ESO\,121-G020, {\em b}) ESO\,428-G033, {\em c}) \esoq{}, {\em d}) ESO\,348-G009.  The three curves which are in all panels correspond to the approaching side (dashed line), receding side (dot-dashed line), and both sides (solid line).  The dotted curves in the ESO\,428-G033 panel show the difference which the inclination uncertainty makes for that galaxy.
\label{fig:hirotcur}}
\end{figure*}

The results of fitting rotation curves to the \hi{} velocity fields of these four galaxies is listed in Table~\ref{tab:rot}. The columns are as follows: (1) galaxy name; (2) systemic velocity; (3) position angle of the galaxy's receding side; (4) inclination angle; (5) maximum rotation velocity; and (6) maximum radius.  The final rotation curves are shown in Fig.~\ref{fig:hirotcur}.  The very low inclination of ESO\,428-G033 produced a high uncertainty in the rotation velocity values as shown by the curves plotted.  In the case of ESO\,348-G009 the curve appears to still be rising at the last points of the rotation curve, suggesting that the \hi{} is not tracing the galaxy out to the radius at which the maximum rotation velocity is reached (so \vmax{} should be considered a lower limit in this case).  Similarly for ESO\,121-G020, where we may just be reaching the point where the curve flattens out.  In the other two galaxies we appear to have reached the flat part of the curve.

\section{Optical Properties}
\label{sec:optp}

\begin{deluxetable}{lccccccc} 
\tabletypesize{\scriptsize}
\tablecaption{2.3m Telescope Optical Results.
  \label{tab:opt}}
\tablewidth{0pt}
\tablehead{ \colhead{Name} & \colhead{Band} & \colhead{$m_{T}$\tablenotemark{a}} & \colhead{$\mu_{0}$\tablenotemark{a}} & \colhead{$\langle \mu \rangle _{\rm eff}$\tablenotemark{a}} & \colhead{$r_{\rm eff}$} & \colhead{$r_{\rm H,0}$} & \colhead{\AG{}} \\

 \colhead{} & \colhead{} & \colhead{(mag)} & \colhead{(mag arcsec$^{-2}$)} & \colhead{(mag arcsec$^{-2}$)} & \colhead{(arcsec)} & \colhead{(arcsec)} & \colhead{(mag)} \\

 \colhead{(1)} & \colhead{(2)} & \colhead{(3)} & \colhead{(4)} & \colhead{(5)} & \colhead{(6)} & \colhead{(7)} & \colhead{(8)}}
\startdata
 \mcg{}        & {\em B} & $15.32 \pm 0.06$ & $22.34 \pm 0.02$ & $23.77 \pm 0.04$ & $19.6 \pm 0.7$ & $51.0 \pm 2.0$ & $0.08 \pm 0.01$ \\
               & {\em V} & $14.91 \pm 0.05$ & $21.89 \pm 0.01$ & $22.98 \pm 0.04$ & $16.4 \pm 0.6$ & -- & $0.06 \pm 0.01$ \\
               & {\em R} & $14.40 \pm 0.05$ & $21.60 \pm 0.01$ & $22.68 \pm 0.05$ & $18.1 \pm 0.7$ & -- & $0.05 \pm 0.01$ \\
               & {\em I} & $13.60 \pm 0.06$ & $21.17 \pm 0.01$ & $22.76 \pm 0.05$ & $27.1 \pm 1.0$ & -- & $0.04 \pm 0.01$ \\ \\

 ESO\,473-G024 & {\em B} & $16.38 \pm 0.06$ & $24.63 \pm 0.04$ & $25.30 \pm 0.04$ & $24.3 \pm 0.8$ & $32.4 \pm 0.8$ & $0.08 \pm 0.01$ \\
               & {\em V} & $15.44 \pm 0.07$ & $23.73 \pm 0.03$ & $24.46 \pm 0.03$ & $25.4 \pm 0.7$ & -- & $0.06 \pm 0.01$ \\
               & {\em R} & $15.19 \pm 0.03$ & $23.51 \pm 0.03$ & $24.18 \pm 0.02$ & $25.1 \pm 0.4$ & -- & $0.05 \pm 0.01$ \\
               & {\em I} & $14.76 \pm 0.07$ & $23.42 \pm 0.08$ & $24.35 \pm 0.02$ & $33.0 \pm 0.9$ & -- & $0.04 \pm 0.01$ \\ \\

 ESO\,121-G020 & {\em B} & $15.27 \pm 0.05$ & $23.36 \pm 0.03$ & $23.95 \pm 0.02$ & $21.7 \pm 0.4$ & $47.0 \pm 2.0$ & $0.17 \pm 0.03$ \\
               & {\em R} & $14.62 \pm 0.04$ & $22.71 \pm 0.03$ & $23.29 \pm 0.02$ & $21.6 \pm 0.5$ & -- & $0.11 \pm 0.02$ \\
 ~~\atg{}      & {\em B} & $17.01 \pm 0.06$ & $22.81 \pm 0.04$ & $23.34 \pm 0.02$ & $7.4 \pm 0.3$ & $18.6 \pm 0.6$ & $0.17 \pm 0.03$ \\
               & {\em R} & $16.36 \pm 0.06$ & $22.07 \pm 0.02$ & $22.72 \pm 0.02$ & $7.5 \pm 0.2$ & -- & $0.11 \pm 0.02$ \\ \\

 ESO\,428-G033 & {\em B} & $16.90 \pm 0.10$ & $23.33 \pm 0.11$ & $24.69 \pm 0.02$ & $14.4 \pm 0.5$ & $37.2 \pm 1.2$ & $1.10 \pm 0.18$ \\
              & {\em V} & $16.13 \pm 0.10$ & $23.12 \pm 0.02$ & $24.23 \pm 0.05$ & $16.7 \pm 0.8$ & -- & $0.85 \pm 0.14$ \\
              & {\em R} & $15.61 \pm 0.08$ & $22.56 \pm 0.02$ & $23.48 \pm 0.05$ & $15.0 \pm 0.9$ & -- & $0.68 \pm 0.11$ \\
              & {\em I} & $15.04 \pm 0.09$ & $22.01 \pm 0.03$ & $22.95 \pm 0.04$ & $15.2 \pm 0.7$ & -- & $0.50 \pm 0.08$ \\ \\

 \esoq{}      & {\em B} & $16.13 \pm 0.07$ & $24.97 \pm 0.03$ & $25.48 \pm 0.02$ & $29.7 \pm 0.6$ & $57.6 \pm 0.6$ & $0.95 \pm 0.15$ \\
              & {\em V} & $14.89 \pm 0.06$ & $23.65 \pm 0.03$ & $24.14 \pm 0.02$ & $28.3 \pm 0.7$ & -- & $0.73 \pm 0.12$ \\
              & {\em R} & $14.38 \pm 0.05$ & $23.16 \pm 0.02$ & $23.64 \pm 0.03$ & $28.4 \pm 0.5$ & -- & $0.59 \pm 0.09$ \\
              & {\em I} & $13.76 \pm 0.06$ & $22.91 \pm 0.04$ & $23.40 \pm 0.03$ & $33.9 \pm 0.8$ & -- & $0.43 \pm 0.07$ \\ \\

 ESO\,572-G009 & {\em B} & $16.79 \pm 0.05$ & $24.96 \pm 0.05$ & $26.03 \pm 0.02$ & $28.1 \pm 0.7$ & $30.6 \pm 1.2$ & $0.16 \pm 0.03$ \\
              & {\em V} & $15.65 \pm 0.07$ & $23.69 \pm 0.02$ & $25.12 \pm 0.02$ & $31.1 \pm 0.8$ & -- & $0.12 \pm 0.02$ \\
              & {\em R} & $15.40 \pm 0.04$ & $23.42 \pm 0.03$ & $24.75 \pm 0.01$ & $29.6 \pm 0.5$ & -- & $0.10 \pm 0.02$ \\ \\

 ESO\,505-G007 & {\em B} & $14.20 \pm 0.06$ & $23.99 \pm 0.04$ & $24.00 \pm 0.01$ & $36.4 \pm 0.7$ & $82.8 \pm 1.2$ & $0.36 \pm 0.06$ \\
              & {\em V} & $14.48 \pm 0.05$ & $23.89 \pm 0.04$ & $23.96 \pm 0.01$ & $31.3 \pm 0.6$ & -- & $0.28 \pm 0.04$ \\
              & {\em R} & $13.97 \pm 0.04$ & $23.46 \pm 0.04$ & $23.66 \pm 0.02$ & $34.6 \pm 0.5$ & -- & $0.22 \pm 0.04$ \\ \\

 IC\,4212     & {\em B} & $14.11 \pm 0.04$ & $22.38 \pm 0.02$ & $24.42 \pm 0.02$ & $46.0 \pm 1.0$ & $98.4 \pm 0.6$ & $0.19 \pm 0.03$ \\
              & {\em V} & $13.69 \pm 0.05$ & $22.00 \pm 0.03$ & $23.98 \pm 0.01$ & $45.7 \pm 0.9$ & -- & $0.14 \pm 0.02$ \\
              & {\em R} & $13.29 \pm 0.06$ & $21.53 \pm 0.02$ & $23.53 \pm 0.02$ & $44.5 \pm 0.9$ & -- & $0.12 \pm 0.02$ \\ \\

 ESO\,348-G009 & {\em B} & $14.81 \pm 0.07$ & $23.78 \pm 0.04$ & $24.79 \pm 0.03$ & $39.5 \pm 0.9$ & $72.0 \pm 0.6$ & $0.06 \pm 0.01$ \\
              & {\em V} & $14.68 \pm 0.06$ & $23.42 \pm 0.03$ & $24.40 \pm 0.03$ & $35.1 \pm 1.0$ & -- & $0.04 \pm 0.01$ \\
              & {\em R} & $14.48 \pm 0.05$ & $23.13 \pm 0.03$ & $24.03 \pm 0.03$ & $32.4 \pm 0.8$ & -- & $0.04 \pm 0.01$ \\
              & {\em I} & $13.75 \pm 0.10$ & $22.75 \pm 0.03$ & $23.72 \pm 0.02$ & $39.3 \pm 1.1$ & -- & $0.03 \pm 0.01$ \\
\enddata
\tablenotetext{a}{Correction for Galactic extinction not applied.}
\end{deluxetable}

\begin{figure*} 
\begin{tabular}{@{}lll@{}}
 \includegraphics[width=5.5cm]{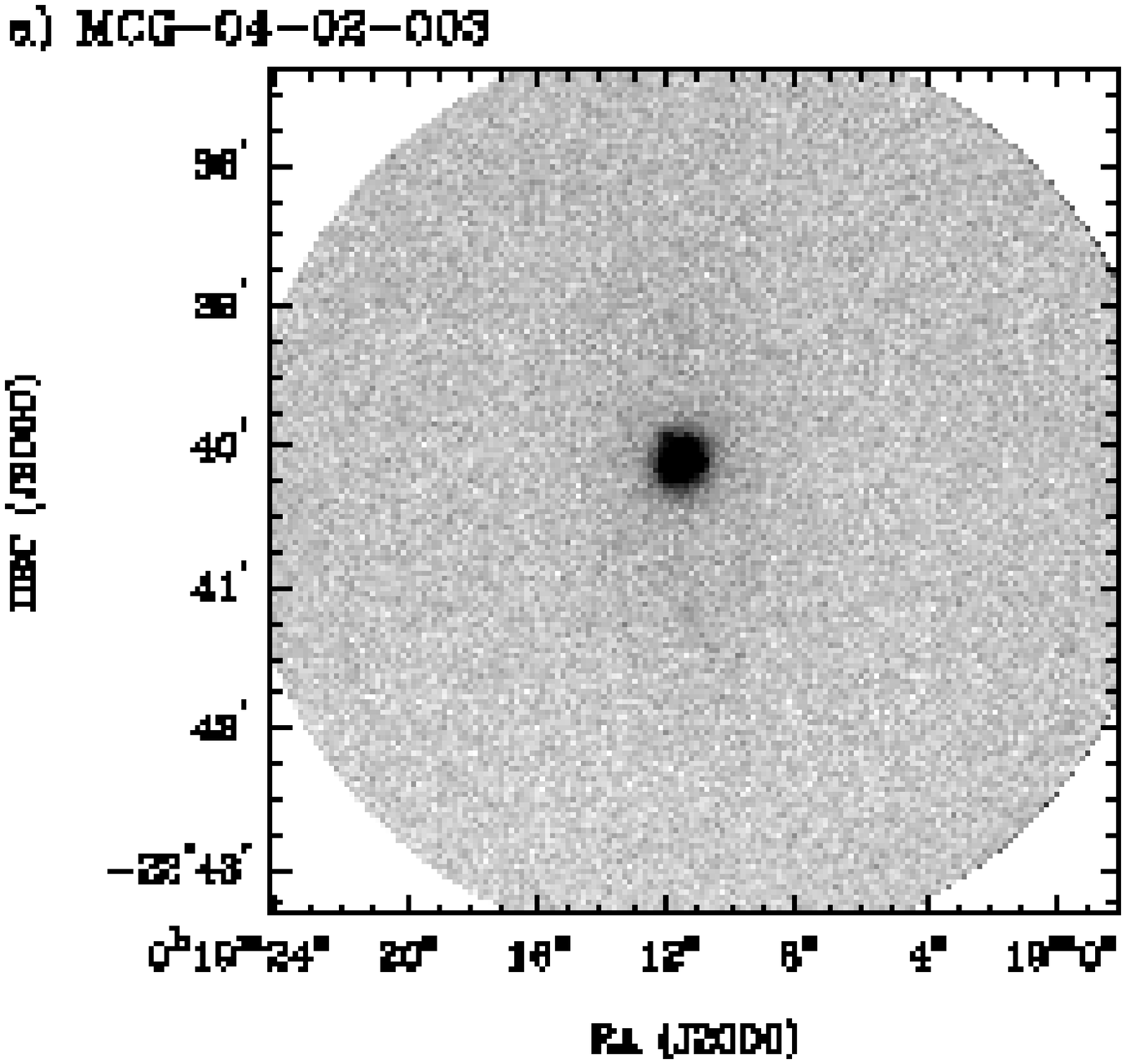} &
 \includegraphics[width=5.5cm]{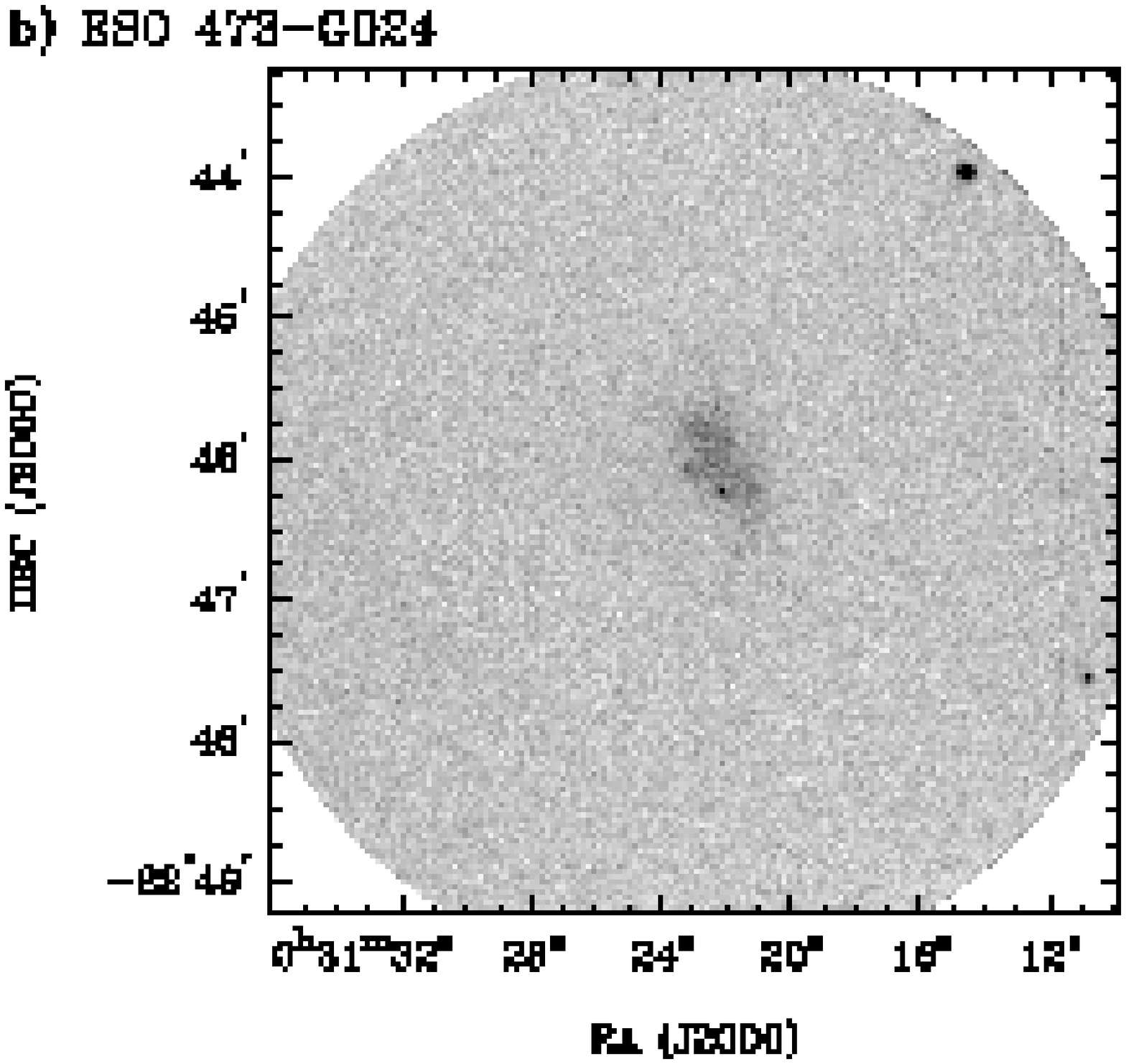} &
 \includegraphics[width=5.5cm]{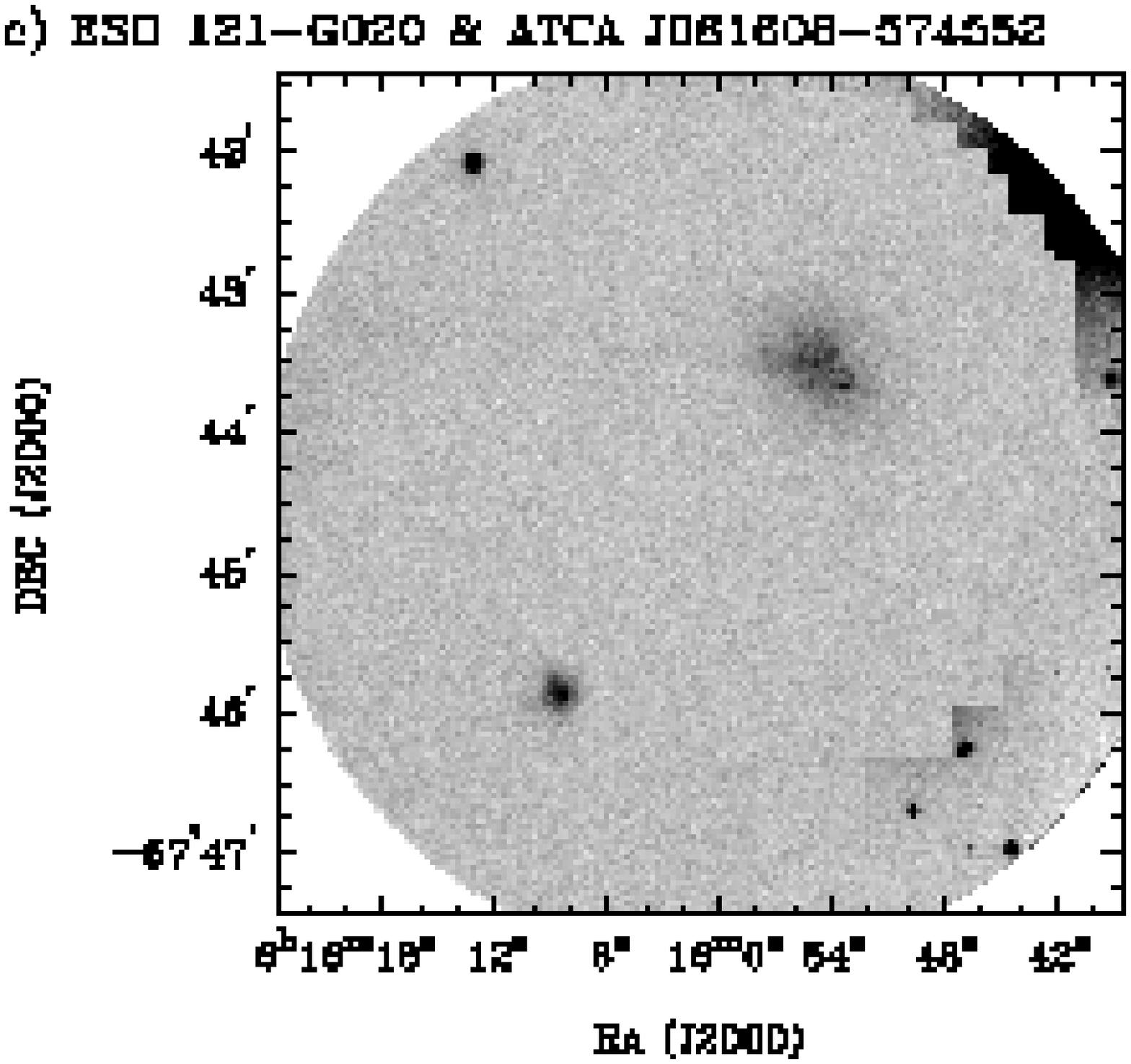} \\ \\

 \includegraphics[width=5.5cm]{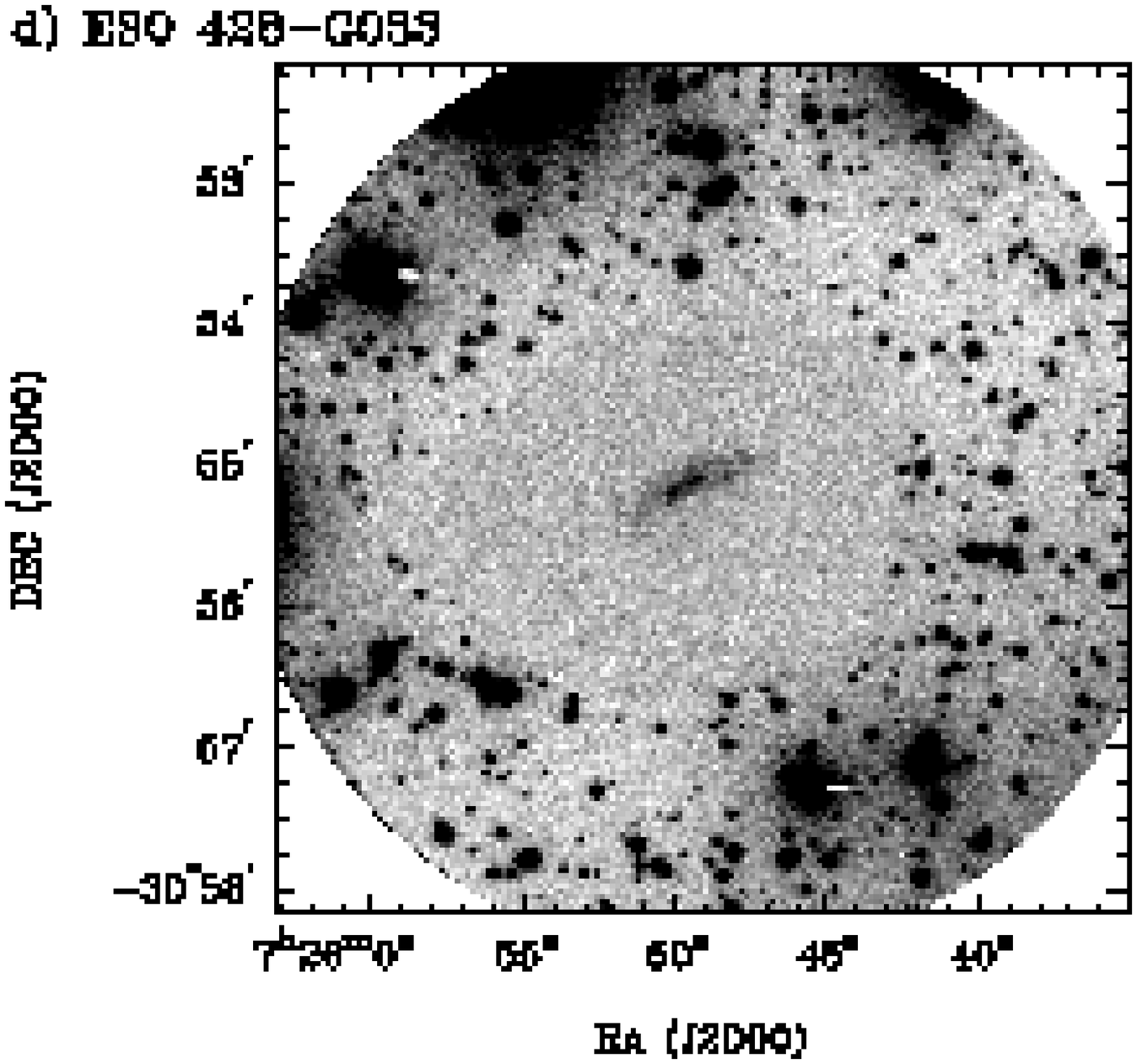} &
 \includegraphics[width=5.5cm]{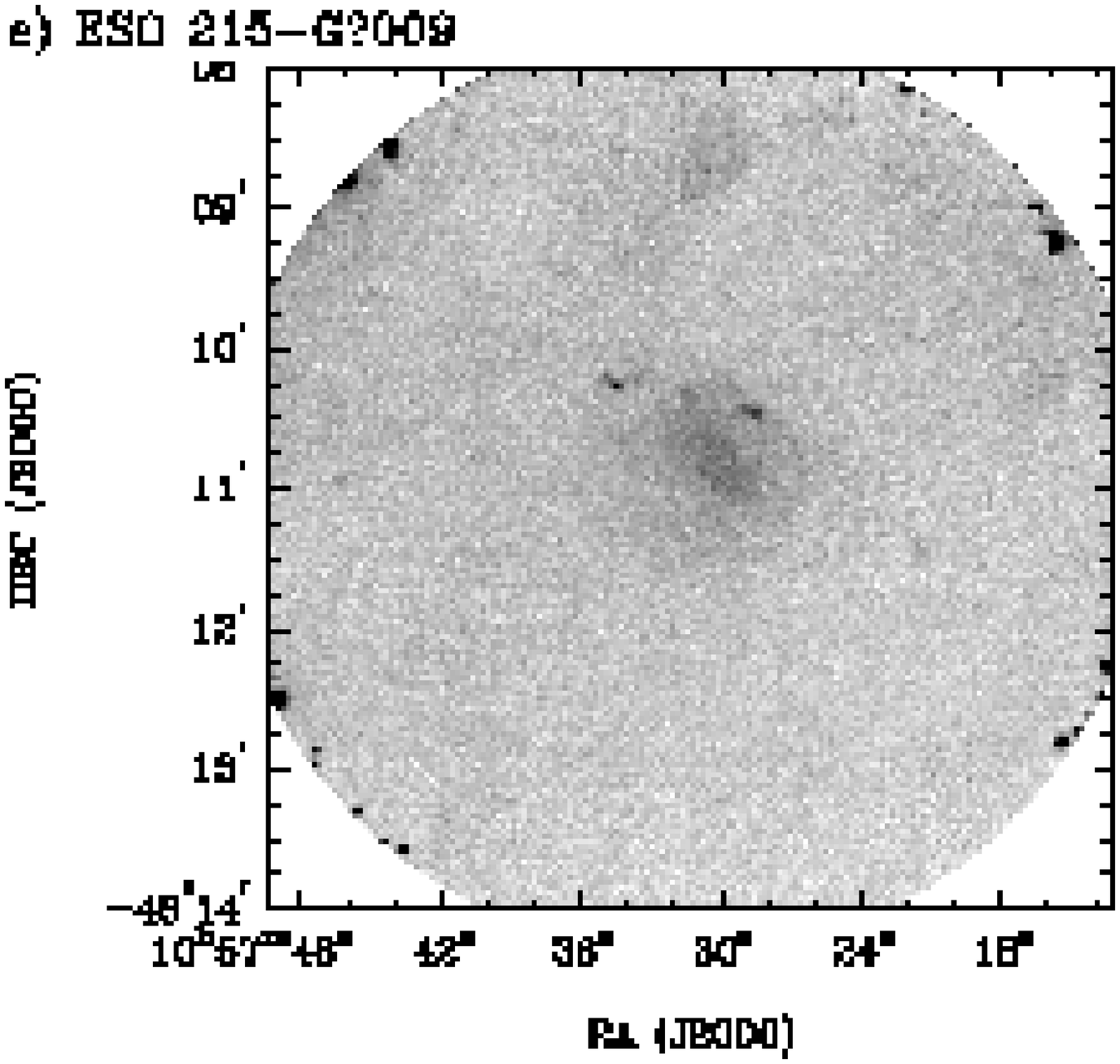} & 
 \includegraphics[width=5.5cm]{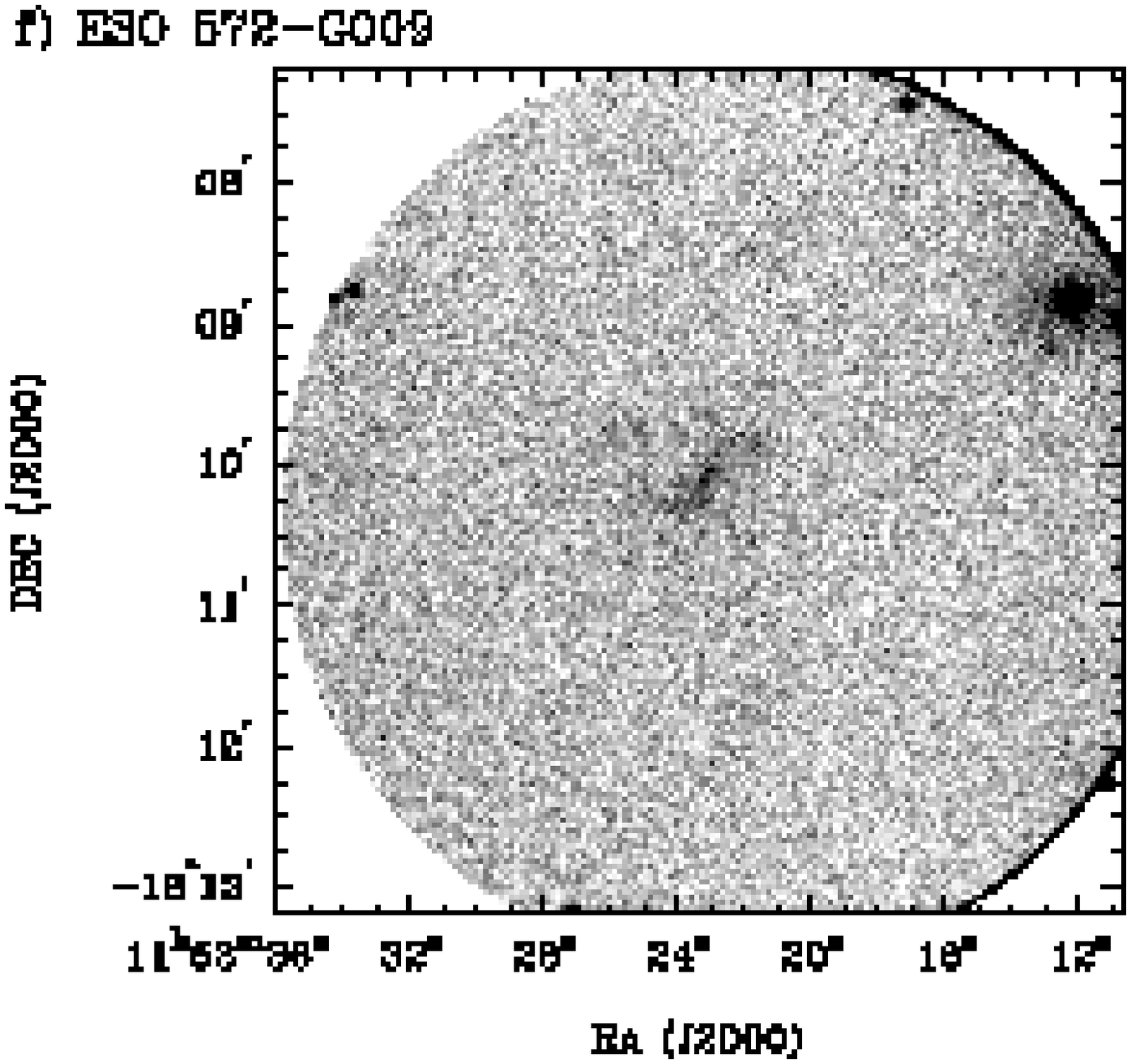} \\ \\

 \includegraphics[width=5.5cm]{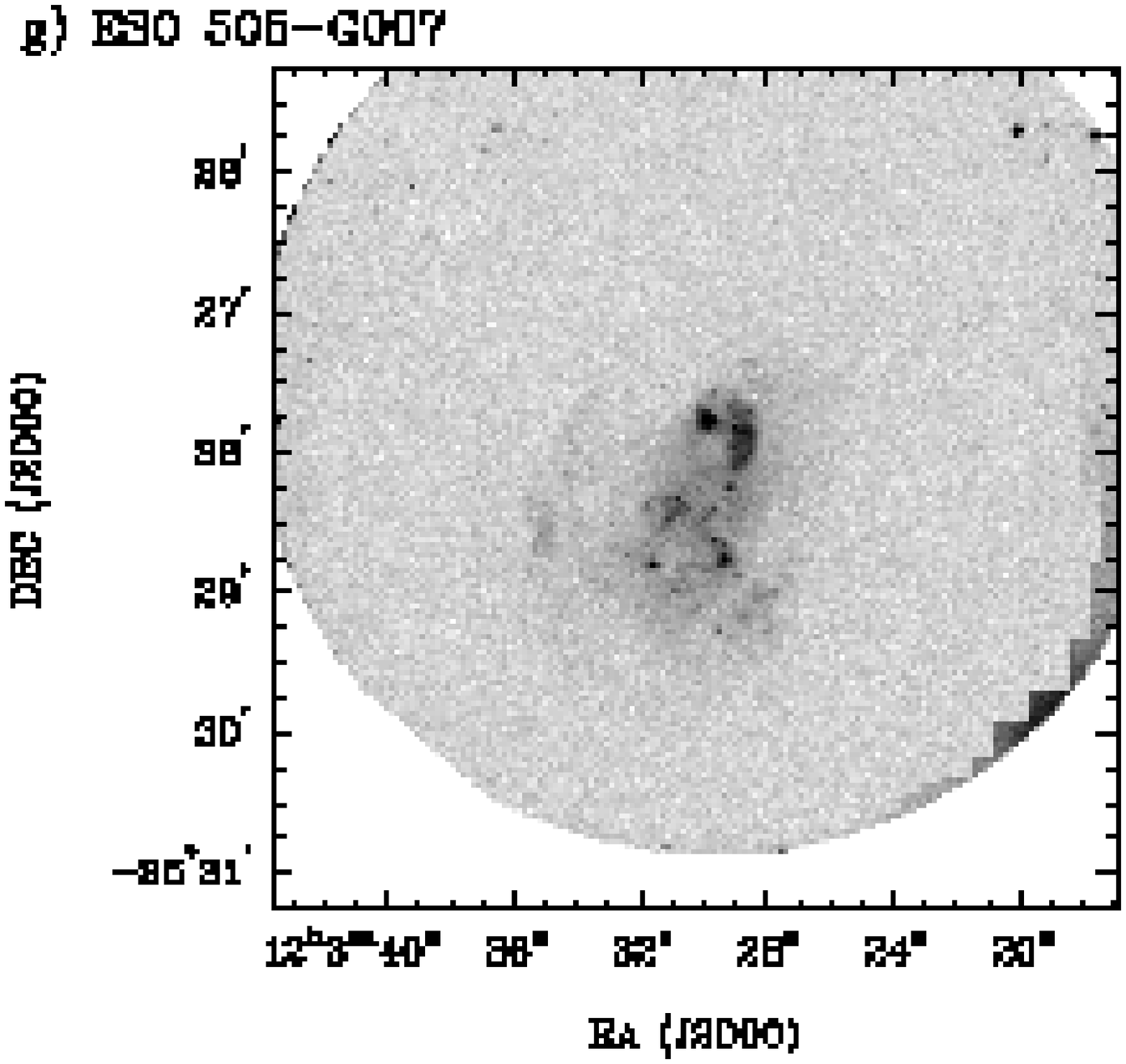} &
 \includegraphics[width=5.5cm]{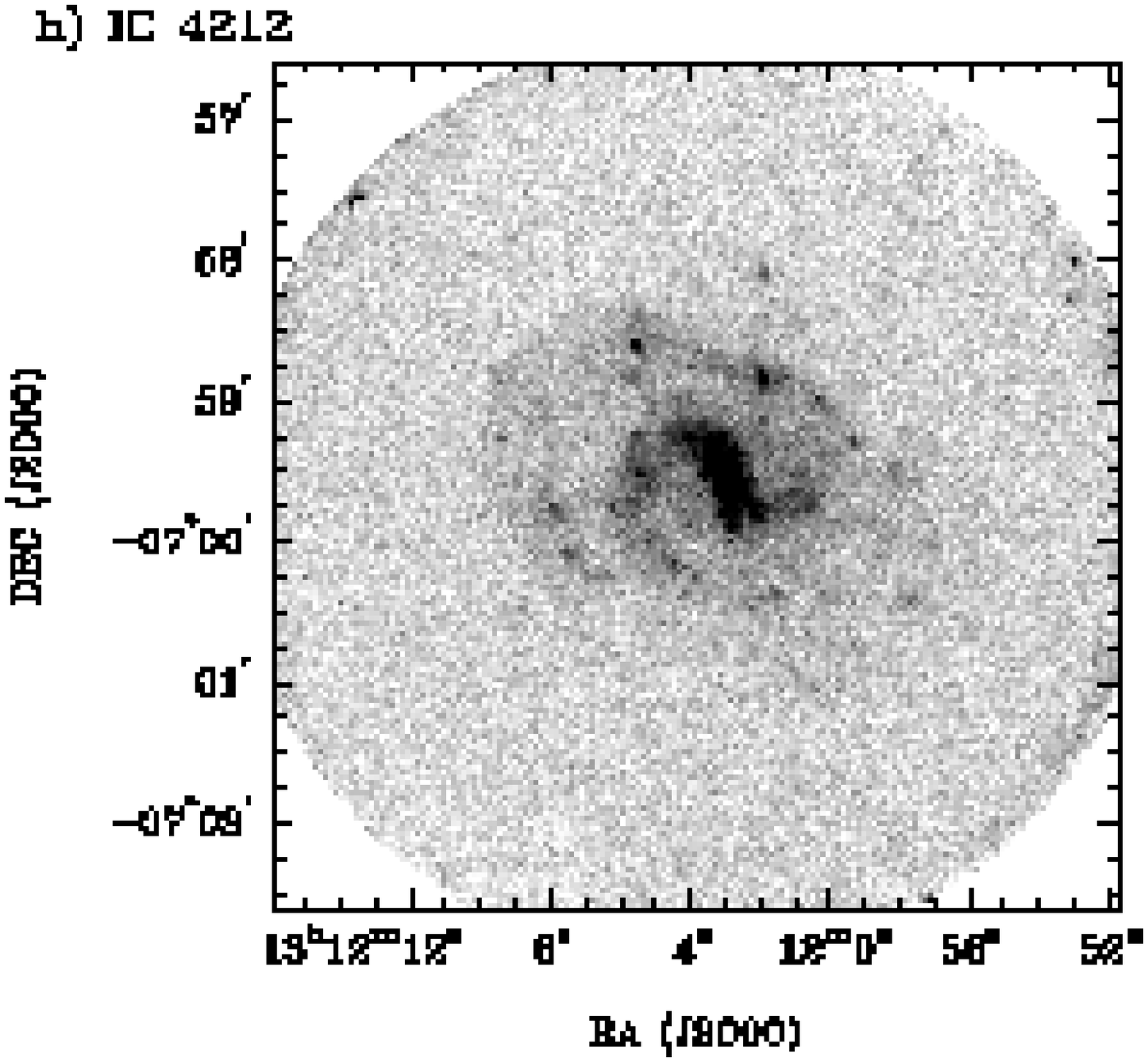} & 
 \includegraphics[width=5.5cm]{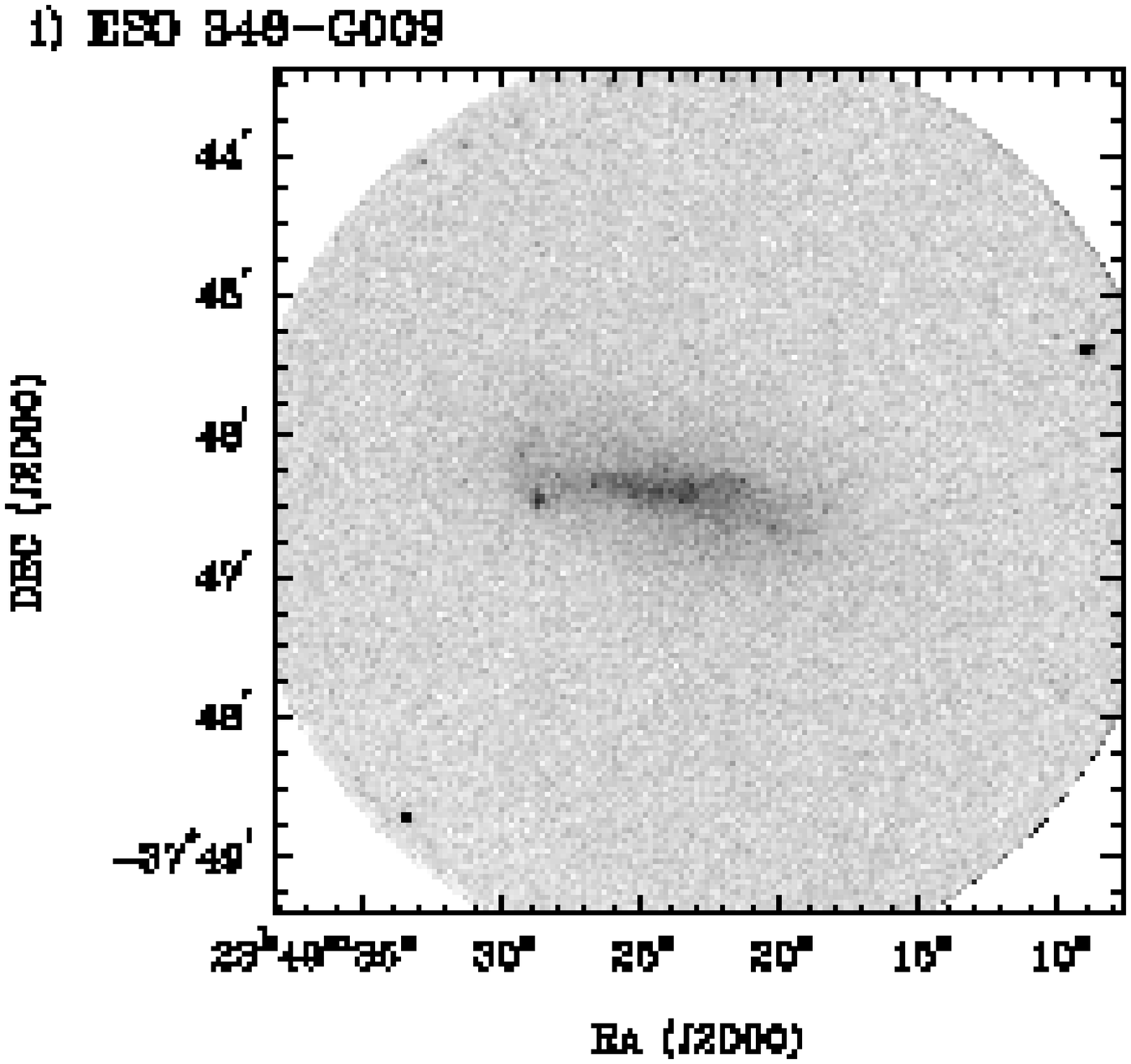} \\
\end{tabular}
\caption{{\em B} images of all galaxies after the foreground stars subtraction process.  The panels show {\em a}) \mcg{}, {\em b}) ESO\,473-G024, {\em c}) ESO\,121-G020 \& \atg{}, {\em d}) ESO\,428-G033, {\em e}) \esoq{}, {\em f}) ESO\,572-G009, {\em g}) ESO\,505-G007, {\em h}) IC\,4212, and {\em i}) ESO\,348-G009.
\label{fig:starsub}}
\end{figure*}

The observed {\em B} images taken on the 2.3m Telescope of all nine galaxies are in Fig.~\ref{fig:optimage}, while Fig.~\ref{fig:starsub} shows the same images after star subtraction.  Star contamination is obviously a problem for \esoq{} and especially ESO\,428-G033, both of which are close to the Galactic Plane, so special care was taken with these images.  Some of the galaxies (notably \mcg{}, \esoq{}, ESO\,505-G007 and ESO\,348-G009) were affected by moderately bright foreground stars sitting on top of the galaxy that had to be removed with care.  As mentioned in \S~\ref{sec:radio}, a companion galaxy to ESO\,121-G020 was found in the \hi{} imaging.  The optical image for ESO\,121-G020 also includes \atg{}, and the photometry for that galaxy is included with the other results here.  For each galaxy a growth curve was measured on the star-subtracted images from the luminosity weighted center in 2 pixel ($\sim$1\farcs2) circular aperture rings to obtain the total intensity and a surface brightness profile.

Our results for the optical {\em BVRI} photometry taken on the 2.3m Telescope are in Table~\ref{tab:opt}.  No Galactic extinction correction was applied to the apparent magnitude and surface brightness.  The columns are as follows: (1) galaxy name; (2) broad band (Cousins) filter used; (3) total apparent magnitude; (4) central surface brightness; (5) effective surface brightness (i.e. the average surface brightness out to the half light radius); (6) half light (effective) radius; (7) radius out to $\mu = 26.6$~mag arcsec$^{-2}$ (i.e. the Holmberg radius in the {\em B} band, extinction corrected); and (8) Galactic extinction correction from SFD98.

\begin{figure*} 
\begin{tabular}{@{}lll@{}}
 \includegraphics[width=5.5cm]{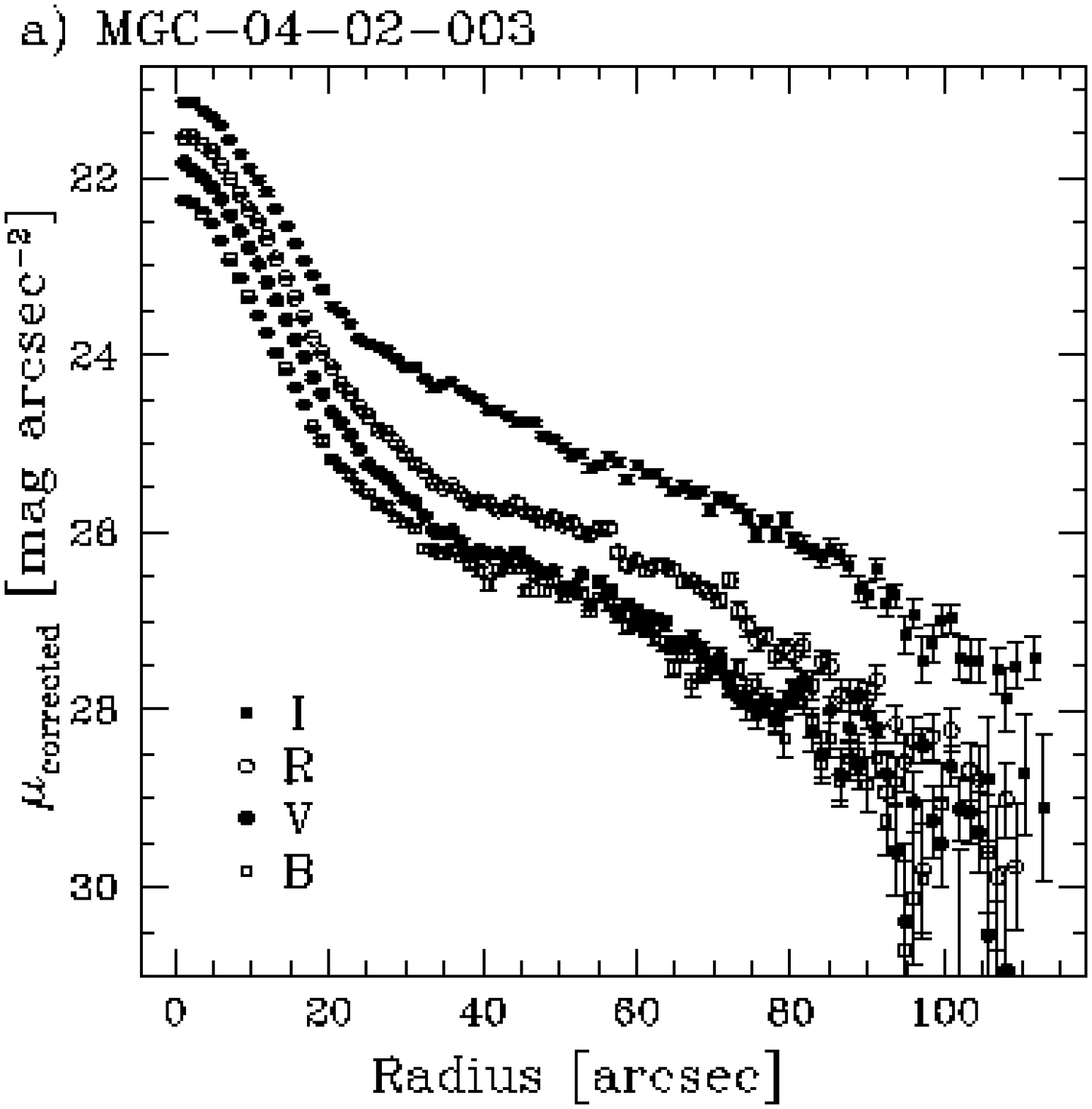} &
 \includegraphics[width=5.5cm]{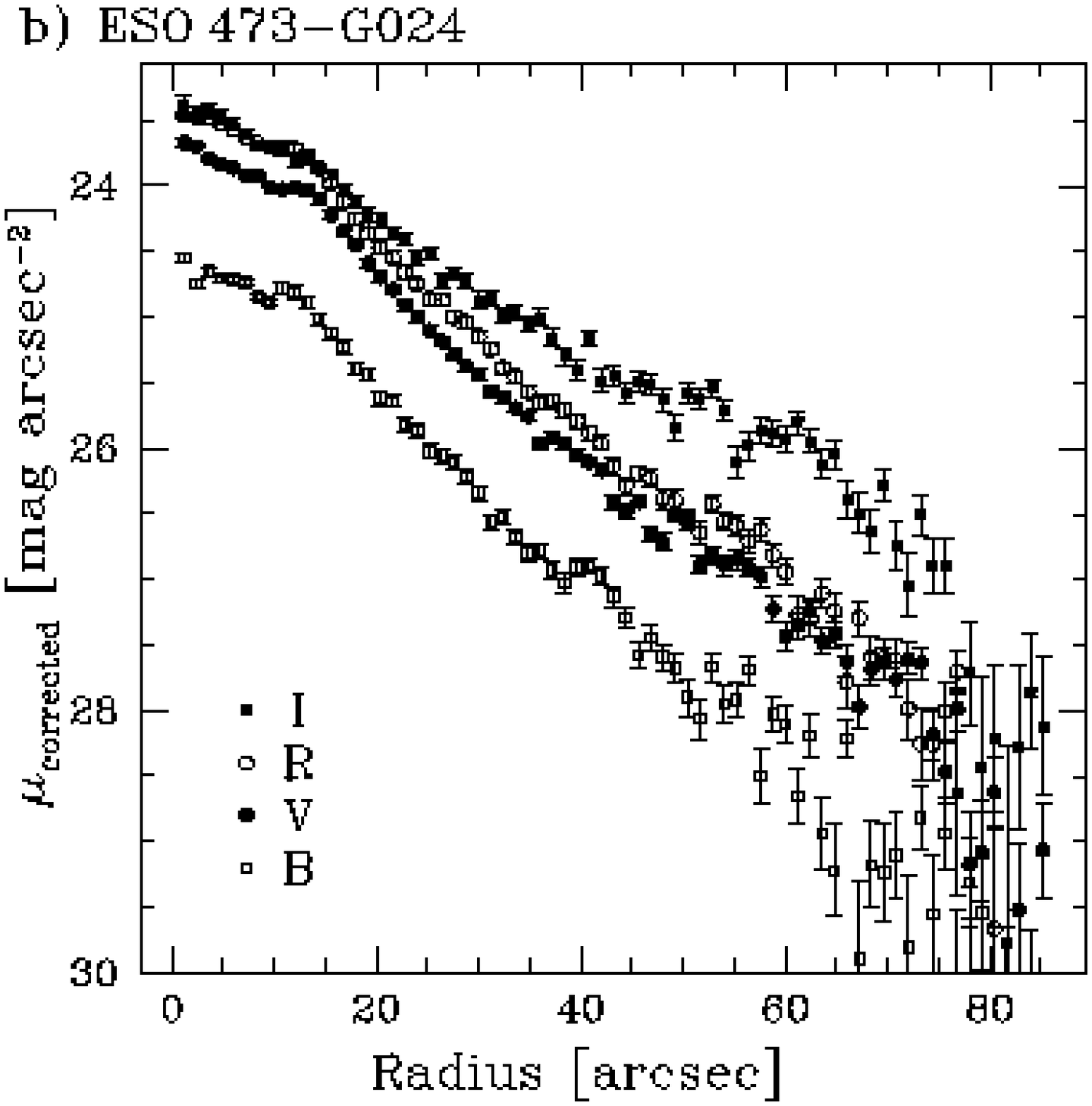} &
 \includegraphics[width=5.5cm]{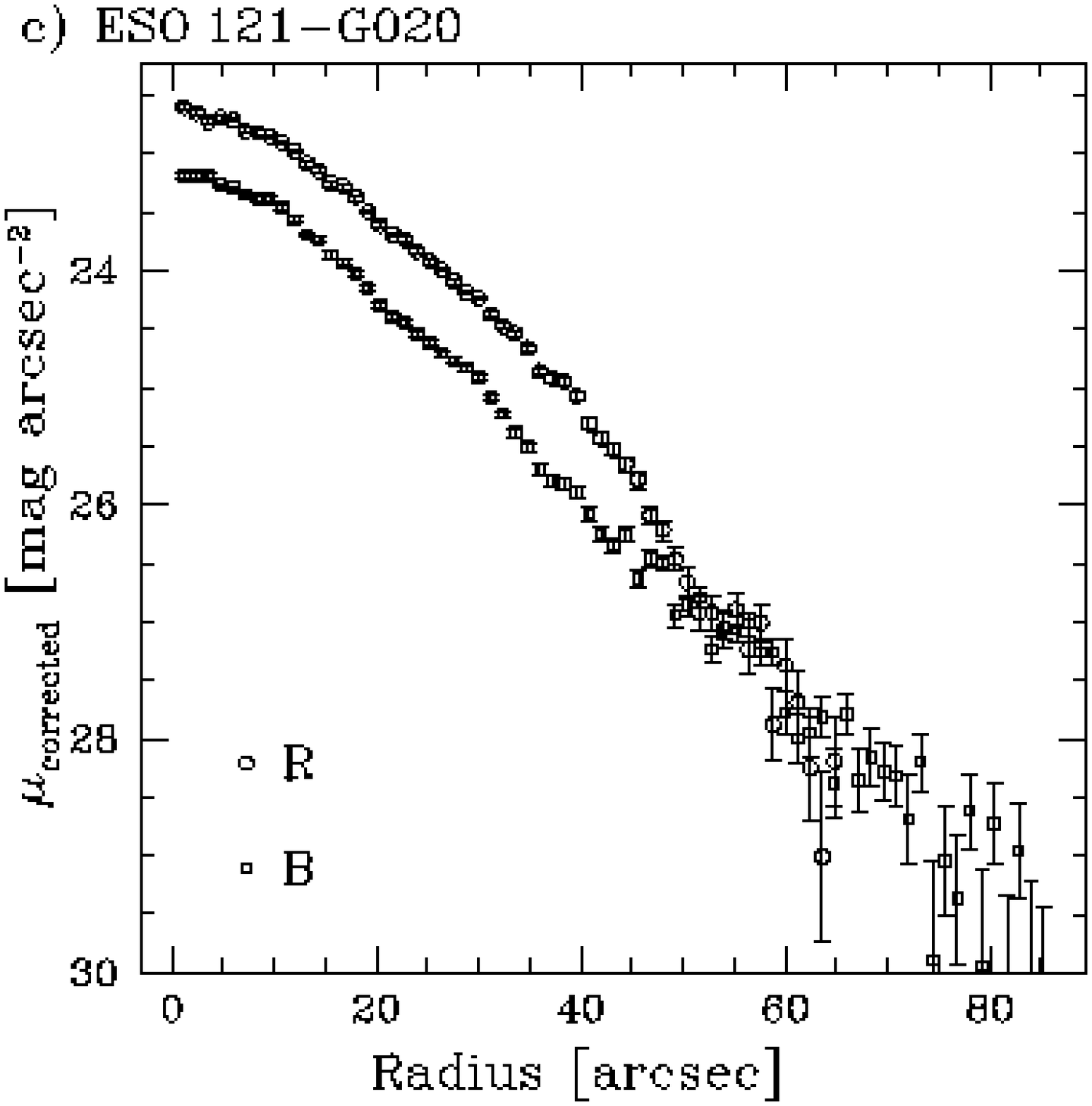} \\

 \includegraphics[width=5.5cm]{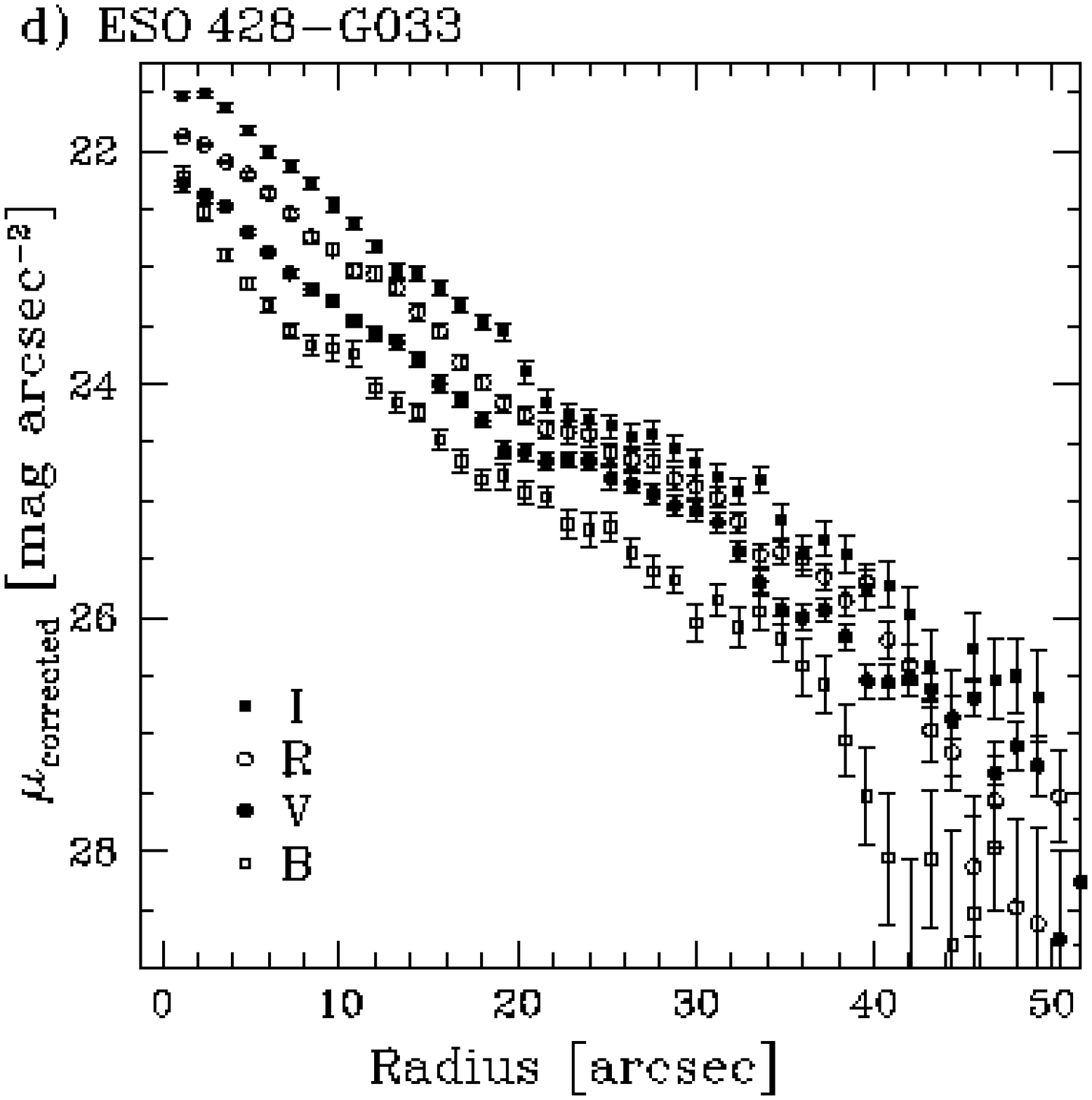} &
 \includegraphics[width=5.5cm]{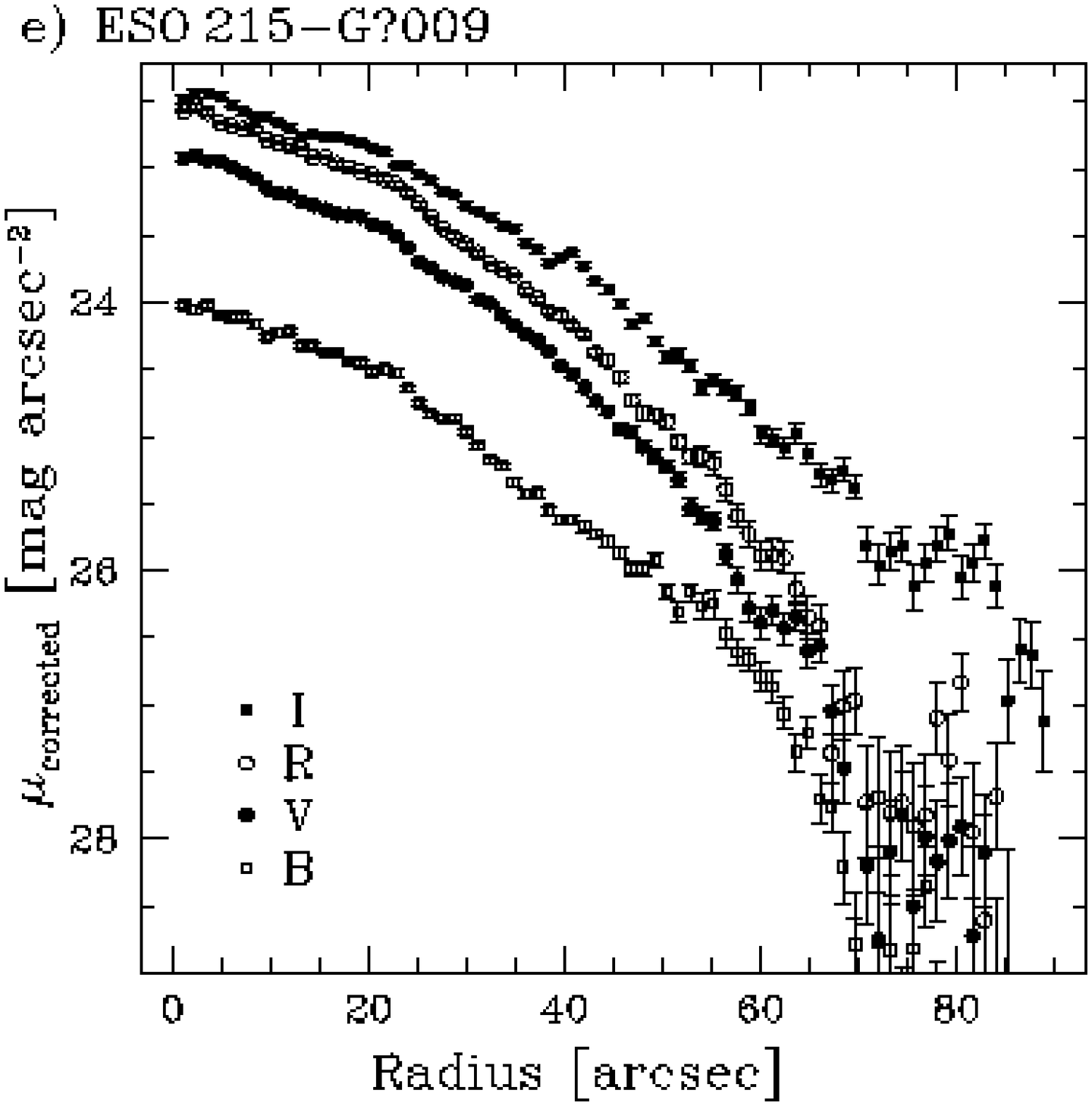} &
 \includegraphics[width=5.5cm]{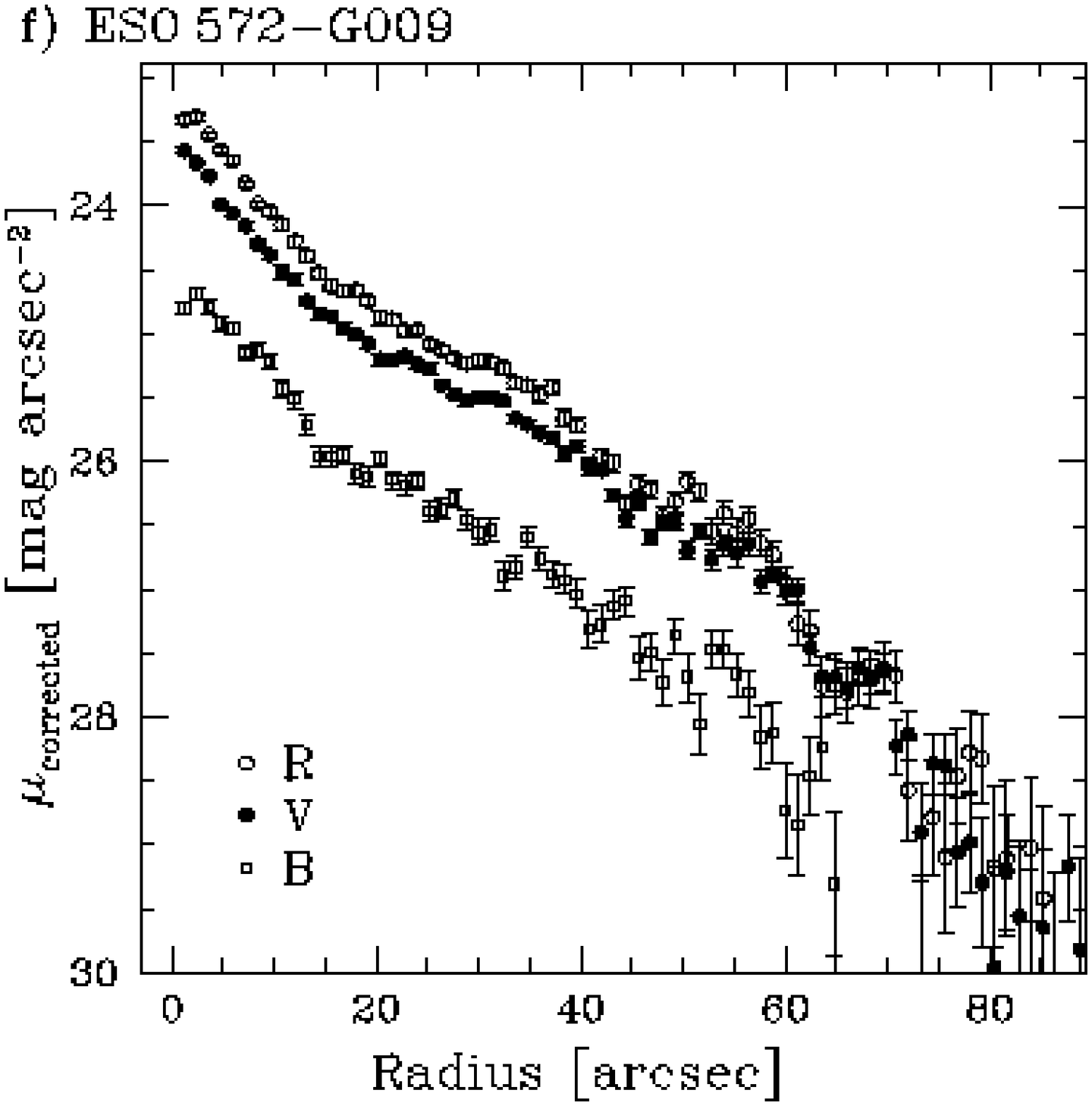} \\

 \includegraphics[width=5.5cm]{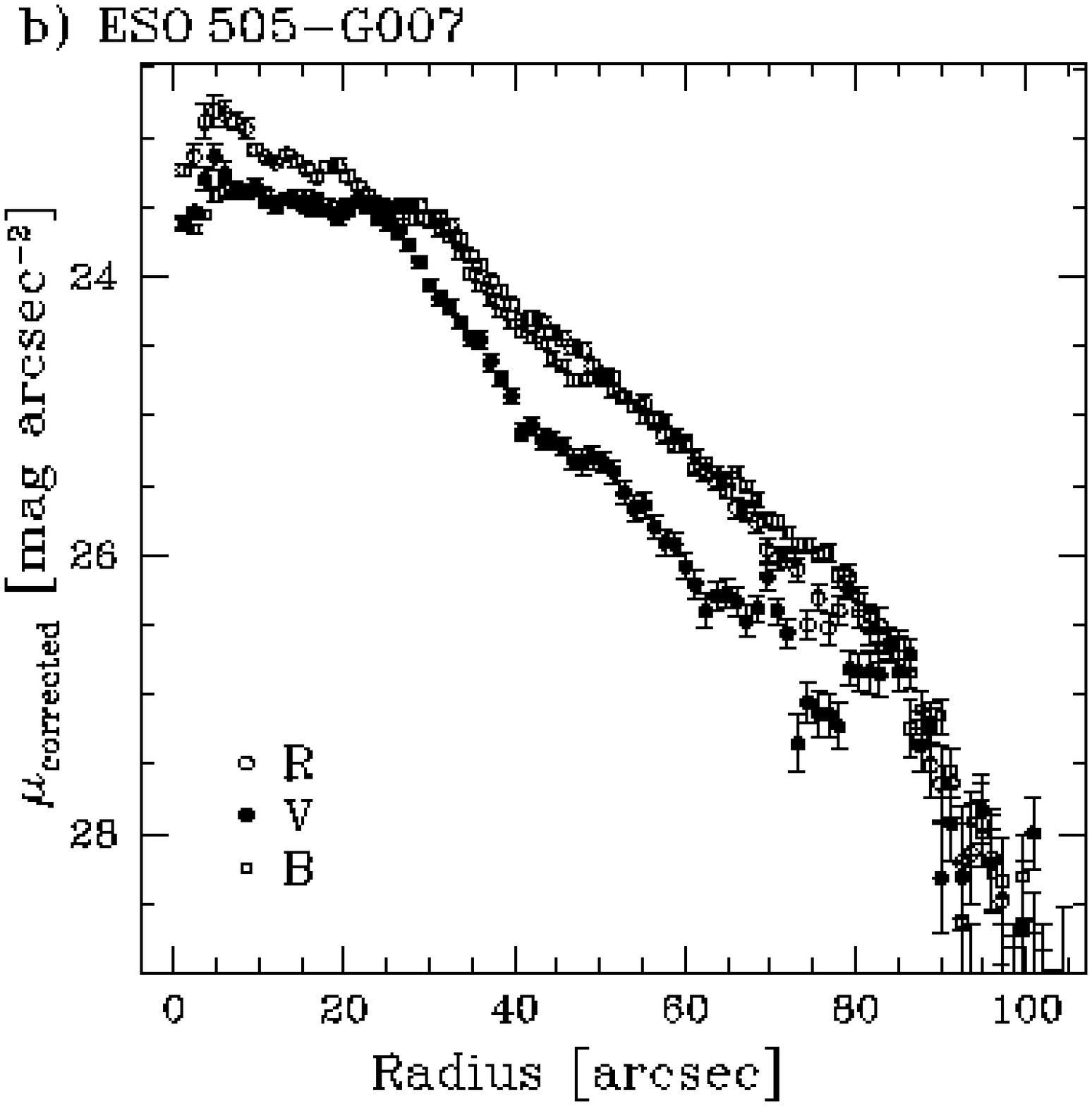} &
 \includegraphics[width=5.5cm]{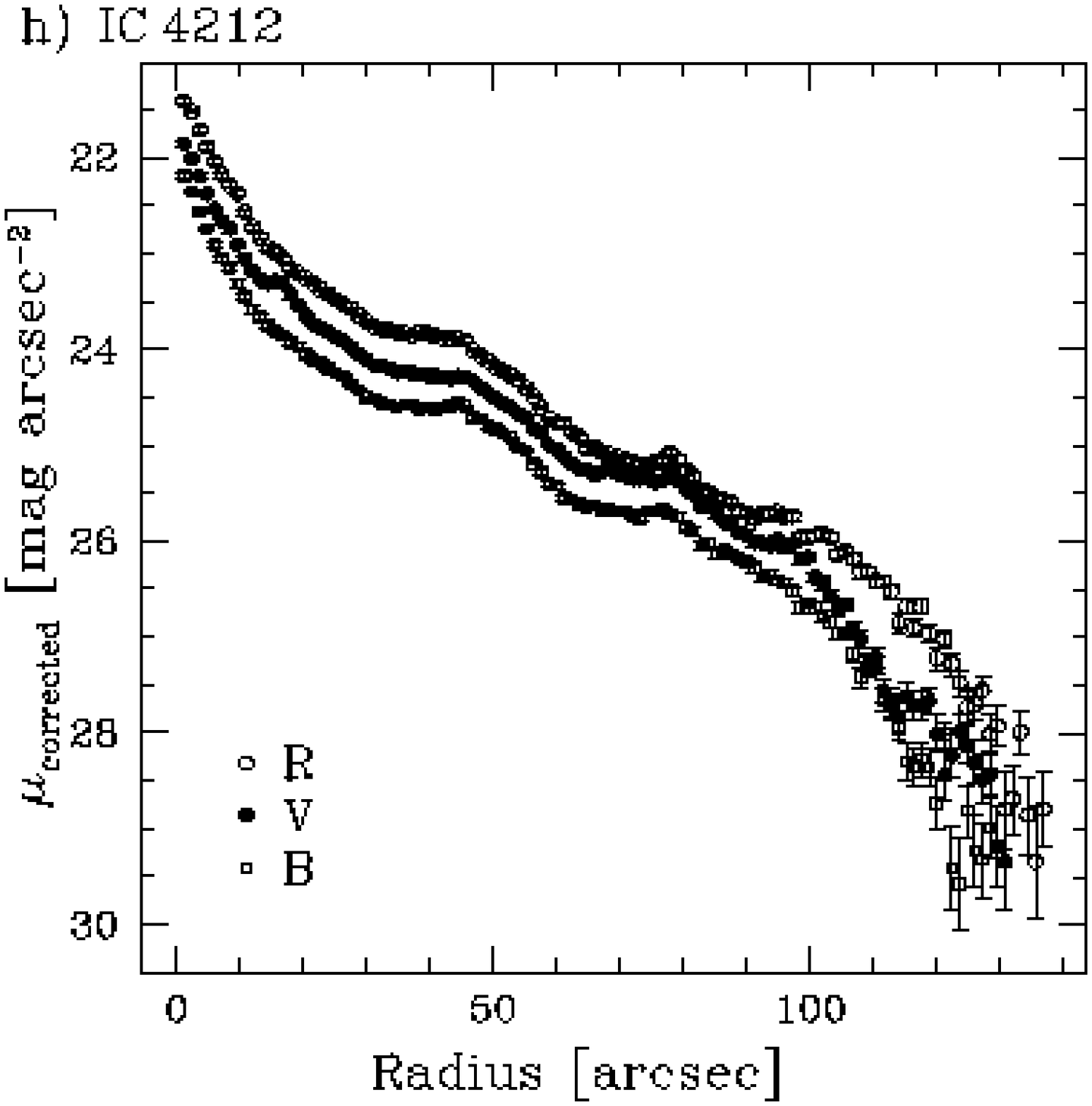} &
 \includegraphics[width=5.5cm]{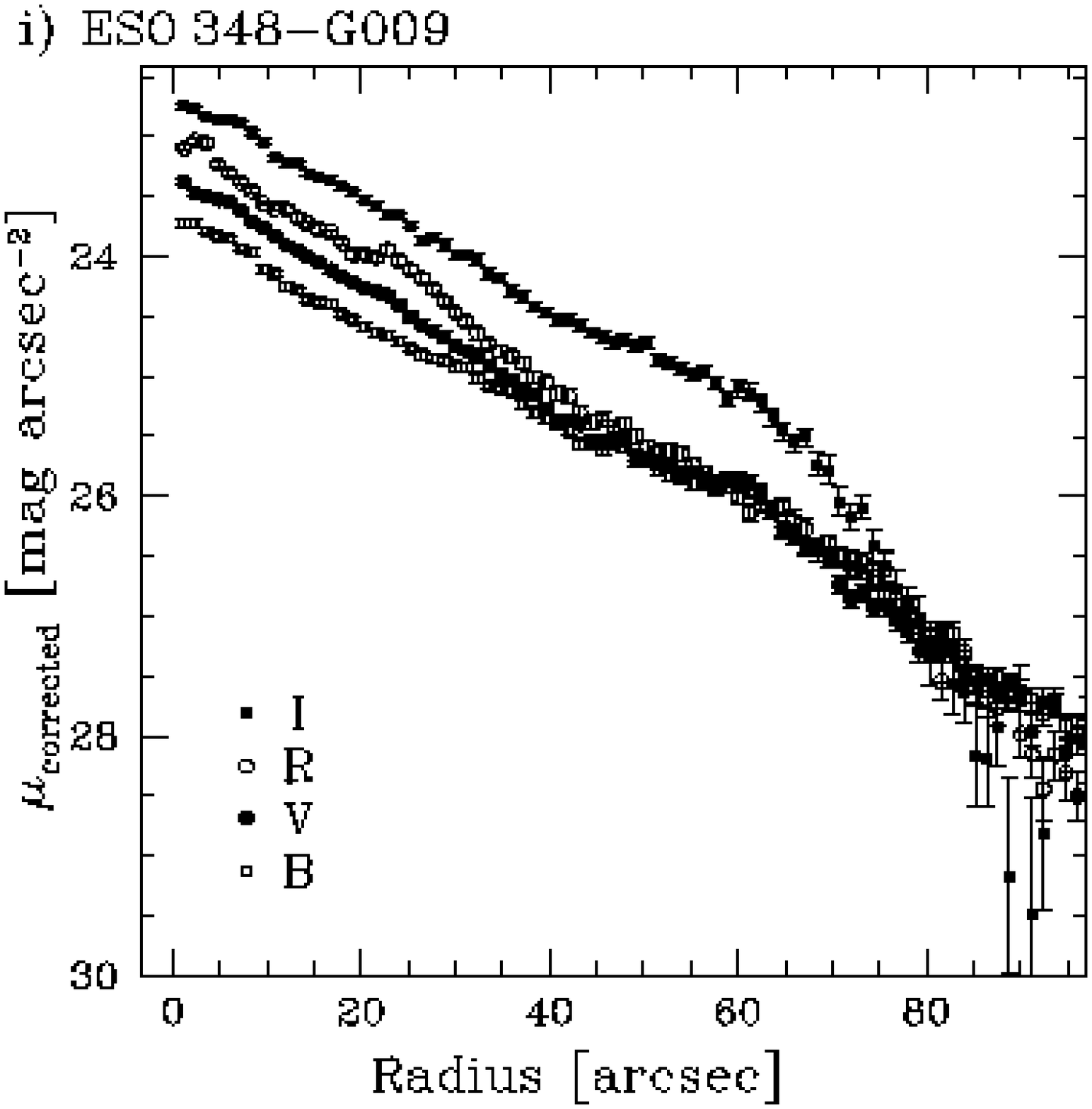} \\
\end{tabular}
\caption{{\em BVRI} surface brightness profiles of the nine galaxies.  The panels show {\em a}) \mcg{}, {\em b}) ESO\,473-G024, {\em c}) ESO\,121-G020, {\em d}) ESO\,428-G033, {\em e}) \esoq{}, {\em f}) ESO\,572-G009, {\em g}) ESO\,505-G007, {\em h}) IC\,4212, and {\em i}) ESO\,348-G009.  All profiles were corrected for Galactic extinction (see Table~\ref{tab:opt}).
\label{fig:sbpro}}
\end{figure*}

\begin{figure*} 
 \includegraphics[width=8cm]{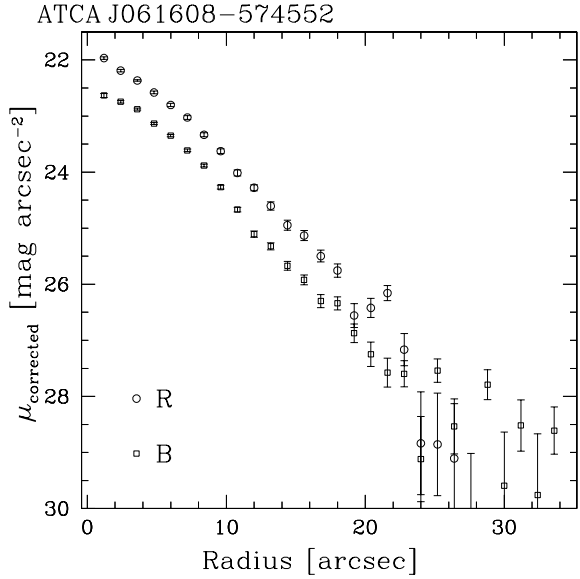}
\caption{{\em BR} surface brightness profiles for \atg{}.  Both profiles were corrected for Galactic extinction (see Table~\ref{tab:opt}).
\label{fig:sbpatca}}
\end{figure*}

Surface brightness profiles for the nine initially selected galaxies in each of the observed bands are in Fig.~\ref{fig:sbpro}, while the profile of \atg{} is in Fig.~\ref{fig:sbpatca}.  All profiles have been corrected for Galactic extinction, but no attempt to correct for inclination was made as it is difficult to calculate given the morphology of many galaxies, and as the correction for edge on galaxies (such as ESO\,348-G009) would be unrealistic without correction for the thickness of the stellar disk.  The profiles for several galaxies show some of the underlying morphology.  \mcg{} (Fig.~\ref{fig:sbpro}{\em a}) has two distinct components, an inner bright bulge region which is prominent in the optical image, and a surrounding low surface brightness disk (almost 4~mag fainter in surface brightness).  IC\,4212 (Fig.~\ref{fig:sbpro}{\em h}) appears to have a small central bulge, and the effect of the spiral arms are evident in the bumps of the profile.  ESO\,572-G009 (Fig.~\ref{fig:sbpro}{\em f}) also seems to have a small central bulge, which is seen in the image (Fig.~\ref{fig:starsub}{\em f}).  ESO\,348-G009 and ESO\,428-G033 (Figs.~\ref{fig:sbpro}{\em d} and {\em i}) exhibit pure exponential disks, while ESO\,473-G024 (Fig.~\ref{fig:sbpro}{\em b}) is similar but with a flatter central region.

The various light profiles of ESO\,505-G007 (Fig.~\ref{fig:sbpro}{\em g}) are flat in the central region as the luminosity weighted centre of this galaxy is a faint region between many denser sites that are probably star formation regions.  The optical structure of this galaxy (see Fig.~\ref{fig:starsub}{\em g}) possibly reflects a highly disturbed galaxy with increased star formation due to interaction with the neighboring galaxy ESO\,505-G008 (the source which contaminates its \hi{} spectrum).  The new galaxy \atg{} notably has a much steeper exponential profile than its companion ESO\,121-G020, as well as a higher central surface brightness.  This denser structure may be the result of recent star formation (there is a bright region in the centre of the galaxy which could be a star formation site), and may have lead to it previously being missed in optical surveys if it were mistaken for a star \citep[it is falsely identified as a star in the USNO star catalogue,][]{mon03}.

For only two galaxies, ESO\,473-G024 and \esoq{}, did we find {\em B} band apparent magnitudes that are in agreement with the values quoted in LEDA (even within their sometimes large error bars).  In the case of all seven other galaxies our value of \mB{} was brighter than that given in LEDA, sometimes by several magnitudes.  The effect this has on the \hi{} mass-to-light ratios and why this occurred so often in our sample will be discussed in the following sections (\S\S~\ref{sec:dis-comp} and \ref{sec:dis-slide}).

\section{Discussion of Individual Galaxies}
\label{sec:dis-indi}

\subsection{Comparing Optical and \hi{} Properties}
\label{sec:dis-comp}

Now that we have more accurate optical and \hi{} measurements we can recalculate many of the physical properties of the galaxies, including \mlr{}.  Table~\ref{tab:summary} summarises new parameters from our observations.  The columns are as follows: (1) galaxy name; (2) {\em B} band absolute magnitude; (3) {\em B} band luminosity; (4) \hi{} mass; (5) \hi{} mass-to-{\em B} band luminosity ratio (\mlr{}); (6) total dynamical mass; (7) \hi{} mass-to-total mass ratio; and (8) total mass-to-{\em B} band luminosity ratio.  The latter three values (\Mtot{}, \mtmr{}, and \tmlr{}) are only given for the four galaxies where we fit rotation curves.

\begin{deluxetable}{lccccccc} 
\tabletypesize{\scriptsize}
\tablecaption{Summary of Derived Galaxy Properties from ATCA and 2.3m Data.
  \label{tab:summary}}
\tablewidth{0pt}
\tablehead{ \colhead{Name} & \colhead{\MB{}} & \colhead{\LB{}} & \colhead{\MHI{}} & \colhead{\mlr{}} & \colhead{\Mtot{}} & \colhead{\mtmr{}} & \colhead{\tmlr{}} \\

   \colhead{}  & \colhead{(mag)} & \colhead{($\times 10^{7}$\Lsun{})} & \colhead{($\times 10^{7}$\Msun{})} & \colhead{(\mmls{})} & \colhead{($\times 10^{9}$\Msun{})} & \colhead{} & \colhead{(\mmls{})} \\
 
 \colhead{(1)} & \colhead{(2)} & \colhead{(3)} & \colhead{(4)} & \colhead{(5)} & \colhead{(6)} & \colhead{(7)} & \colhead{(8)}}
\startdata
 \mcg{}        & $-14.65 \pm 0.06$ & $11.3 \pm 0.6$ & $34.5 \pm 1.3$ & $3.0 \pm 0.3$ \\ \\

 ESO\,473-G024 & $-13.10 \pm 0.06$ &  $2.7 \pm 0.1$ &  $7.8 \pm 1.2$ & $2.8 \pm 0.6$ \\ \\

 ESO\,121-G020 & $-12.97 \pm 0.06$ & $2.39 \pm 0.13$ & $3.61 \pm 0.12$ & $1.50 \pm 0.13$ & $0.16 \pm 0.05$ & $0.23 \pm 0.11$ & $7 \pm 4$ \\
 ~~\atg{}      & $-11.23 \pm 0.07$ & $0.48 \pm 0.03$ & $1.07 \pm 0.08$ & $2.2 \pm 0.3$  \\ \\

 ESO\,428-G033 &  $-16.7 \pm 0.2$  &   $28 \pm 5$   &  $130 \pm 3$   & $4.5 \pm 0.9$ & $110^{+190}_{-70}$ & $0.012^{+0.020}_{-0.008}$ & $\sim380$ \\ \\

 \esoq{}       &  $-12.9 \pm 0.2$  &  $2.3 \pm 0.4$ & $50.8 \pm 1.7$ &  $22 \pm 4$   & $4.5 \pm 1.6$ & $0.11 \pm 0.04$ & $200 \pm 110$ \\ \\

 ESO\,572-G009 & $-14.92 \pm 0.06$ & $14.5 \pm 0.8$ &   $71 \pm 13$  &  $4.8 \pm 1.1$ \\ \\

 ESO\,505-G007 & $-17.75 \pm 0.08$ &  $196 \pm 14$  &  $210 \pm 30$   & $1.1 \pm 0.2$ \\ \\

 IC\,4212      & $-17.36 \pm 0.05$ &  $137 \pm 6$   &  $356 \pm 8$   & $2.55 \pm 0.17$ \\ \\

 ESO\,348-G009 & $-14.87 \pm 0.07$ & $13.8 \pm 0.9$ & $21.8 \pm 0.5$ & $1.56 \pm 0.16$ & $3.8 \pm 1.1$ & $0.06 \pm 0.02$ & $27 \pm 10$ \\
\enddata
\end{deluxetable}

\subsubsection{\mcg{}}

Our {\em B} band apparent magnitude for this galaxy is $\sim0.5$~mag brighter than that expressed in LEDA, which was based on the APM plate scan results of \citet{mad90}.  The galaxy has a distinctive optical morphology (Fig.~\ref{fig:starsub}{\em a}).  A shallow CCD image shows only a small, bright, centrally concentrated circular region like a BCD.  However, deeper observations reveal a second, low surface brightness disk component extending well beyond this central bulge, which is clearly visible in our surface brightness profile for this galaxy (Fig.~\ref{fig:sbpro}{\em a}).  Difficulty in measuring the full extent of this faint disk could account for the disagreement between our measurement and the previous result of \citet{mad90}.  The \hi{} flux density result on the other hand is in excellent agreement with the BGC and the observations of \citet{fou90}, suggesting we have recovered almost all the Parkes flux density.  The result of our combined \hi{} and optical values is that the \hi{} mass-to-light ratio drops moderately to $3.0 \pm 0.3$\mls{}, still higher than ``typical'' late-type galaxies but not unusually so.  Despite the large beam for the \mcg{} observations, we can clearly see the \hi{} envelope is extended in a North-South direction.  This is in the same direction as the outer low surface brightness optical disk, and the two correlate quite well in both shape and extent.

\subsubsection{ESO\,473-G024}

Of the nine sample galaxies, ESO\,473-G024 had the best agreement with the {\em B} band apparent magnitude quoted by LEDA, also agreeing with the results in \citet{lau89} and \citet{mad90}.  Our \hi{} flux density is in agreement with both the BGC and \citet{fou90}.  Consequently, \mlr{} remained at a moderate ratio of $2.8 \pm 0.6$\mls{}, one of the higher ratios of this sample after accurate measurements.  It is a well studied dwarf irregular from the Sculptor group, and the \mlr{} has been noted before \citep{ski03a,ski03b}.  In the optical the galaxy is extended in the North-South direction, and has a number of potential star formation regions (see Fig.~\ref{fig:starsub}{\em b}), some of which were studied by \citet[][ see also \S~\ref{sec:dis-implicat}]{ski03a,ski03b}.

\subsubsection{ESO\,121-G020 and \atg{}}

The unexpected discovery of the companion galaxy \atg{} to the South East of ESO\,121-G020 must affect the \mlr{} previously calculated for this galaxy, as the BGC \hi{} flux density measurement for HIPASS\,J0615--57 includes the \hi{} emission from both galaxies.  But, more importantly, our optical {\em B} magnitude determined here is $\sim1.7$~mag brighter than the value in LEDA, although it is closer to the \citet{lau89} total magnitude ($15.85 \pm 0.09$~mag).  The new ratio for the combined system is $1.62 \pm 0.18$\mls{}, and for the individual galaxies it is $1.50 \pm 0.13$\mls{} for ESO\,121-G020 and $2.2 \pm 0.3$\mls{} for \atg{}.  Down to our sensitivity limits there is neither an \hi{} nor a stellar bridge between the two galaxies detected.

\subsubsection{ESO\,428-G033}
\label{sec:dis-compd}

At only 6\fdg9 from the Galactic Plane, ESO\,428-G033 suffers from relatively high Galactic extinction \citep[$1.10 \pm 0.18$~mag in {\em B},][]{sch98} and foreground star contamination (see Fig.~\ref{fig:optimage}{\em d}).  Our {\em B} band magnitude of $16.90 \pm 0.10$~mag is $\sim1$~mag brighter than the LEDA value (which has a high uncertainty of 0.5~mag), but agrees with the \citet{lau89} measurement ($16.83 \pm 0.09$~mag).  Despite this, with our new data the galaxy still has a high \hi{} mass-to-light ratio of $4.5 \pm 0.9$\mls{}, which is in good agreement with the results of \citet{kra92}.  The optical morphology of the galaxy is difficult to determine due to the high foreground star obscuration.  We can only discern an elongated stellar feature (see Fig.~\ref{fig:starsub}{\em d}), which could be a galaxy disk.  Alternatively, this could be a central bar, while the fainter spiral arms remain undetected.  The \hi{} distribution of ESO\,428-G033 is nearly circular, indicating a disk that is close to face on, which favors the optical feature being a bar rather than a disk.  The position angle of velocity field is aligned with the optical emission.

\subsubsection{\esoq{}}

This low surface brightness dwarf irregular galaxy was discussed in detail in \pI{}, where we confirmed that it did have the unusually high \mlr{} as initially suggested by the combination of the BGC results and the magnitudes listed in LEDA.  It is included here for comparison with the other galaxies and we refer the reader to the previous work for more details on it.  After obtaining new estimates of \mB{} and \FHI{} for all our sample galaxies, \esoq{} remains as the stand out galaxy with \mlr{} = $22 \pm 4$\mls{}.  To our knowledge it has one of the highest \hi{} mass-to-light ratios that has been confirmed by accurate measurement to date for any galaxy system, being approximately double the ratio of the best example in the literature DDO\,154 \citep{car89}, and about four times the ratio of our next highest sample objects (ESO\,572-G009 and ESO\,428-G033).  Like DDO\,154 and another known high \mlr{} galaxy NGC\,3741 \citep{beg05}, \esoq{} has a highly extended \hi{} envelope, over six times the optical Holmberg radius.  \citet{beg05} also point out that all three galaxies are isolated and have low tidal indexes from nearby galaxies.  The nearest neighbor we can identify to \esoq{} is approximately 1.7 Mpc away in the Centaurus A Group.

\subsubsection{ESO\,572-G009}

Although our \hi{} flux density was in good agreement with the BGC and \citet{fou90}, our apparent {\em B} magnitude is about half a magnitude brighter than that given in LEDA and that of \citet{lau89}.  Despite this correction ESO\,572-G009 remains one of the few galaxies in our sample to retain a high \mlr{} at $4.8 \pm 1.1$\mls{}.  Morphologically it is a faint low surface brightness galaxy and appears to have two stellar components, a brighter cigar shaped central region (likely to be a central bar) surrounded by a fainter disk that extends only a short distance (see Fig.~\ref{fig:starsub}{\em f}).

\subsubsection{ESO\,505-G007}

A previous measurement of the integrated \hi{} flux density taken on the Effelsberg Radio Telescope by \citet[ $19.3 \pm 2.2$\jks]{ric87} agrees with both the BGC's and our ATCA results.  However, literature \mB{} measurements vary wildly \citep[both $17.64 \pm 0.09$~mag and $15.95 \pm 0.09$~mag from][ dependent on the isophotal level it is measured to]{lau89}.  Our measurement is $\sim$3.5~mag brighter than that listed in LEDA (corresponding to a 25 times more luminous).  The severe underestimate of the {\em B} band apparent magnitude quoted by LEDA for this galaxy has a dramatic effect on the \hi{} mass-to-light ratio.  From the highest ratio of the 789 BGC galaxies with {\em B} magnitudes in LEDA, \mlr{} for ESO\,505-G007 has dropped down to a more typical ratio of $1.18 \pm 0.12$\mls{}, the lowest of our sample galaxies.  We note that optically ESO\,505-G007 has an unusual irregular morphology (see Fig.~\ref{fig:starsub}{\em g}), with several large clumpy structures and some ragged spiral arm-like features, which could suggest recent disruption and star formation triggered by interaction with ESO\,505-G008.

\subsubsection{IC\,4212}

IC\,4212 is an unusual galaxy to have in our sample since its optical morphology is that of a face on spiral rather than the dwarf irregular we might expect for the magnitude given in LEDA.  It has two bright, loosely wound arms, several fainter arms and a small bright central bar (see Fig.~\ref{fig:starsub}{\em h}).  The uncertainty in the apparent magnitude given in LEDA of $\pm 1.1$~mag is quite large.  Our optical measurements find that IC\,4212 is a much brighter galaxy than suggested by the \mB{} value LEDA lists (over two magnitudes).  This means that the \mlr{} is not the extreme value initally suggested (the third highest ratio of the 789 BGC galaxies).  However, IC\,4212 retains at a ratio of $2.55 \pm 0.17$\mls{}, which relatively high for a galaxy with such a distinct spiral structure, typical values for late type spirals being less than 1\mls{} \citep{rob94}.  We measure a deconvolved \hi{} diameter of $\sim$200\arcsec{}, suggesting it may extend significantly beyond the optical disk.

\subsubsection{ESO\,348-G009}

The apparent {\em B} magnitude we measured for ESO\,348-G009 was again brighter than the value listed in LEDA, while the \hi{} flux density is consistent with the BGC result.  This means we end up with an \hi{} mass-to-light ratio of $1.56 \pm 0.16$\mls{}, down from the preliminary value of $9 \pm 7$\mls{}.  The optical image shows an edge on disk galaxy extending East-West, with some clumpy structures visible along the length of the disk (see Fig.~\ref{fig:starsub}{\em i}).  The position angle of the \hi{} velocity field is aligned closely with the stellar disk.

\subsection{Distance Uncertainties}
\label{sec:dis-dist}

As was discussed in \pI{}, the use of distances calculated from Local Group velocities for galaxies in the Local Universe can be problematic as the peculiar velocities in nearby groups are potentially of similar order to the redshifts themselves, and the local Hubble flow can differ from the cosmological expansion.  Evidence from studies on the Sculptor group \citep{jer98} suggests that the local velocity-distance relationship is much steeper in the direction of this group than for galaxies further out due to the probable gravitational influence of the Local Group.  If we use the Hubble constant of \citet[ H$_0$ = 119\kms\,Mpc$^{-1}$]{jer98} for the three Sculptor Group members in our sample \citep[membership confirmed using][]{cot97} then the Local Group velocity distances to these galaxies would be less than listed in Table~\ref{tab:prop}, with 6.0 Mpc for \mcg{}, 4.8 Mpc for ESO\,473-G024, and 5.3 Mpc for ESO\,348-G009.  This would put all three on the far side of the Sculptor group \citep{jer98}, and would mean that the distance dependent quantities in Table~\ref{tab:summary} (everything except \mlr{}) would require adjustment.

\section{Discussion} 
\label{sec:dis}

\subsection{The revised \hi{} mass-to-light ratios}
\label{sec:dis-slide}

\begin{figure*} 
  \includegraphics[width=8cm]{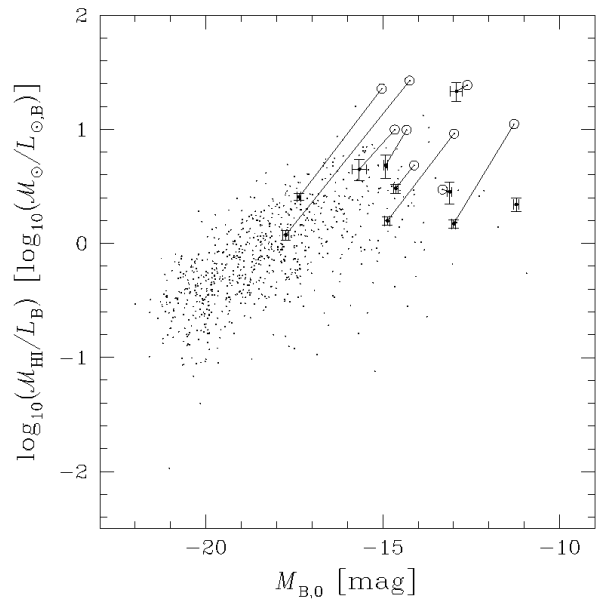}
\caption{\hi{} mass-to-light ratio versus absolute {\em B} magnitude, with the 789 galaxies shown in Fig.~\ref{fig:mlmbt} marked with small points.  The nine galaxies in this paper are shown with open circles (preliminary values as in Fig.~\ref{fig:mlmbt}) and large points with error bars (revised values as derived in this paper).  Lines link the preliminary to the revised values, and the point by itself is \atg{}.
\label{fig:slidenine}}
\end{figure*}

The plot in Fig.~\ref{fig:slidenine} shows \mlr{} versus \MB{} (as in Fig.~\ref{fig:mlmbt}) with the new positions of the nine target galaxies resulting from our observations.  The lines connect our results (large points with error bars) to the initial estimates (open circles).  While the new measurements resulted in lower \mlr{} values for all selected galaxies, the decrease is particularly significant for galaxies which had preliminary values of \mlr{} $> 5$\mls{}.  While disappointing, our result is not too surprising given that we selected galaxies from the BGC with initially the most extreme \mlr{} values and large uncertainties in their optical magnitudes, so we preferentially selected galaxies with underestimated \mB{}.  For many of our target galaxies optical magnitudes exist that agree with our results \citep[e.g.][]{lau89}, but the mean magnitudes available from LEDA were generally highly underestimated.

All but one of the selected galaxies, \esoq{}, now have revised \hi{} mass-to-light ratios in the range $\sim$1-5\mls{}.  Such revisions are by no means uncommon among claims of high \mlr{} galaxies as we discussed in \S~6.2 of \pI{}, and as seen in \citet{vzee97} and \citet{chu02}.  The faint luminosity of these galaxies and the difficulty in getting high quality data in both the optical and radio regime make examples of high \mlr{} galaxies difficult to find.  Only a few other galaxies with confirmed high \mlr{} are in the literature, most notably DDO\,154 \citep[9.4\mls{}][]{car89,hof93}, UGCA\,292 \citep[7.0\mls{}][]{you03}, and NGC\,3741 \citep[5.8\mls{}][]{beg05}.

While we must be careful of small number statistics, the trend of all the findings of current studies strongly suggest that there do not appear to be large numbers of ``dim'' galaxies like \esoq{} in the local Universe.  Therefore high \mlr{} galaxies cannot account for much of the discrepancy between observations and theoretical predictions of low mass galaxy numbers.  But there are other possible ways that galaxies could be missed observationally, and several suggestions how they might be detected.  Some methods have been proposed for finding ``empty'' dark matter halos, such as the suggestion to use the Milky Way halo microlensing statistics to look for dark matter satellite influence \citep{wid98}, or to analyse the gravitational lensing of quasars to determine the dark matter sub-halos of the lensing object \citep{moo99,dal02}.  However, the existence of halos without baryons is still highly speculative.

True ``dark'' galaxies in the form of isolated, rotating, extragalactic \hi{} clouds have so far proven elusive \citep{rya02,kor04,doy05}.  Some isolated \hi{} sources have been found in HIPASS \citep{kil00,ryd01,ryd04} and were interpreted as high velocity clouds or tidal debris by the respective authors.  A recent claim of a ``dark galaxy'' close to the one-armed spiral NGC\,4254, in the outskirts of the Virgo Cluster \citep{min05} also appears to be tidal debris \citep{bek05}; see also \citep{oos05}.  \citet{tay05} have discussed theoretically that an isolated \hi{} cloud which formed without a stellar component is likely to be unstable to star formation, and therefore would not remain dark.

\subsection{The importance of dust extinction} 
\label{sec:dis-extinct}

As well as the accuracy of the {\em B} band photometry and \hi{} flux density it is also important to discuss the one contributor to \mlr{} that is beyond the scope of our observations, the dust extinction due to both our Galaxy (see Table~\ref{tab:opt}) and the host galaxy (``internal'' extinction).  As we noted in \pI{} Galactic extinction is particularly important for \esoq{} due to its sky position only 10\fdg5 from the Galactic Plane.  ESO\,428-G033 at $b$ = -6\fdg9 is the only other galaxy that is subject to similar Galactic extinction (\AB{} = $1.10 \pm 0.18$~mag, SFD98). 

\begin{figure*} 
\begin{tabular}{cc}
 \includegraphics[width=8cm]{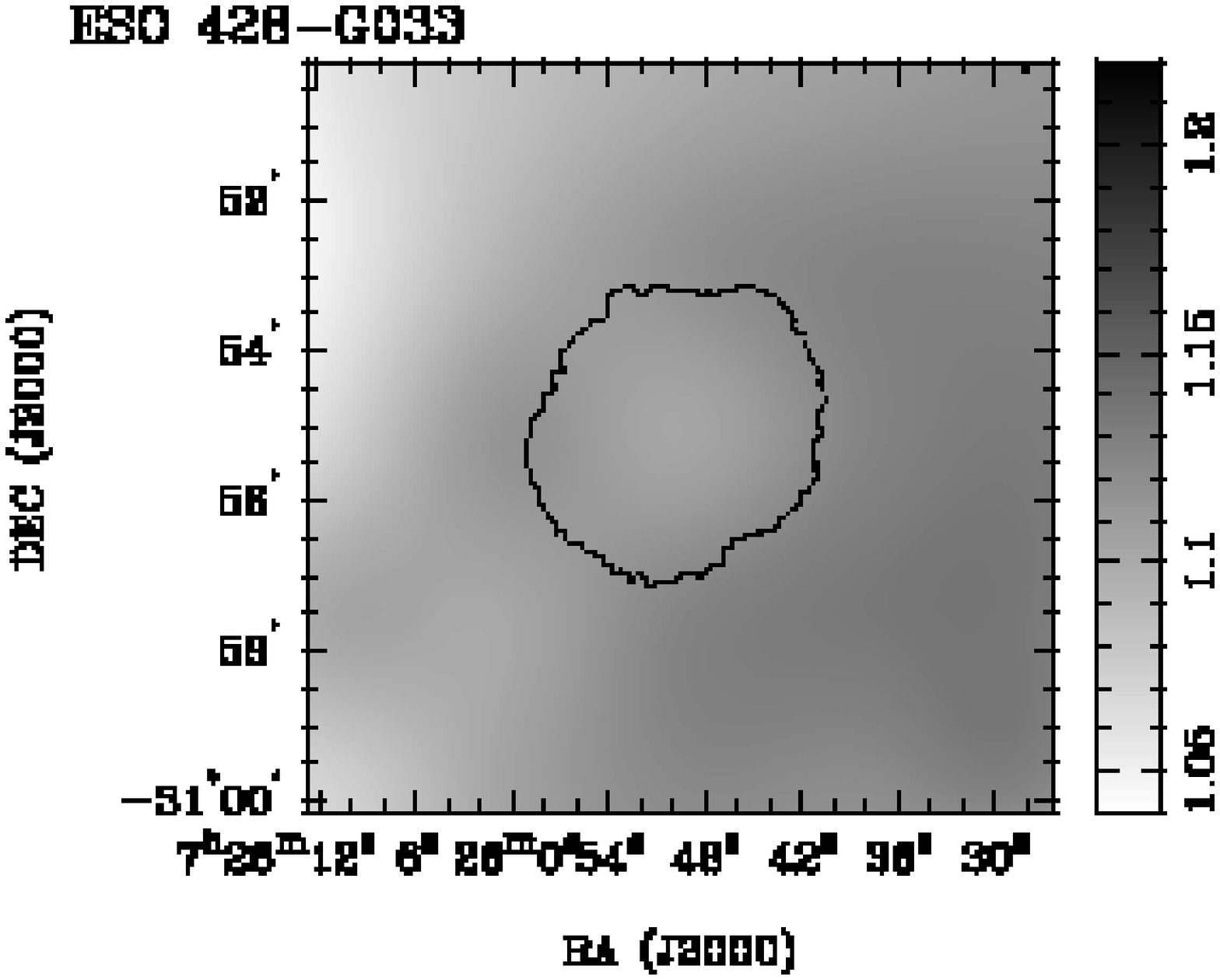} &
 \includegraphics[width=8cm]{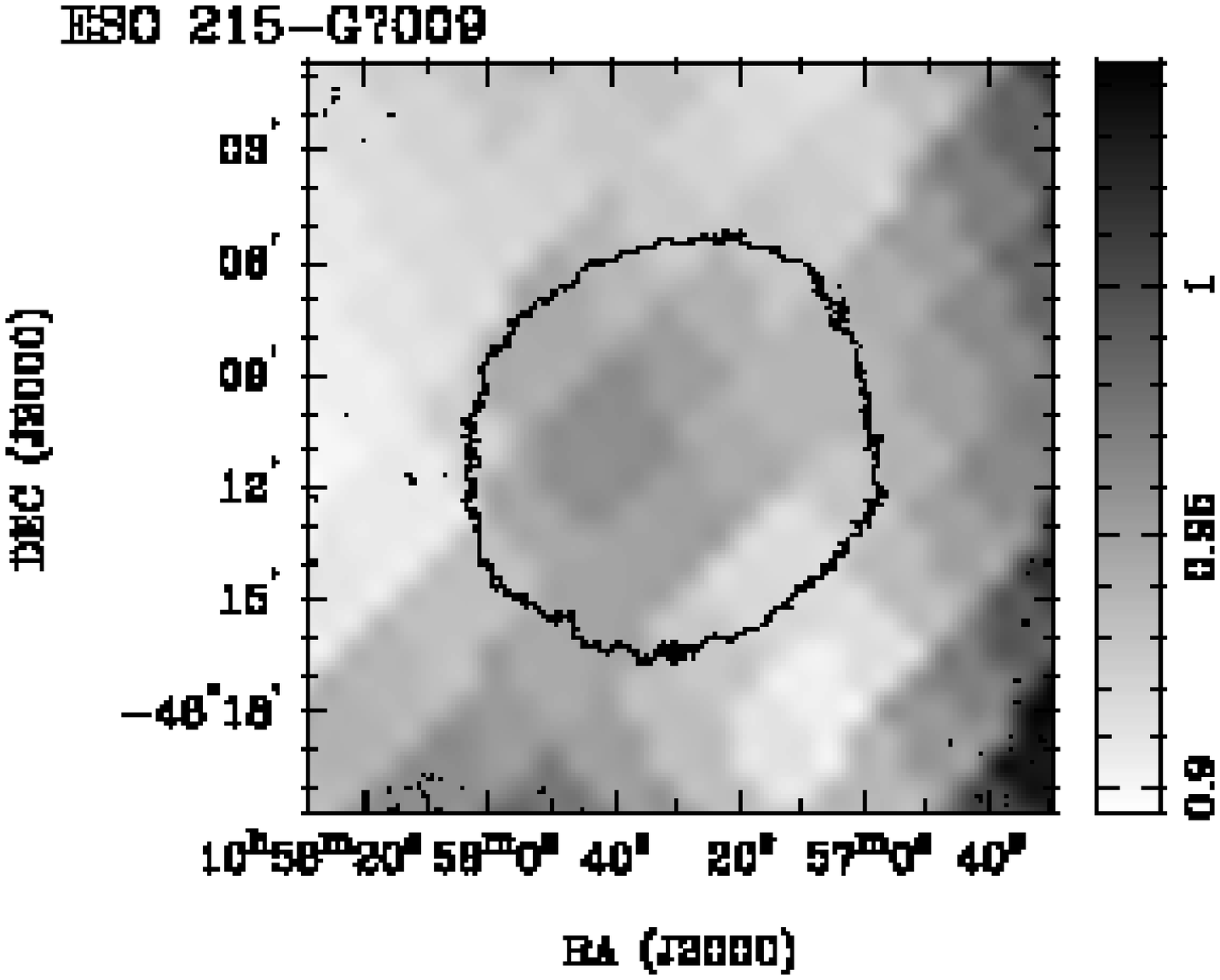} \\
\end{tabular}
\caption{The \citet{sch98} Galactic extinction maps for ESO\,428-G033 (left) and \esoq{} (right) in \AB{} (units of magnitudes).  The contours marked on each image correspond to the lowest \hi{} contours for each galaxy as shown in Fig.~\ref{fig:himap}.
\label{fig:schlegel}}
\end{figure*}

The uncertainties in the SFD98 Galactic extinction grow proportionally to the value.  This means that it only contributes a significant fraction to the total error in \mlr{} for \esoq{} and ESO\,428-G033.  Close to the Galactic Plane the dust distribution can be patchy.  Fig.~\ref{fig:schlegel} shows the SFD98 dust extinction maps in terms of \AB{} for the regions around ESO\,428-G033 and \esoq{}, with the last \hi{} contour from Fig.~\ref{fig:himap} superimposed for reference.  Both maps show relatively low variation in the amount of extinction over the field around the galaxy, less than the uncertainty in \AB{} in both cases.  This suggests that the SFD98 value at the position of both galaxies is an accurate representation of the true Galactic extinction.  ESO\,428-G033 was studied by \citet{kra92} as part of an investigation of galaxies in a region of the Galactic Plane with reduced dust extinction.  Their results are in agreement with our measurements, and in general they found that the properties of their sample objects were typical of other samples of nearby galaxies despite the relatively high extinction.  It is worth noting that even if the Galactic extinction was much higher at the position of \esoq{}, say by 0.5~mag, the galaxy would still have a very high \mlr{} of $14 \pm 4$\mls{}.

While we can at least get a relatively accurate estimate of Galactic extinction, dust extinction internal to the host galaxy itself is much harder to quantify.  Dust extinction in late-type galaxies is poorly understood but thought to be lower than in early-type spirals due to low metalicity and ineffective dust accretion processes \citep{dwe98,hir99}.  We would expect a disk galaxy to have significant extinction when viewed close to edge on, like ESO\,348-G009 ($i = 80$\degr{}$\pm 5$\degr{}, see Table~\ref{tab:rot}), due to the geometry of the dust distribution.  We discussed \esoq{}'s possible internal extinction in \pI{}, concluding that it is most likely low for this close to face on galaxy.  The galaxy ESO\,121-G020 may have a high inclination angle of $78$\degr{}$\pm 5$\degr{} (Table~\ref{tab:rot}), in contrast to the optical dimensions, and a moderate amount of internal extinction.  The optical emission of the galaxy ESO\,428-G033 is highly obscured by foreground stars.  Due to the current inability to estimate internal extinction we have not accounted for it in our \mlr{} calculations.

\subsection{Physical Characteristics of High \mlr{} Galaxies Implications for Their Existence}
\label{sec:dis-implicat}

Several common elements between the galaxies with the most extreme \mlr{}'s are becoming more evident.  The three best examples, \esoq{} ($22 \pm 4$\mls{}, \pI{}), DDO\,154 \citep[9.4\mls{},][]{car89,hof93}, and NGC\,3741 \citep[5.8\mls{},][]{beg05}, all have \hi{} envelopes that are 5 to 8 times the optical Holmberg radius, as IC\,4212 may also have.  This may be because the gas is at a low density and is in a stable state, as was seen for \esoq{} in \pI{}.  All these galaxies also have low tidal indexes \citep{kar04}, indicating {\em they are isolated in space} and have little external stimulation to form stars.  Despite having very low stellar content for their dynamical masses, the baryonic masses of the galaxies are always of the order of $\sim10\%$ of the total dynamical mass, which is a typical fraction seen in galaxies from $L_{*}$ to the dwarf regime \citep{beg05}.  This suggests {\em that high \mlr{} galaxies are not lacking the baryons to create stars, but are underluminous as they lack either the internal or external stimulation for further star formation}.

\citet{ski03b} obtained spectra of \hii{} regions within five Sculptor group galaxies including two in our sample, ESO\,473-G024 and ESO\,348-G009.  For ESO\,473-G024, for which we found a moderately high \mlr{}, they were able to produce oxygen and nitrogen abundances.  The oxygen abundance indicated a low metalicity which is typical of other late type galaxies.  Normally dwarf galaxies with a similar low metalicity have low nitrogen to oxygen ratios in a narrow range around an average of $\log{N/O} \simeq -1.6$ \citep{izo99}.  It is thought this is because these galaxies are undergoing their first burst of star formation, and that the nitrogen from this burst has not yet had time to dissipate into the ISM \citep[coming from type {\sc ii} supernovae of intermediate mass stars, while oxygen comes from higher mass stars, see][]{ski03b}.  However, \citet{ski03b} found that the N/O ratio for ESO\,473-G024 was relatively high.  Most importantly for our study, they compared these results to a study of the high \mlr{} galaxy DDO\,154 by \citet{ken01} which shows the same trend for N/O.  In both galaxies nitrogen from any past star formation events has had time to disperse and there is no current burst of star formation to reduce the N/O ratio, neither galaxy having a particularly high current star formation rate \citep[][ respectively]{ski03a,ken01} and like \esoq{} they may be considered quiescent galaxies.  In fact, their current SFR and luminosity \citep[which can be used as a rough estimate of the average past star formation rate,][]{tin80} are similar to what was found for \esoq{} in \pI{}.  This suggests that {\em high \hi{} mass-to-light ratio galaxies may have been forming stars at a low, constant rate over their lifetimes.}  In order to understand further the significance of this result we would need to expand this study to spectra of \hii{} regions in other high \mlr{} galaxies (especially \esoq{}) and look at other ratios which may indicate the timing of star formation events.  Metalicity may also be an important element in determining the fraction of baryons which remain in gas form \citep{tay05}.

A close look at the plots of \mlr{} versus \MB{} (Figs.~\ref{fig:mlmbt} and \ref{fig:slidenine}) shows that there is potentially an upper envelope to a galaxy's \hi{} mass-to-light ratio at a given luminosity.  Low luminosity galaxies appear to be able to have a higher portion of their detectable baryons in the form of neutral hydrogen than galaxies around $L_{*}$, where the baryonic mass is dominated by stars even for the most gas rich galaxies.  What this suggests is that {\em there is a minimum quantity of stars a galaxy will form that goes as a function of initial baryonic mass}.  Support for this idea can also be found in the theoretical work of \citet{tay05}.  Whether or not a galaxy forms more than this minimum is likely to be a function of such factors as environment and the galaxy's dark matter properties.

\citet{tay05} developed models to determine whether a neutral gas disk without stars (a ``dark galaxy'') could remain dynamically stable or if some gas will collapse and form stars.  They found that without an internal radiation field the majority of the gas in the disk will become gravothermally unstable, even for galaxies with very low baryonic masses (down to $5\times10^{6}$\Msun{}).  They also found that the fraction of unstable gas decreases as the baryonic mass decreases.  This may provide an explanation to why we see the slope in the upper envelope for \mlr{}, the lower mass galaxies only having to convert a much smaller fraction of their baryons to stars in order to become stable.  Galaxies such as \esoq{} and DDO\,154 are close to our upper envelope, and may define the extreme cases of galaxies which have formed only the minimum number of star required in order to remain stable and have not experienced any other events which may trigger star formation.  In this way they might be used to distinguish between various models for galaxy collapse by defining the minimum star formation required for stability.  We will further explore this possibility, and other properties that vary with \mlr{}, with a larger sample of 37 late type dwarf galaxies in an upcoming paper.

\section{Conclusions}
\label{sec:conc}

We obtained accurate optical CCD apparent magnitudes and \hi{} flux densities for nine late type dwarf galaxies and recalculated their \hi{} mass-to-light ratios.  The new \mlr{} values are:
\begin{itemize}
\item $22 \pm 4$\mls{} for \esoq{}, 
\item $4.8 \pm 1.1$\mls{} for ESO\,572-G009, 
\item $4.5 \pm 0.9$\mls{} for ESO\,428-G033, 
\item $3.0 \pm 0.3$\mls{} for \mcg{}, 
\item $2.8 \pm 0.6$\mls{} for ESO\,473-G024, 
\item $2.6 \pm 0.2$\mls{} for IC\,4212, 
\item $1.6 \pm 0.2$\mls{} for ESO\,348-G009, 
\item $1.5 \pm 0.1$\mls{} for ESO\,121-G020, and 
\item $1.2 \pm 0.1$\mls{} for ESO\,505-G007. 
\end{itemize}
Many of these \hi{} mass-to-light ratios are significantly below the initial estimates, due to inaccurate magnitude estimates in the literature. This strongly emphasises the importance of having accurate observations in both the \hi{} line and the optical.  Based on the new \hi{} mass-to-light ratio distribution we conclude that genuine ``dim'' galaxies with high ratios (\mlr{}$>$5\mls{}) are rare in the local Universe.

A previously uncatalogued companion galaxy to ESO\,121-G020 was found at a projected distance of 3\arcmin{}.  \atg{} has an \hi{} mass of $\sim10^{7}$\Msun{} and \mlr{} of $2.2 \pm 0.3$\mls{}.  This was the only such companion detected, and is well within the beam of the Multibeam instrument used by the HIPASS survey.  Despite our low resolution \hi{} observations we were able to separate the galaxy ESO\,505-G007 from its close neighbor ESO\,505-G008 and determined \hi{} flux densities of $21 \pm 3$ and $8 \pm 3$\jks{}, respectively.

The best examples of high \mlr{} dwarf galaxies in the literature all have highly extended \hi{} disks, are spatially isolated and have normal baryonic content for their dynamical masses.  The galaxies are not lacking the baryons to create stars, but are underluminous as they lack either the internal or external stimulation for further star formation.  Future examination of element abundances within star formation sites of high \mlr{} galaxies may reveal important clues about their star formation history.  Recent observations \citep{ski03b,ken01} support the idea that high \mlr{} galaxies may have been forming stars at a low, constant rate over their lifetimes as proposed in \pI{}.  There may be a minimum quantity of stars a galaxy will form that depends on the initial baryonic mass, which is supported by the theoretical work of \citet{tay05}.  If this is true then maybe high \hi{} mass-to-light ratio galaxies have over their lifetimes only formed the minimum number of stars necessary to maintain the stability of their \hi{} gas disk.

\section*{Acknowledgments}

We are grateful for the assistance of Ken Freeman and Lister Staveley-Smith in this project, especially for their assistance with observations.  We would like to thank Erwin de Blok for his help with various aspects of the \hi{} data reduction and interpretation.  We would also like to thank Marilena Salvo and Gayandhi de Silva for their observing assistance.  Our thanks also go to the anonymous referee for their useful comments, especially regarding the \hi{} spectra.  The 2.3-meter Telescope is run by the Australian National University as part of Research School of Astronomy and Astrophysics.  The Australia Telescope Compact Array and the Parkes Radio Telescope are part of the Australia Telescope that is funded by the Commonwealth of Australia for operation as a National Facility managed by CSIRO.  This research has made use of the NASA/IPAC Extragalactic Database (NED), which is operated by the Jet Propulsion Laboratory, California Institute of Technology, under contract with the National Aeronautics and Space Administration.  The Digitized Sky Survey (DSS) was produced at the Space Telescope Science Institute under U.S. Government grant NAG W-2166, based on photographic data obtained using the UK Schmidt Telescope.

\clearpage

\end{document}